\pdfoutput=1
\documentclass[12pt,a4paper]{article}
\usepackage{ifthen}
\newboolean{pdflatex}
\setboolean{pdflatex}{true}
\newboolean{articletitles}
\setboolean{articletitles}{true}
\newboolean{uprightparticles}
\setboolean{uprightparticles}{false}
\def\paperauthors{LHCb collaboration}
\def\paperasciititle{Angular analysis of the decay B+ -> K*+ mu+ mu- decay}
\def\papertitle{Angular analysis\\ of the $B^{+}\rightarrow K^{\ast+}\mup\mun$ decay}
\def\paperkeywords{{High Energy Physics}, {LHCb}}
\def\papercopyright{\the\year\ CERN for the benefit of the LHCb collaboration}
\def\paperlicence{CC BY 4.0 licence}
\def\paperlicenceurl{https://creativecommons.org/licenses/by/4.0/}

\usepackage[top=1in, bottom=1.25in, left=1in, right=1in]{geometry}

\usepackage{microtype}
\usepackage{lineno}  % for line numbering during review
\usepackage{xspace} % To avoid problems with missing or double spaces after
\usepackage{caption} %these three command get the figure and table captions automatically small

\usepackage{graphicx}  % to include figures (can also use other packages)
\usepackage{color}
\usepackage{colortbl}
\graphicspath{{./figs/}} % Make Latex search fig subdir for figures
\usepackage{amsmath} % Adds a large collection of math symbols
\usepackage{amssymb}
\usepackage{amsfonts}
\usepackage{upgreek} % Adds in support for greek letters in roman typeset

\newcommand*\patchAmsMathEnvironmentForLineno[1]{%
\expandafter\let\csname old#1\expandafter\endcsname\csname #1\endcsname
\expandafter\let\csname oldend#1\expandafter\endcsname\csname
end#1\endcsname
 \renewenvironment{#1}%
   {\linenomath\csname old#1\endcsname}%
   {\csname oldend#1\endcsname\endlinenomath}%
}
\newcommand*\patchBothAmsMathEnvironmentsForLineno[1]{%
  \patchAmsMathEnvironmentForLineno{#1}%
  \patchAmsMathEnvironmentForLineno{#1*}%
}
\AtBeginDocument{%
\patchBothAmsMathEnvironmentsForLineno{equation}%
\patchBothAmsMathEnvironmentsForLineno{align}%
\patchBothAmsMathEnvironmentsForLineno{flalign}%
\patchBothAmsMathEnvironmentsForLineno{alignat}%
\patchBothAmsMathEnvironmentsForLineno{gather}%
\patchBothAmsMathEnvironmentsForLineno{multline}%
\patchBothAmsMathEnvironmentsForLineno{eqnarray}%
}

\usepackage{hyperxmp}

\usepackage[pdftex,
            pdfauthor={\paperauthors},
            pdftitle={\paperasciititle},
            pdfkeywords={\paperkeywords},
            pdfcopyright={Copyright (C) \papercopyright},
            pdflicenseurl={\paperlicenceurl}]{hyperref}
\usepackage[colorinlistoftodos,textsize=scriptsize]{todonotes}

\usepackage[bottom,flushmargin,hang,multiple]{footmisc}

\usepackage[all]{hypcap} % Internal hyperlinks to floats.

\usepackage{xspace} 
\usepackage{upgreek}

\def\lhcb   {\mbox{LHCb}\xspace}

\def\babar  {\mbox{BaBar}\xspace}
\def\belle  {\mbox{Belle}\xspace}

\def\lhc    {\mbox{LHC}\xspace}

\def\MagUp {\mbox{\em Mag\kern -0.05em Up}\xspace}

\ifthenelse{\boolean{uprightparticles}}%
{

 \def\Pmu         {\ensuremath{\upmu}\xspace}

 \def\Ppi         {\ensuremath{\uppi}\xspace}

 \def\Ppsi        {\ensuremath{\uppsi}\xspace}

 \def\PDelta      {\ensuremath{\Delta}\xspace}                 
 \def\PXi         {\ensuremath{\Xi}\xspace}                 
 \def\PLambda     {\ensuremath{\Lambda}\xspace}                 
 \def\PSigma      {\ensuremath{\Sigma}\xspace}                 
 \def\POmega      {\ensuremath{\Omega}\xspace}                 
 \def\PUpsilon    {\ensuremath{\Upsilon}\xspace}

 \def\PB      {\ensuremath{\mathrm{B}}\xspace}                 
                  
 \def\PD      {\ensuremath{\mathrm{D}}\xspace}

 \def\PJ      {\ensuremath{\mathrm{J}}\xspace}                 
 \def\PK      {\ensuremath{\mathrm{K}}\xspace}

 \def\PW      {\ensuremath{\mathrm{W}}\xspace}

 \def\Pb      {\ensuremath{\mathrm{b}}\xspace}                 
 \def\Pc      {\ensuremath{\mathrm{c}}\xspace}

 \def\Pi      {\ensuremath{\mathrm{i}}\xspace}

 \def\Pp      {\ensuremath{\mathrm{p}}\xspace}

 \def\Ps      {\ensuremath{\mathrm{s}}\xspace}                 
                  
 \def\Pu      {\ensuremath{\mathrm{u}}\xspace}

 \def\thebaroffset{0.0em}
}
{

 \def\Pmu         {\ensuremath{\mu}\xspace}

 \def\Ppi         {\ensuremath{\pi}\xspace}

 \def\Ppsi        {\ensuremath{\psi}\xspace}                 
                  
 \mathchardef\PDelta="7101
 \mathchardef\PXi="7104
 \mathchardef\PLambda="7103
 \mathchardef\PSigma="7106
 \mathchardef\POmega="710A
 \mathchardef\PUpsilon="7107
                  
 \def\PB      {\ensuremath{B}\xspace}                 
                  
 \def\PD      {\ensuremath{D}\xspace}

 \def\PJ      {\ensuremath{J}\xspace}                 
 \def\PK      {\ensuremath{K}\xspace}

 \def\PW      {\ensuremath{W}\xspace}

 \def\Pb      {\ensuremath{b}\xspace}                 
 \def\Pc      {\ensuremath{c}\xspace}

 \def\Pi      {\ensuremath{i}\xspace}

 \def\Pp      {\ensuremath{p}\xspace}

 \def\Ps      {\ensuremath{s}\xspace}                 
                  
 \def\Pu      {\ensuremath{u}\xspace}

 \def\thebaroffset{0.18em}
}
\newcommand{\offsetoverline}[2][\thebaroffset]{\kern #1\overline{\kern -#1 #2}}%

\makeatletter
\ifcase \@ptsize \relax% 10pt
  \newcommand{\miniscule}{\@setfontsize\miniscule{4}{5}}% \tiny: 5/6
\or% 11pt
  \newcommand{\miniscule}{\@setfontsize\miniscule{5}{6}}% \tiny: 6/7
\or% 12pt
  \newcommand{\miniscule}{\@setfontsize\miniscule{5}{6}}% \tiny: 6/7
\fi
\makeatother

\DeclareRobustCommand{\optbar}[1]{\shortstack{{\miniscule (\rule[.5ex]{1.25em}{.18mm})}
  \\ [-.7ex] $#1$}}

\def\mup        {{\ensuremath{\Pmu^+}}\xspace}
\def\mun        {{\ensuremath{\Pmu^-}}\xspace} % muon negative (\mum is taken)

\def\mumu       {{\ensuremath{\Pmu^+\Pmu^-}}\xspace}

\def\ellm       {{\ensuremath{\ell^-}}\xspace}
\def\ellp       {{\ensuremath{\ell^+}}\xspace}

\def\Wp     {{\ensuremath{\PW^+}}\xspace}

\def\uquark    {{\ensuremath{\Pu}}\xspace}
\def\uquarkbar {{\ensuremath{\overline \uquark}}\xspace}

\def\squark    {{\ensuremath{\Ps}}\xspace}

\def\cquark    {{\ensuremath{\Pc}}\xspace}

\def\bquark    {{\ensuremath{\Pb}}\xspace}
\def\bquarkbar {{\ensuremath{\overline \bquark}}\xspace}

\def\pion   {{\ensuremath{\Ppi}}\xspace}

\def\pip    {{\ensuremath{\pion^+}}\xspace}
\def\pim    {{\ensuremath{\pion^-}}\xspace}

\def\kaon    {{\ensuremath{\PK}}\xspace}
\def\KorKbar {\kern \thebaroffset\optbar{\kern -\thebaroffset \PK}{}\xspace}

\def\KS      {{\ensuremath{\kaon^0_{\mathrm{S}}}}\xspace}

\def\Kstarz  {{\ensuremath{\kaon^{*0}}}\xspace}

\def\Kstar   {{\ensuremath{\kaon^*}}\xspace}

\def\Kstarp  {{\ensuremath{\kaon^{*+}}}\xspace}
\def\Kstarm  {{\ensuremath{\kaon^{*-}}}\xspace}

\def\D       {{\ensuremath{\PD}}\xspace}

\def\DorDbar {\kern \thebaroffset\optbar{\kern -\thebaroffset \PD}\xspace}

\def\Dp      {{\ensuremath{\D^+}}\xspace}
\def\Dm      {{\ensuremath{\D^-}}\xspace}

\def\DpDm    {\ensuremath{\Dp {\kern -0.16em \Dm}}\xspace}

\def\B       {{\ensuremath{\PB}}\xspace}

\def\BorBbar {\kern \thebaroffset\optbar{\kern -\thebaroffset \PB}\xspace}

\def\Bd      {{\ensuremath{\B^0}}\xspace}

\def\BdorBdbar {\kern \thebaroffset\optbar{\kern -\thebaroffset \Bd}\xspace}
\def\Bu      {{\ensuremath{\B^+}}\xspace}
\def\Bub     {{\ensuremath{\B^-}}\xspace}

\def\Bs      {{\ensuremath{\B^0_\squark}}\xspace}

\def\BsorBsbar {\kern \thebaroffset\optbar{\kern -\thebaroffset \Bs}\xspace}

\def\jpsi     {{\ensuremath{{\PJ\mskip -3mu/\mskip -2mu\Ppsi}}}\xspace}
\def\psitwos  {{\ensuremath{\Ppsi{(2S)}}}\xspace}

\def\Y#1S{\ensuremath{\PUpsilon{(#1S)}}\xspace}

\def\proton      {{\ensuremath{\Pp}}\xspace}

\def\LorLbar     {\kern \thebaroffset\optbar{\kern -\thebaroffset \PLambda}\xspace}

\newcommand{\decay}[2]{\ensuremath{#1\!\to #2}\xspace} 

\def\to                 {\ensuremath{\rightarrow}\xspace}

\def\qsq       {{\ensuremath{q^2}}\xspace}

\def\CP                {{\ensuremath{C\!P}}\xspace}

\def\BdToKstmm    {\decay{\Bd}{\Kstarz\mup\mun}}

\def\AFB      {\ensuremath{A_{\mathrm{FB}}}\xspace}
\def\FL       {\ensuremath{F_{\mathrm{L}}}\xspace}
\def\AT#1     {\ensuremath{A_{\mathrm{T}}^{#1}}\xspace}           % 2

\def\ctl       {\ensuremath{\cos{\theta_\ell}}\xspace}
\def\ctk       {\ensuremath{\cos{\theta_K}}\xspace}

\def\C#1      {\ensuremath{\mathcal{C}_{#1}}\xspace}                       % 9
\def\Cp#1     {\ensuremath{\mathcal{C}_{#1}^{'}}\xspace}                    % 7
\def\Ceff#1   {\ensuremath{\mathcal{C}_{#1}^{\mathrm{(eff)}}}\xspace}        % 9  
\def\Cpeff#1  {\ensuremath{\mathcal{C}_{#1}^{'\mathrm{(eff)}}}\xspace}       % 7
\def\Ope#1    {\ensuremath{\mathcal{O}_{#1}}\xspace}                       % 2
\def\Opep#1   {\ensuremath{\mathcal{O}_{#1}^{'}}\xspace}                    % 7

\newcommand{\nospaceunit}[1]{\ensuremath{\text{#1}}}       
\newcommand{\aunit}[1]{\ensuremath{\text{\,#1}}}       
\newcommand{\unit}[1]{\aunit{#1}\xspace}                   % {kg}   

\newcommand{\tev}{\aunit{Te\kern -0.1em V}\xspace}
\newcommand{\gev}{\aunit{Ge\kern -0.1em V}\xspace}
\newcommand{\mev}{\aunit{Me\kern -0.1em V}\xspace}
\newcommand{\kev}{\aunit{ke\kern -0.1em V}\xspace}
\newcommand{\ev}{\aunit{e\kern -0.1em V}\xspace}
 
\newcommand{\mevc}{\ensuremath{\aunit{Me\kern -0.1em V\!/}c}\xspace}
\newcommand{\gevc}{\ensuremath{\aunit{Ge\kern -0.1em V\!/}c}\xspace}
\newcommand{\mevcc}{\ensuremath{\aunit{Me\kern -0.1em V\!/}c^2}\xspace}
\newcommand{\gevcc}{\ensuremath{\aunit{Ge\kern -0.1em V\!/}c^2}\xspace}
\newcommand{\gevgevcccc}{\ensuremath{\gev^2\!/c^4}\xspace} % for q^2

\def\mum  {\ensuremath{\,\upmu\nospaceunit{m}}\xspace}

\def\fb   {\ensuremath{\aunit{fb}}\xspace}
\def\invfb   {\ensuremath{\fb^{-1}}\xspace}

\def\deriv {\ensuremath{\mathrm{d}}}

\def\gsim{{~\raise.15em\hbox{$>$}\kern-.85em
          \lower.35em\hbox{$\sim$}~}\xspace}
\def\lsim{{~\raise.15em\hbox{$<$}\kern-.85em
          \lower.35em\hbox{$\sim$}~}\xspace}

\def\pt         {\ensuremath{p_{\mathrm{T}}}\xspace}

\def\ptot       {\ensuremath{p}\xspace}

\def\evtgen     {\mbox{\textsc{EvtGen}}\xspace}

\def\geant      {\mbox{\textsc{Geant4}}\xspace}

\def\photos     {\mbox{\textsc{Photos}}\xspace}

\def\pythia     {\mbox{\textsc{Pythia}}\xspace}

\def\tell1  {TELL1\xspace}
\def\ukl1   {UKL1\xspace}

\newcommand{\phz}{\phantom{0}}
\usepackage{cite} % Allows for ranges in citations
\usepackage{mciteplus}
\usepackage{stackengine}
\def\thetal {\ensuremath{\theta_{\ell}}\xspace}
\def\thetak {\ensuremath{\theta_{K}}\xspace}
\def\FL {\ensuremath{F_{\rm L}}\xspace}
\def\FS {\ensuremath{F_{\rm S}}\xspace}
\def\AFB {\ensuremath{A_{\rm FB}}\xspace}

\begin{document}

%%%%%%%%%%%%%%%%%%%%%%%%%
%%%%% Title     %%%%%%%%%
%%%%%%%%%%%%%%%%%%%%%%%%%
\renewcommand{\thefootnote}{\fnsymbol{footnote}}
\setcounter{footnote}{1}
\begin{titlepage}
\pagenumbering{roman}

% Header ---------------------------------------------------
\vspace*{-1.5cm}
\centerline{\large EUROPEAN ORGANIZATION FOR NUCLEAR RESEARCH (CERN)}
\vspace*{1.5cm}
\noindent
\begin{tabular*}{\linewidth}{lc@{\extracolsep{\fill}}r@{\extracolsep{0pt}}}
\ifthenelse{\boolean{pdflatex}}% Logo format choice
{\vspace*{-1.5cm}\mbox{\!\!\!\includegraphics[width=.14\textwidth]{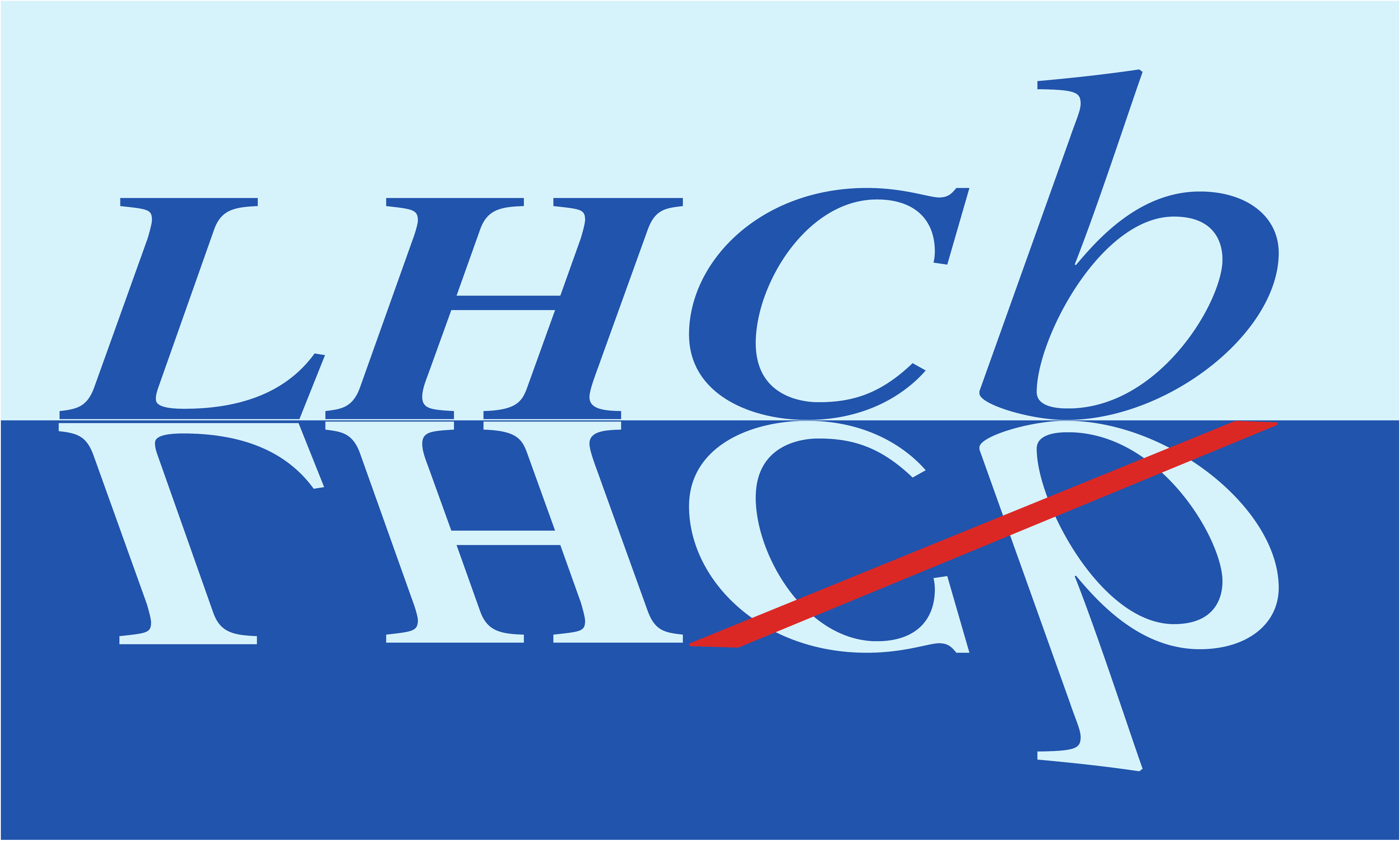}} & &}%
{\vspace*{-1.2cm}\mbox{\!\!\!\includegraphics[width=.12\textwidth]{figs/lhcb-logo.eps}} & &}%
\\
 & & CERN-EP-2020-239 \\  % ID 
 & & LHCb-PAPER-2020-041 \\  % ID 
 & & April 22, 2021 \\
 & & \\
\end{tabular*}

\vspace*{4.0cm}

% Title --------------------------------------------------
{\normalfont\bfseries\boldmath\huge
\begin{center}
  \papertitle 
\end{center}
}

\vspace*{2.0cm}

% Authors -------------------------------------------------
\begin{center}
\paperauthors\footnote{Authors are listed at the end of this Letter.}
\end{center}

\vspace{\fill}

% Abstract -----------------------------------------------
\begin{abstract}
  \noindent
   We present an angular analysis of the \decay{\Bu}{\Kstarp(\to\KS\pip)\mup\mun} decay using 9\invfb of \proton\proton collision data collected with the \lhcb experiment. For the first time, the full set of \CP-averaged angular observables is measured in intervals of the dimuon invariant mass squared. Local deviations from Standard Model predictions are observed, similar to those in previous LHCb analyses of the isospin-partner \decay{\Bd}{\Kstarz\mup\mun} decay. The global tension is dependent on which effective couplings are considered and on the choice of theory nuisance parameters.
  
\end{abstract}

\vspace*{2.0cm}

\begin{center}
  Published in
  \href{https://doi.org/10.1103/PhysRevLett.126.161802}{Phys.~Rev.~Lett.~\textbf{126}, 161802 (2021)}
\end{center}

\vspace{\fill}

{\footnotesize 
\centerline{\copyright~\papercopyright. \href{\paperlicenceurl}{\paperlicence}.}}
\vspace*{2mm}

\end{titlepage}

\newpage
\setcounter{page}{2}
\mbox{~}

\renewcommand{\thefootnote}{\arabic{footnote}}
\setcounter{footnote}{0}
\cleardoublepage

%%%%%%%%%%%%%%%%%%%%%%%%%
%%%%% Main text %%%%%%%%%
%%%%%%%%%%%%%%%%%%%%%%%%%

\pagestyle{plain}
\setcounter{page}{1}
\pagenumbering{arabic}
Transitions between \bquark quarks and \squark quarks with the emission of two charged leptons, $\ellp\ellm$, only proceed through loop-level processes. Such decays are therefore sensitive to possible contributions from heavy mediators that are inaccessible to direct-production searches. Recent studies of \decay{\bquark}{\squark\ellp\ellm} branching fractions~\cite{Aubert:2006vb,LHCb-PAPER-2014-006,LHCb-PAPER-2015-009,LHCb-PAPER-2015-023,LHCb-PAPER-2016-012}, angular distributions~\cite{Aubert:2006vb,Aaltonen:2011ja,LHCb-PAPER-2015-023, LHCb-PAPER-2015-051,Wehle:2016yoi,Sirunyan:2017dhj,Aaboud:2018krd,LHCb-PAPER-2018-029,LHCb-PAPER-2020-002,cmscollaboration2020angular} and ratios of branching fractions between decays with different flavours of lepton pairs~\cite{Lees:2012tva,LHCb-PAPER-2017-013,LHCb-PAPER-2019-009, choudhury2020test,bellecollaboration2020test} show discrepancies with respect to the predictions of the Standard Model~(SM). While these deviations can be consistently explained by the presence of contributions from additional vector or axial-vector currents~\cite{Altmannshofer:2014cfa,Hiller:2014yaa,Gripaios:2014tna,Varzielas:2015iva,Crivellin:2015mga,Celis:2015ara,Falkowski:2015zwa,Barbieri:2016las,Crivellin:2017zlb,Sala:2017ihs, Ko:2017lzd,Sheng:2018vvm,Hiller:2018wbv,Alguero:2019ptt,Aebischer:2019mlg,Arbey:2019duh,Ciuchini:2019usw,Kowalska:2019ley,Alok:2019ufo}, effects from uncertainties related to hadronic form factors or long-distance contributions cannot be ruled out~\cite{Jager:2014rwa,Lyon:2014hpa,Ciuchini:2015qxb, Bobeth:2017vxj, Gubernari:2020eft}.

The \decay{\B}{\Kstar\mumu} decay, where \Kstar denotes the $\Kstar(892)$ meson, has been the subject of extensive studies~\cite{LHCb-PAPER-2013-037,LHCb-PAPER-2013-019,LHCb-PAPER-2015-051,LHCb-PAPER-2020-002}. A large number of these decays are recorded at the \lhc experiments and the flavour of the \B meson can be identified from the \decay{\Kstar}{\PK\Ppi} decay products. This allows the full set of angular observables of the \decay{\B}{\Kstar\mumu} decay to be studied. A recent study~\cite{LHCb-PAPER-2020-002} of the \decay{\Bd}{\Kstarz\mumu} decay channel by the \lhcb collaboration confirmed the tension in the angular observables with respect to the SM predictions.

This letter reports the first measurement of the complete set of angular observables in the isospin partner decay \decay{\Bu}{\Kstarp\mumu}, with the \Kstarp meson reconstructed through the decay chain \decay{\Kstarp}{\KS\pip} with \decay{\KS}{\pip\pim}. Charge-conjugation is implied throughout this letter. This decay is mediated by the same underlying processes as the \decay{\Bd}{\Kstarz\mumu} decay, while potentially receiving additional contributions from \decay{\bquarkbar}{\uquarkbar\Wp} transitions, leading to the emission of a \Kstarp meson~\cite{Ball:2007}. Furthermore, any deviation from isospin symmetry, as reported previously in the \decay{\B}{\Kstar\gamma} decay~\cite{Horiguchi:2017ntw}, could result in a difference in the angular distributions between the isospin partners. In the SM, however, isospin-breaking effects are expected to be small.
The analysis uses the data set collected by the \lhcb collaboration in the years 2011, 2012 (Run 1) and 2015--2018 (Run 2), at centre-of-mass energies of 7, 8 and $13\tev$, respectively. The data set corresponds to an integrated luminosity of $9\invfb$.

The \lhcb detector~\cite{LHCb-DP-2008-001,LHCb-DP-2014-002} is a single-arm forward spectrometer covering the \mbox{pseudorapidity} range $2<\eta <5$, designed for the study of particles containing \bquark or \cquark quarks. 
The detector includes a high-precision tracking system consisting of a silicon-strip vertex detector surrounding the $pp$ interaction region~\cite{LHCb-DP-2014-001}, a large-area silicon-strip detector located upstream of a dipole magnet with a bending power of about $4\unit{Tm}$, and three stations of silicon-strip detectors and straw drift tubes~\cite{LHCb-DP-2013-003,LHCb-DP-2017-001} placed downstream of the magnet.
The tracking system provides a measurement of the momentum, \ptot, of charged particles with a relative uncertainty that varies from 0.5\% at low momentum to 1.0\% at $200\gevc$.
The minimum distance of a track to a primary \proton\proton collision vertex (PV), the impact parameter, is measured with a resolution of $(15+29/\pt)\mum$, where \pt is the component of the momentum transverse to the beam, in\,\gevc.
Different types of charged hadrons are distinguished using information from two ring-imaging Cherenkov detectors~\cite{LHCb-DP-2012-003}. 
Photons, electrons and hadrons are identified by a calorimeter system consisting of scintillating-pad and preshower detectors, an electromagnetic and a hadronic calorimeter. Muons are identified by a system composed of alternating layers of iron and multiwire proportional chambers~\cite{LHCb-DP-2012-002}.
The online event selection is performed by a trigger~\cite{LHCb-DP-2012-004,LHCb-DP-2019-001}, which consists of a hardware stage, based on information from the calorimeter and muon systems, followed by a software stage, which applies a full event
reconstruction.

Simulated decays are used to model the effects of the reconstruction and the candidate selection. In the simulation, $pp$ collisions are generated using \pythia~\cite{Sjostrand:2007gs,*Sjostrand:2006za} with a specific \lhcb configuration~\cite{LHCb-PROC-2010-056}. Decays of unstable particles are described by \evtgen~\cite{Lange:2001uf}, in which final-state radiation is generated using \photos~\cite{Golonka:2005pn}. The interaction of the generated particles with the detector, and its response, are implemented using the \geant toolkit~\cite{Allison:2006ve, *Agostinelli:2002hh}, as described in Ref.~\cite{LHCb-PROC-2011-006}.
 Corrections to the simulation are applied to account for mismodelling in the \pt spectrum of the \Bu~mesons and the multiplicity of tracks in the event. The corrections are obtained from a background-subtracted data sample of \decay{\Bu}{\left(\jpsi\to\mumu\right)\Kstarp} decays.

In the first two stages of the trigger, the event is selected based on kinematical and geometrical properties of the muons. In the last trigger stage, dimuon or topological trigger algorithms are used to select the \Bu candidate.
The \decay{\KS}{\pip\pim} decays are reconstructed in two different categories: the \emph{long} category involves short-lived \KS~candidates for which the pions are reconstructed in the vertex detector; the \emph{downstream} category comprises \KS~candidates that decay later such that track segments of the pions can only be reconstructed in tracking detectors downstream of the vertex locator. The \KS~candidates reconstructed in the long category have better mass, momentum and vertex resolution than those in the downstream category, where the latter has a larger sample size than the former. The \KS candidates are required to have an invariant mass within $30\mevcc$ of the known \KS mass~\cite{Zyla:2020zbs}.

The \decay{\Kstarp}{\KS\pip} decay is reconstructed by combining a \KS candidate with a charged pion and requiring their invariant mass to be within $100\mevcc$ of the world average of the \Kstarp~mass~\cite{Zyla:2020zbs}.
The \decay{\Bu}{\Kstarp\mumu} candidates are formed by combining the \Kstarp candidate with two well-identified, oppositely charged muons. The \Bu~candidates are required to have an invariant mass, $m(\KS\pip\mumu)$, in the range $5150\text{--}6000\mevcc$. The lower value of the mass window is chosen to reject background from partially reconstructed \decay{\B}{\KS\pip\pi\mumu} decays.
Dimuon pairs having invariant mass squared, \qsq, around the $\phi(1020)$ (\mbox{$0.98<\qsq<1.1\gevgevcccc$}), \jpsi (\mbox{$8.0<\qsq<11.0\gevgevcccc$}) and \psitwos (\mbox{$12.5<\qsq<15.0\gevgevcccc$}) resonances are vetoed.
All tracks in the final state are required to have a significant impact parameter with respect to any PV and the \Bu~candidate decay vertex needs to be well displaced from any PV in the event. A kinematical fit~\cite{Hulsbergen:2005pu} is performed to the full decay chain, in which the reconstructed \KS mass is constrained to the known value~\cite{Zyla:2020zbs}. 

Decays of \Bd mesons to the \KS\mumu final state with a pion added can form a peaking structure in the \Bu mass window. Therefore, candidates with an invariant mass $m(\KS\mumu)$ within $50\mevcc$ of the known \Bd mass are vetoed. 
Background originating from \decay{\Bu}{\left(\jpsi\to\mumu\right)\Kstarp} decays is probed by testing the \KS\pip and dimuon invariant masses formed by exchanging the particle hypotheses between the pion from the \Kstarp meson decay and the muon with the same charge. The candidates with a dimuon mass within $50\mevcc$ of the \jpsi meson mass and a \KS\pip invariant mass within $30\mevcc$ of the \Kstarp mass are then rejected.
The background from \B decays with two hadrons misidentified as muons is negligible.

\begin{figure}[t]
  \begin{center}
    \includegraphics[width=0.6\linewidth]{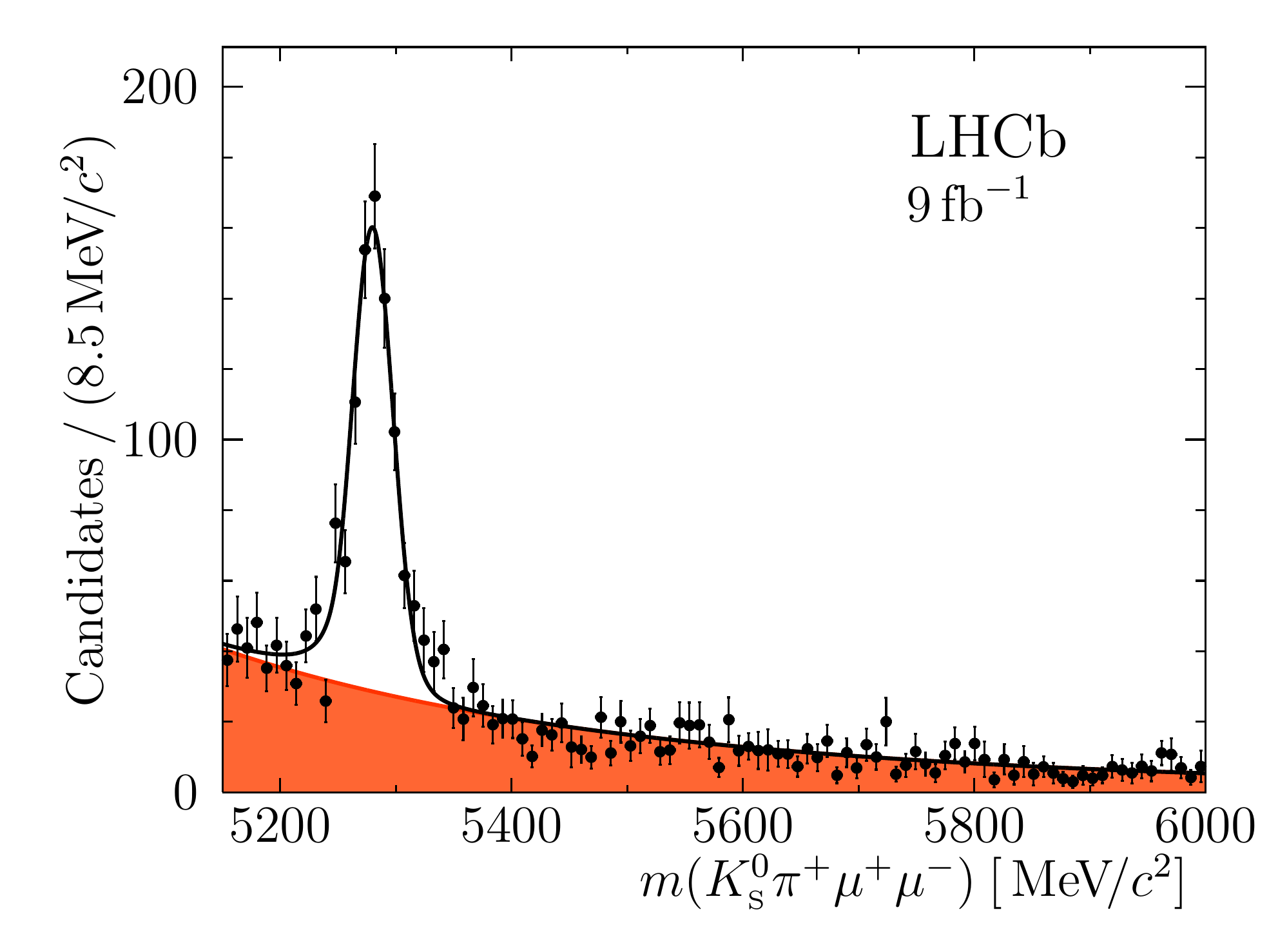}
    \vspace*{-0.5cm}
  \end{center}
  \caption{Distribution of the $\KS\pip\mumu$ invariant mass. The black points represent the full data set, while the solid curve shows the fit result. The background component is represented by the orange shaded area.}
  \label{fig:massfit}
\end{figure}

To increase the signal-to-background ratio, a multivariate classification is employed. The data are split into four subsets, according to the Run 1 and Run 2 data-taking periods and the category of the \KS meson. A boosted decision tree with gradient boosting~\cite{Breiman, Freund:1997xna} from the TMVA toolkit~\cite{Hocker:2007ht} is then trained on each data set individually, using simulated events as a proxy for signal and candidates with $m(\KS\pip\mumu)$ larger than $5400\mevcc$ as a proxy for background. The variables include kinematical and topological properties of the final state or intermediate particles, the quality of the vertex of the \Bu candidate, and an isolation criterion related to the asymmetry in \pt between all tracks inside a cone around the flight directions of the \Bu candidates and the tracks associated to the \Bu decay products~\cite{LHCb-PAPER-2018-036}.
Figure~\ref{fig:massfit} shows the \Bu-candidate invariant mass distribution, $m(\KS\pip\mumu)$, for all the selected data.
A fit model with a double-sided Crystal Ball function for the signal and an exponential function for the background component is overlaid.
The number of \decay{\Bu}{\Kstarp\mumu} signal candidates from this fit is $737\pm34$, where the uncertainty is statistical only.

Ignoring the natural width of the \Kstarp meson, the decay \decay{\Bu}{\Kstarp\mumu} can be fully described by four variables: \qsq and the set of three angles $\vec{\Omega}=\left(\thetal,\thetak,\phi\right)$. The angle between the \mup (\mun) and the direction opposite to that of the \Bu (\Bub) in the rest frame of the dimuon system is denoted \thetal. The angle between the direction of the \KS and the \Bu (\Bub) in the rest frame of the \Kstarp (\Kstarm) system is denoted \thetak. The angle $\phi$ is the angle between the plane defined by the momenta of the muon pair and the plane defined by the kaon and pion momenta in the \Bu (\Bub) rest frame. A full description of the angular basis is given in Ref.~\cite{LHCb-PAPER-2013-019}.

Averaging over \Bu and \Bub decays, with rates respectively denoted $\Gamma$ and $\bar{\Gamma}$, the differential decay rate of the \decay{\Bu}{\Kstarp\mumu} decay with the $\KS\pip$ system in a P-wave configuration is

\begin{equation}
\begin{split}
\left.\frac{1}{\deriv(\Gamma+\bar{\Gamma})/\deriv \qsq}\,\frac{\deriv^4(\Gamma+\bar{\Gamma})}{\deriv\qsq\,\deriv\vec{\Omega}}\right|_{\rm P} =
\frac{9}{32\pi} \Big[
  & \tfrac{3}{4} (1-\FL)\sin^2\thetak + \FL\cos^2\thetak \\
 \phantom{\Big[} +& \tfrac{1}{4}(1-\FL)\sin^2\thetak\cos 2\thetal\\
 \phantom{\Big[}-& \FL \cos^2\thetak\cos 2\thetal + {S_3}\sin^2\thetak \sin^2\thetal \cos 2\phi\\
\phantom{\Big[}+& {S_4} \sin 2\thetak \sin 2\thetal \cos\phi + {S_5}\sin 2\thetak \sin \thetal \cos \phi\\
\phantom{\Big[}+& \tfrac{4}{3} \AFB \sin^2\thetak \cos\thetal + {S_7} \sin 2\thetak \sin\thetal \sin\phi\\
+& {S_8} \sin 2\thetak \sin 2\thetal \sin\phi + {S_9}\sin^2\thetak \sin^2\thetal \sin 2\phi  \Big],
\end{split}
\label{eq:pdfpwave}
\end{equation}

\noindent where \FL is the fraction of the longitudinally polarised \Kstarp mesons, \AFB is the forward-backward asymmetry of the dimuon system and $S_i$ are other \CP-averaged observables~\cite{LHCb-PAPER-2015-051}.

The $\KS\pip$ system can also be in an S-wave configuration, which modifies the differential decay rate to
\begin{equation}
\begin{split}
\left.\frac{1}{\deriv(\Gamma+\bar{\Gamma})/\deriv \qsq}\,\frac{\deriv^4(\Gamma+\bar{\Gamma})}{\deriv\qsq\,\deriv\vec{\Omega}}\right|_{{\rm P}+{\rm S}} ~=~ &
(1-\FS) \left.\frac{1}{\deriv(\Gamma+\bar{\Gamma})/\deriv \qsq}\,\frac{\deriv^4(\Gamma+\bar{\Gamma})}{\deriv\qsq\,\deriv\vec{\Omega}}\right|_{\rm P}\\
& + \frac{3}{16\pi} \FS\sin^{2}\theta_{l}  \\
& + \frac{9}{32\pi} ( S_{11}  + S_{13} \cos 2\theta_{l} ) \cos\theta_{K}  \\
& + \frac{9}{32\pi} ( S_{14} \sin 2\theta_{l} + S_{15} \sin\theta_{l} ) \sin\theta_{K}\cos\phi  \\
& + \frac{9}{32\pi} ( S_{16} \sin\theta_{l} + S_{17} \sin 2\theta_{l} ) \sin\theta_{K}\sin\phi \, ,
\end{split}
\label{eq:pdfswave}
\end{equation}

\noindent where \FS denotes the S-wave fraction and the coefficients $S_{11}$, $S_{13}$--$S_{17}$ arise from interference between the S- and P-wave amplitudes. 
Throughout this letter, \FS and the interference coefficients are treated as nuisance parameters. In addition to the observable basis comprising \FL, \AFB and $S_{3} \text{--} S_{9}$, a basis with so-called optimised observables, denoted $P_{i}^{(\prime)}$, for which the leading form-factor uncertainties cancel~\cite{DescotesGenon:2012zf}, is used. The notation for the $P_{i}^{(\prime)}$ observables is defined in Ref.~\cite{LHCb-PAPER-2013-037}.

Due to the limited number of signal candidates, the observables cannot all be measured simultaneously. A folding procedure is therefore employed that uses symmetries of the differential decay rate in the angles to cancel some observables, reducing the number of free parameters in the fit. By performing different folds, all angular observables can be studied, without any loss in precision. Five different folds are used to study the observables \AFB and $S_{9}$ ($P_2$ and $P_3$), $S_{4}$ ($P^{\prime}_4$), $S_{5}$ ($P^{\prime}_5$), $S_{7}$ ($P^{\prime}_6$) and $S_{8}$ ($P^{\prime}_8)$, respectively. The observables \FL and $S_{3}$ ($P_1$) are measured in each fold. 
This procedure is detailed in Ref.~\cite{DeCian:1605179} and was previously used in Refs.~\cite{LHCb-PAPER-2013-037, LHCb-PAPER-2013-019, Wehle:2016yoi, Sirunyan:2017dhj,Aaboud:2018krd}. The values of \FL and $S_{3}$ ($P_1$) are taken from the same fold that is used to extract the value of $S_{8}$ ($P^{\prime}_8$), as the number of free parameters in the fit is the smallest in this fold.

The angular observables are extracted using an unbinned maximum-likelihood fit to the \Bu candidate mass and the three decay angles in intervals of \qsq. The eight narrow and two wide \qsq intervals are identical to those in Refs.~\cite{LHCb-PAPER-2015-051,LHCb-PAPER-2020-002}. The angular distributions are fitted with the function described in Eq.~(\ref{eq:pdfswave}) for the signal, and with second-order polynomials in \ctk and \ctl for the background. The background in the $\phi$ angle is uniform. No significant correlation is observed between the angular background distributions in the \Bu candidate mass sidebands, justifying a factorisation of the background description in the three decay angles.

The reconstruction and selection efficiency varies over the angular and \qsq phase space. This acceptance effect is parametrised before folding using the sum over the product of four one-dimensional Legendre polynomials, each depending on one angle or \qsq. 
This is analogous to the procedure used in Ref.~\cite{LHCb-PAPER-2020-002}.
The effect is corrected using weights derived from simulation.
The weight then corresponds to the inverse of the efficiency. No dependence of the acceptance effect on the \Kstarp candidate mass is observed. 

Given the low signal yield and narrow \qsq intervals, the S-wave fraction~\FS cannot be determined with sufficient precision to guarantee unbiased results for the P-wave angular observables. Therefore, a two-dimensional unbinned maximum-likelihood fit to $m(\KS\pip\mumu)$ and the \Kstarp candidate mass $m(\KS\pip)$ is first performed in three \qsq intervals: $1.1\text{--}8.0$, $11.0\text{--}12.5$ and $15.0\text{--}19.0\gevgevcccc$.
The $m(\KS\pip\mumu)$ distribution is fitted using the signal and background model described above.
The \Kstarp candidate mass is fitted using a relativistic Breit-Wigner function to describe the P-wave component, the LASS parametrisation to describe the S-wave component~\cite{Aston:1987ir} and a linear function to describe the combinatorial background. S- and P-wave interference terms are neglected in this treatment. The value of \FS in the default narrow \qsq intervals is then computed by multiplying the value of \FS in the broad intervals with the ratio between \FL in the narrow and broad intervals. This procedure assumes a similar \qsq dependence of the longitudinal component of the P wave and the S wave and is broadly compatible with the results from Ref.~\cite{LHCb-PAPER-2016-012}. Given the weak dependence of the P-wave observables on the value of \FS, this procedure ensures unbiased results without relying on values of \FS from an external measurement. 
Pseudoexperiments indicate that determining \FS in this manner induces at most a bias of 13\% of the statistical uncertainty on the angular observables. This is treated as a systematic uncertainty.
All values of \FS are measured to be positive and compatible with the results in Ref.~\cite{LHCb-PAPER-2016-012}.

Fitting the folded data set only provides statistical correlations between observables measured in the same fold. In order to obtain the correlations between all observables, the bootstrapping technique~\cite{efron:1979} is used to produce a large number of pseudodata sets. The measurement of the observables in each fold of these pseudodata sets enables computing the correlations between observables in different folds. The statistical precision of the elements of the correlation matrix is determined to be around $0.11$.
In order to ensure correct coverage in the presence of physical boundaries of the observables, the statistical uncertainty for each observable in each \qsq interval for the signal channel is evaluated using the Feldman-Cousins technique~\cite{1998PhRvD..57.3873F}.

The full analysis procedure with acceptance correction, extraction of \FS and extraction of the angular observables, is tested on a sample of \decay{\Bu}{\jpsi\Kstarp} decays with the same selection as applied to the signal channel, but requiring the dimuon invariant mass squared to be in the range $8.68\text{--}10.09\gevgevcccc$. The results are found to be in good agreement with previous measurements from the \babar\cite{Aubert:2007hz}, \belle\cite{Itoh:2005ks} and \lhcb\cite{LHCb-PAPER-2013-023} experiments.

Several sources of systematic uncertainties are considered and their sizes are estimated using pseudoexperiments. Various contributions to the overall systematic uncertainty are related to the correction of acceptance effects. They include the limited size of the simulation sample and the parametrisation of the acceptance function. Other systematic uncertainties are related to the correction of differences between data and simulation, the model of the \Bu candidate mass distribution and angular background, the impact of the \decay{\Bd}{\KS\mumu} veto on the mass distribution of the combinatorial background, the angular resolution and the effect of constraining the value of \FS with a two-dimensional fit.
Pseudoexperiments are used to assess a possible bias introduced by the fit procedure. The pseudodata samples are generated based on the result of the fit to data or on the predictions from either the SM or a new physics scenario favoured by the LHCb measurement from Ref.~\cite{LHCb-PAPER-2020-002} with the real part of the Wilson coefficient $C_9$ shifted by $-1$ with respect to SM predictions. Here, $C_9$ is the strength of the vector coupling in an effective field theory of \bquark quark to \squark quark transitions. The largest bias observed is 33\% of the statistical uncertainty for $S_4$ in the \qsq interval $4.0\text{--}6.0\gevgevcccc$. Given that the biases can depend on the values of the observables themselves, the largest biases observed among the three pseudodata samples are taken as systematic uncertainties.
The potential exchange of the \pip mesons from the decays of the \Kstarp and \KS candidates and the angular background description differing between the upper and lower mass sidebands are both considered as further sources of systematic uncertainty. Both effects are found to be negligible.
All systematic uncertainties are added in quadrature and their total size is reported together with the numerical results of the observables in Sec.~\ref{sec:SandPTables} of the Supplemental Material. A summary of the contributions from the various sources is given in Table~\ref{tab:systematics} of the Supplemental Material. The statistical uncertainty dominates for all \qsq intervals and all observables, which implies that correlations with the results from Ref.~\cite{LHCb-PAPER-2020-002} are negligible.

\begin{figure}[htb]
  \begin{center}
    \includegraphics[width=0.49\linewidth]{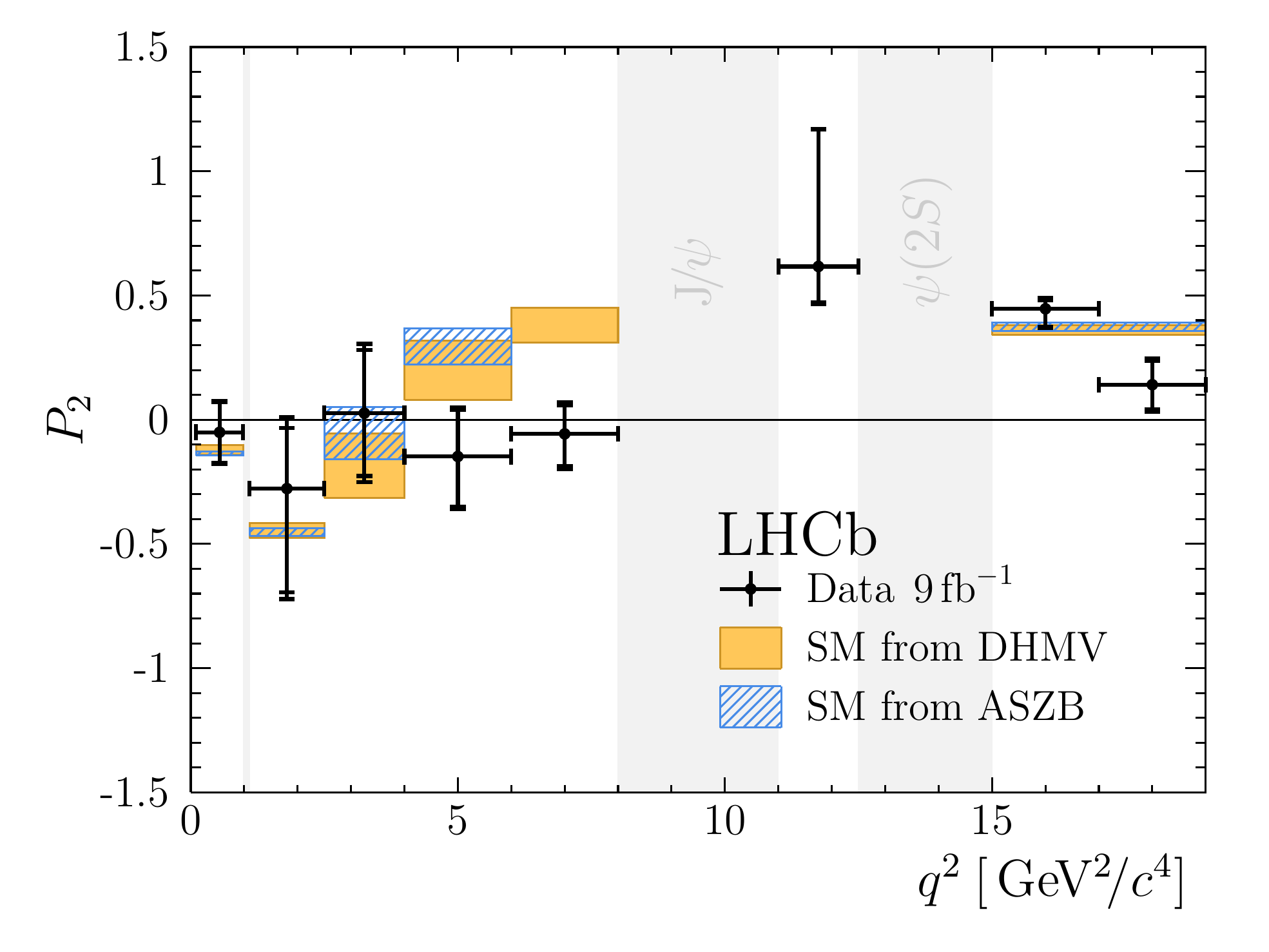}
    \includegraphics[width=0.49\linewidth]{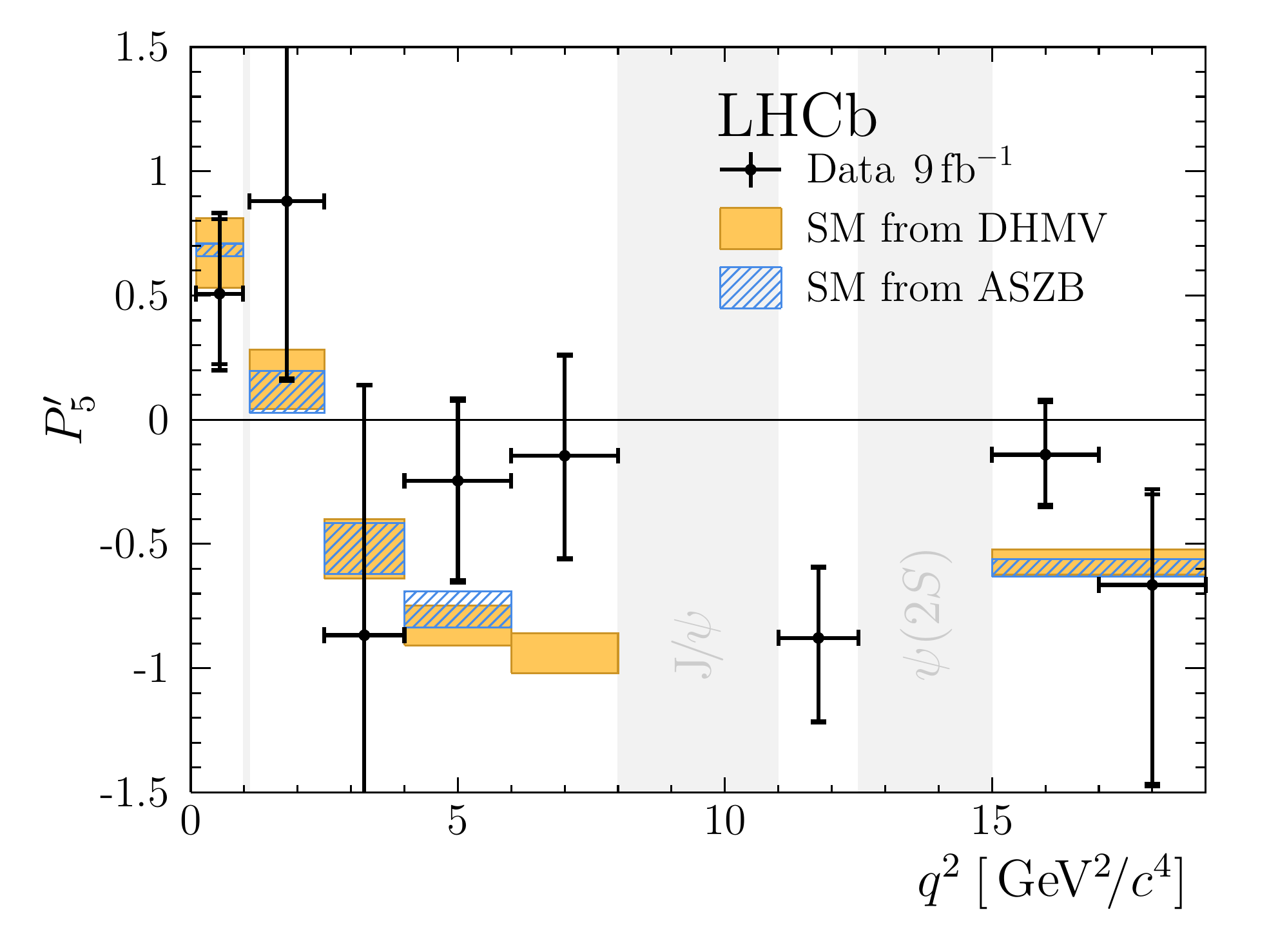}
    \vspace*{-0.5cm}
  \end{center}
  \caption{ The \CP-averaged observables (left) $P_2$ and (right) $P^\prime_5$ in intervals of \qsq. The first (second) error bars represent the statistical (total) uncertainties. The theoretical predictions in blue are based on Ref.~\cite{Altmannshofer:2014rta} with hadronic form factors taken from Refs.~\cite{Straub:2015ica, Horgan:2013hoa,Horgan:2015vla} and are obtained with the \textsc{Flavio} software package~\cite{Straub:2018kue} (version 2.0.0). The theoretical predictions in orange are based on Refs.~\cite{Descotes-Genon2016,Capdevila2018} with hadronic form factors from Ref.~\cite{Khodjamirian:2010vf}. The grey bands indicate the regions of excluded $\phi(1020)$, \jpsi and \psitwos resonances.}
  \label{fig:observablesMain}
\end{figure}

The results of the angular fits for the observables \mbox{$P_2 = \tfrac{2}{3}\AFB/(1-\FL)$} and \mbox{$P_{5}^{\prime} = S_5/\sqrt{\FL(1-\FL)}$} are shown in Fig.~\ref{fig:observablesMain}. They are compared with the two SM predictions taken from Ref.~\cite{Altmannshofer:2014rta} with hadronic form factors from Refs.~\cite{Straub:2015ica,Horgan:2013hoa,Horgan:2015vla}, and from Refs.~\cite{Descotes-Genon2016,Capdevila2018} with hadronic form factors from Ref.~\cite{Khodjamirian:2010vf}. 
The rest of the observables are presented in Figs.~\ref{fig:Sobservables} and \ref{fig:Pobservables} in the Supplemental Material to this letter. The numerical results of the angular fits to the data are presented in Tables~\ref{tab:Sobservables} and \ref{tab:Pobservables}, where values for the two wide \qsq intervals are also given. The correlations are given in Tables~\ref{tab:correlationsSBin0}--\ref{tab:correlationsSBinB} and \ref{tab:correlationsPBin0}--\ref{tab:correlationsPBinB} for the $S_{i}$ and $P_{i}^{(\prime)}$ observables, respectively. 

The majority of observables show good agreement with the SM predictions, \FL and \AFB agree well with the measurements in Ref.~\cite{cmscollaboration2020angular}.
The largest local discrepancy is in the measurement of $P_2$ in the $6.0\text{--}8.0\gevgevcccc$ interval, where a deviation of $3.0\sigma$ with respect to the SM prediction is observed.
The pattern of deviations from the SM predictions in the observables $S_5$ ($P_5^{\prime}$) and \AFB ($P_2$) broadly agrees with the deviations observed in the \BdToKstmm channel.

The \textsc{Flavio} package~\cite{Straub:2018kue} (version 2.0.0) is used to perform a fit to the angular observables varying the parameter $\mathrm{Re}(C_9)$, which is motivated by Refs.~\cite{LHCb-PAPER-2015-051,LHCb-PAPER-2020-002}.
In order to minimise the theoretical uncertainties related to contributions from virtual charm-quark loops~\cite{Khodjamirian:2010vf} and broad charmonium resonances~\cite{Grinstein:2004vb, Beylich:2011aq, Brass:2016efg}, the narrow \qsq intervals up to $6.0\gevgevcccc$ plus the wide \qsq interval \mbox{$15.0<\qsq<19.0\gevgevcccc$} are included in the fit.
The default \textsc{Flavio} SM nuisance parameters are used, including form-factor parameters and subleading corrections to account for long-distance QCD interference effects with the charmonium decay modes~\cite{Straub:2015ica,Altmannshofer:2014rta}.
The best-fit point results in a shift with respect to the SM value of $\mathrm{Re}(C_9)$ of $-1.9$ and gives a tension with the SM of $3.1\sigma$. However, the tension observed depends on the \qsq intervals considered, which effective couplings are varied and the handling of the SM nuisance parameters.

In summary, using the complete \proton\proton data set collected with the \lhcb experiment in Runs 1 and 2, the full set of angular observables for the decay \decay{\Bu}{\Kstarp\mumu} is measured for the first time. The results confirm the global tension with respect to the SM predictions previously reported in the decay \BdToKstmm.

\section*{Acknowledgements}
%
% These Acknowledgements valid from 3-May-2019
%
\noindent We express our gratitude to our colleagues in the CERN
accelerator departments for the excellent performance of the LHC. We
thank the technical and administrative staff at the LHCb
institutes.
We acknowledge support from CERN and from the national agencies:
CAPES, CNPq, FAPERJ and FINEP (Brazil); 
MOST and NSFC (China); 
CNRS/IN2P3 (France); 
BMBF, DFG and MPG (Germany); 
INFN (Italy); 
NWO (Netherlands); 
MNiSW and NCN (Poland); 
MEN/IFA (Romania); 
MSHE (Russia); 
MICINN (Spain); 
SNSF and SER (Switzerland); 
NASU (Ukraine); 
STFC (United Kingdom); 
DOE NP and NSF (USA).
We acknowledge the computing resources that are provided by CERN, IN2P3
(France), KIT and DESY (Germany), INFN (Italy), SURF (Netherlands),
PIC (Spain), GridPP (United Kingdom), RRCKI and Yandex
LLC (Russia), CSCS (Switzerland), IFIN-HH (Romania), CBPF (Brazil),
PL-GRID (Poland) and OSC (USA).
We are indebted to the communities behind the multiple open-source
software packages on which we depend.
Individual groups or members have received support from
AvH Foundation (Germany);
EPLANET, Marie Sk\l{}odowska-Curie Actions and ERC (European Union);
A*MIDEX, ANR, Labex P2IO and OCEVU, and R\'{e}gion Auvergne-Rh\^{o}ne-Alpes (France);
Key Research Program of Frontier Sciences of CAS, CAS PIFI, CAS CCEPP, 
Fundamental Research Funds for the Central Universities, 
and Sci. \& Tech. Program of Guangzhou (China);
%Key Research Program of Frontier Sciences of CAS, CAS PIFI,
%Thousand Talents Program, and Sci. \& Tech. Program of Guangzhou (China);
RFBR, RSF and Yandex LLC (Russia);
GVA, XuntaGal and GENCAT (Spain);
the Royal Society
and the Leverhulme Trust (United Kingdom).

\newpage
\section*{Supplemental Material}

This supplemental material includes additional information to that already provided in
the main letter. 

\begin{figure}[h]
  \begin{center}
    \includegraphics[width=0.3\linewidth]{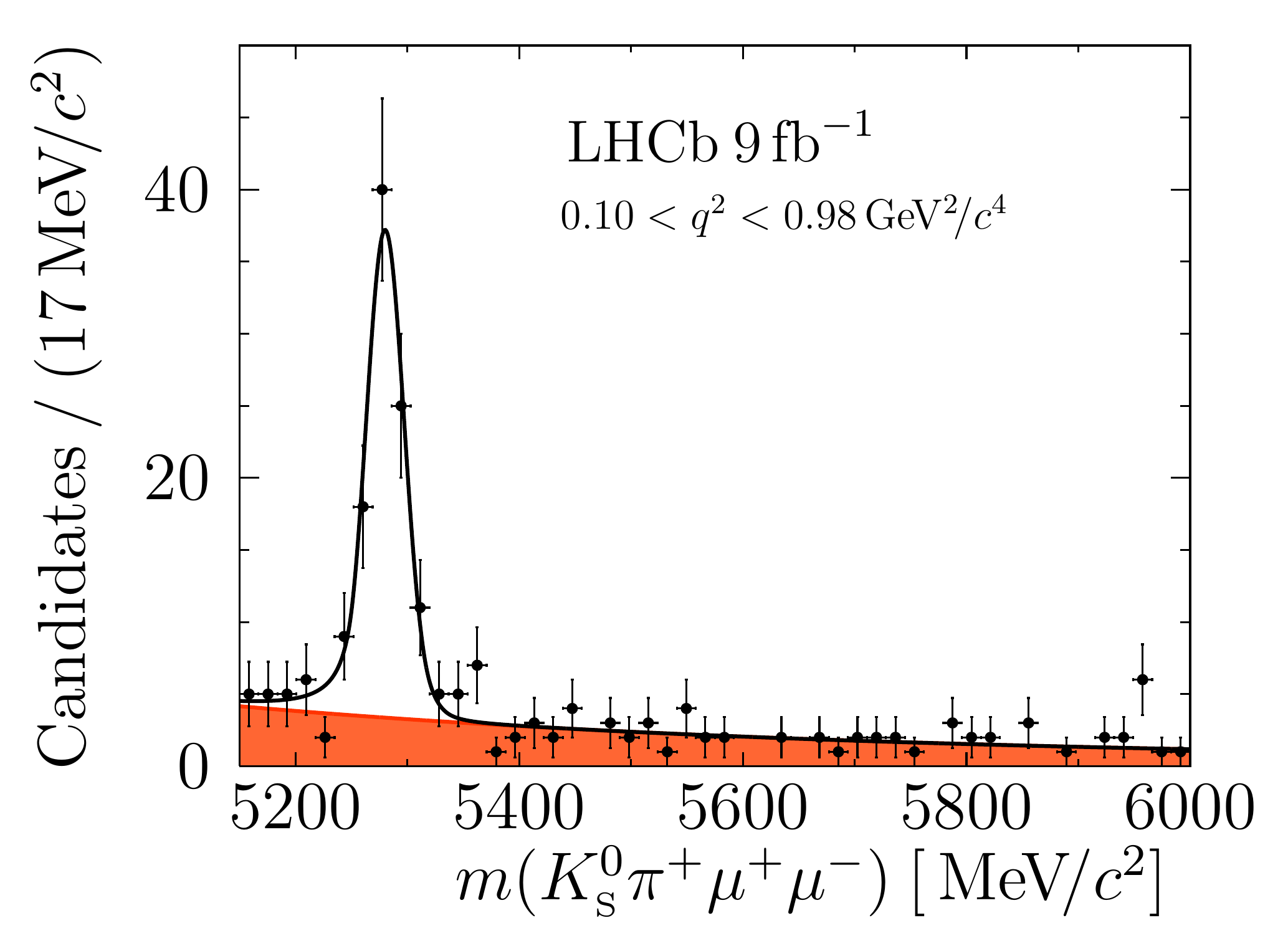}
    \includegraphics[width=0.3\linewidth]{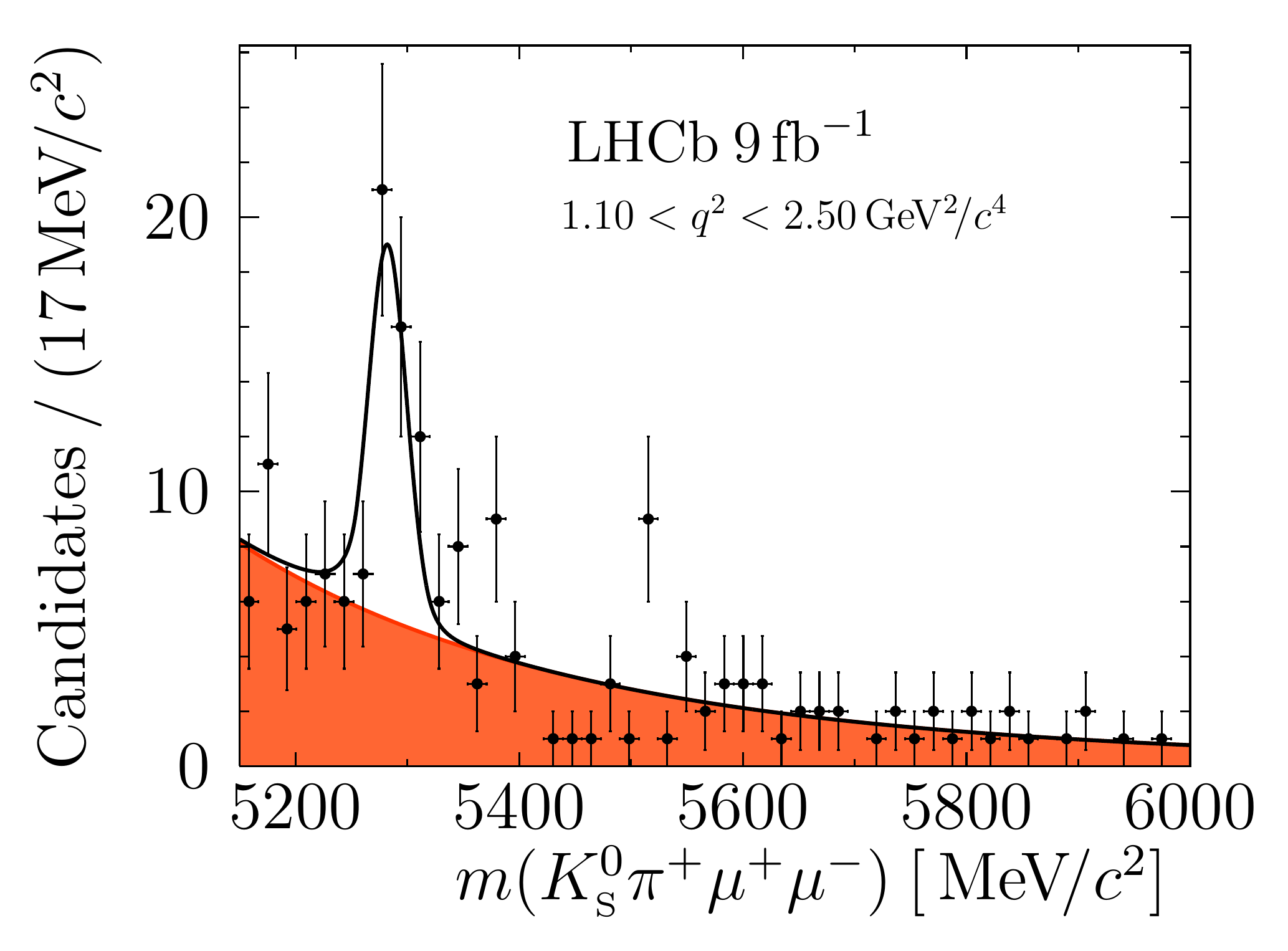}
    \includegraphics[width=0.3\linewidth]{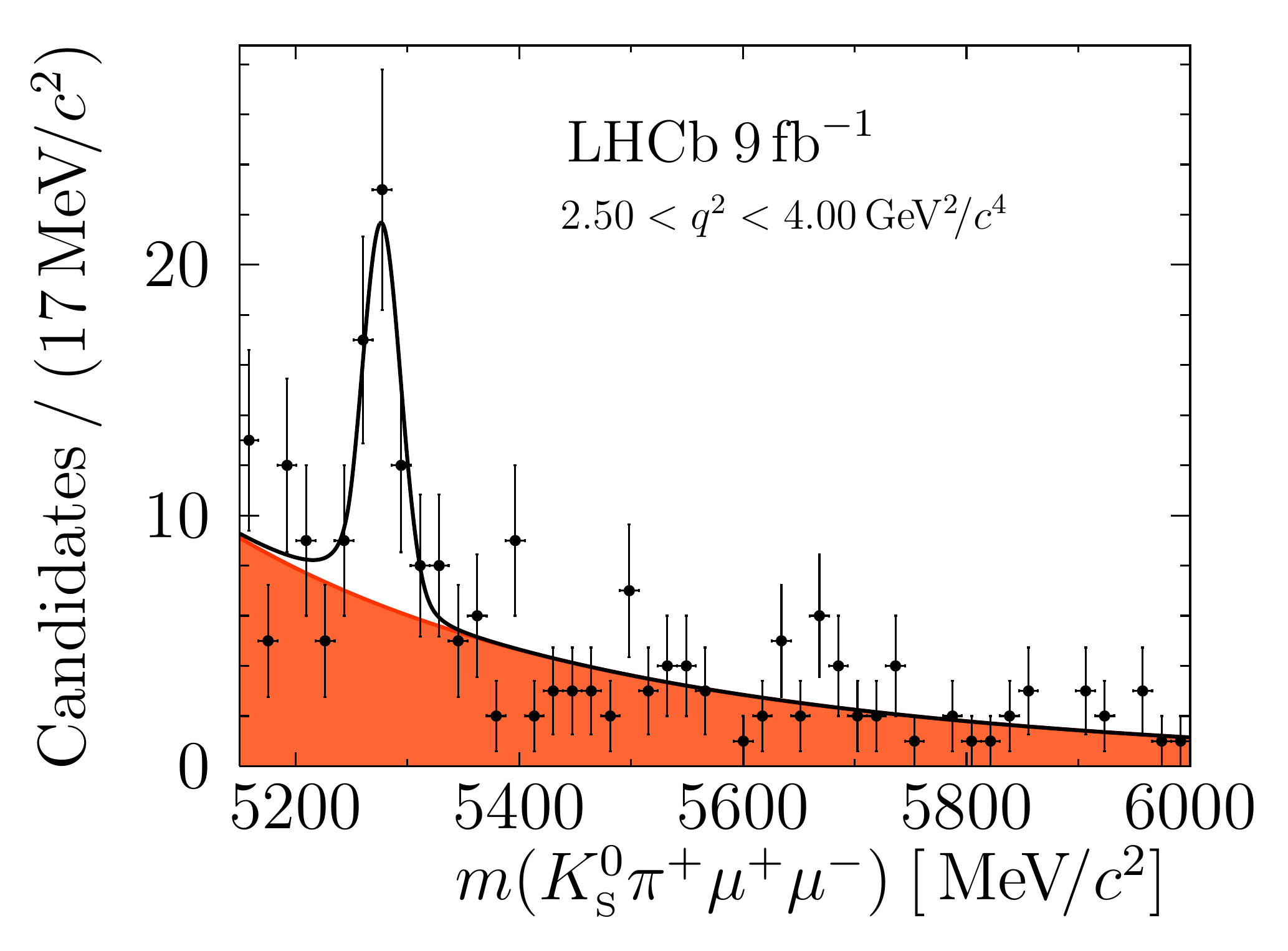}\\
    \vspace*{0.2cm}
    \includegraphics[width=0.3\linewidth]{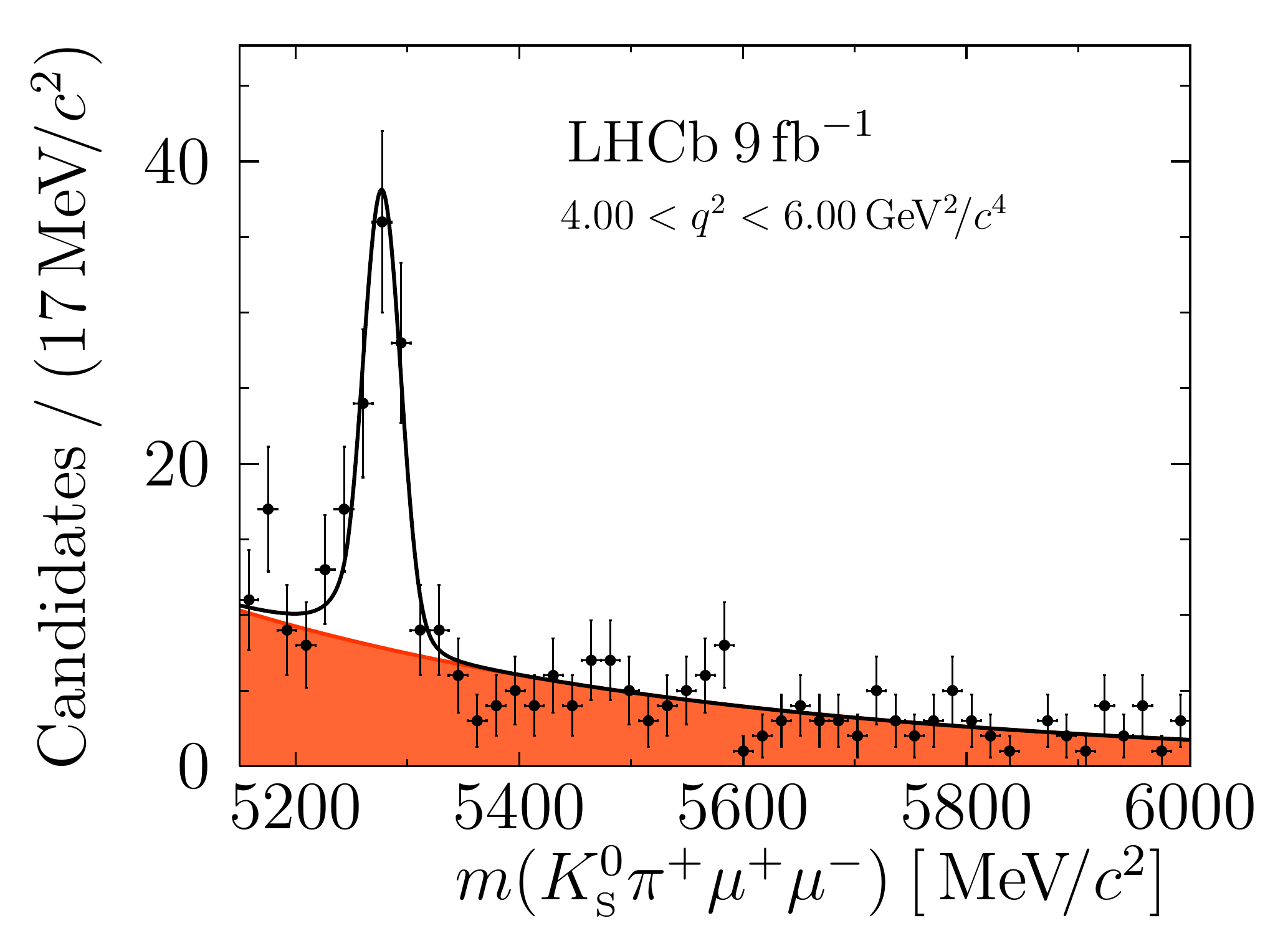}
    \includegraphics[width=0.3\linewidth]{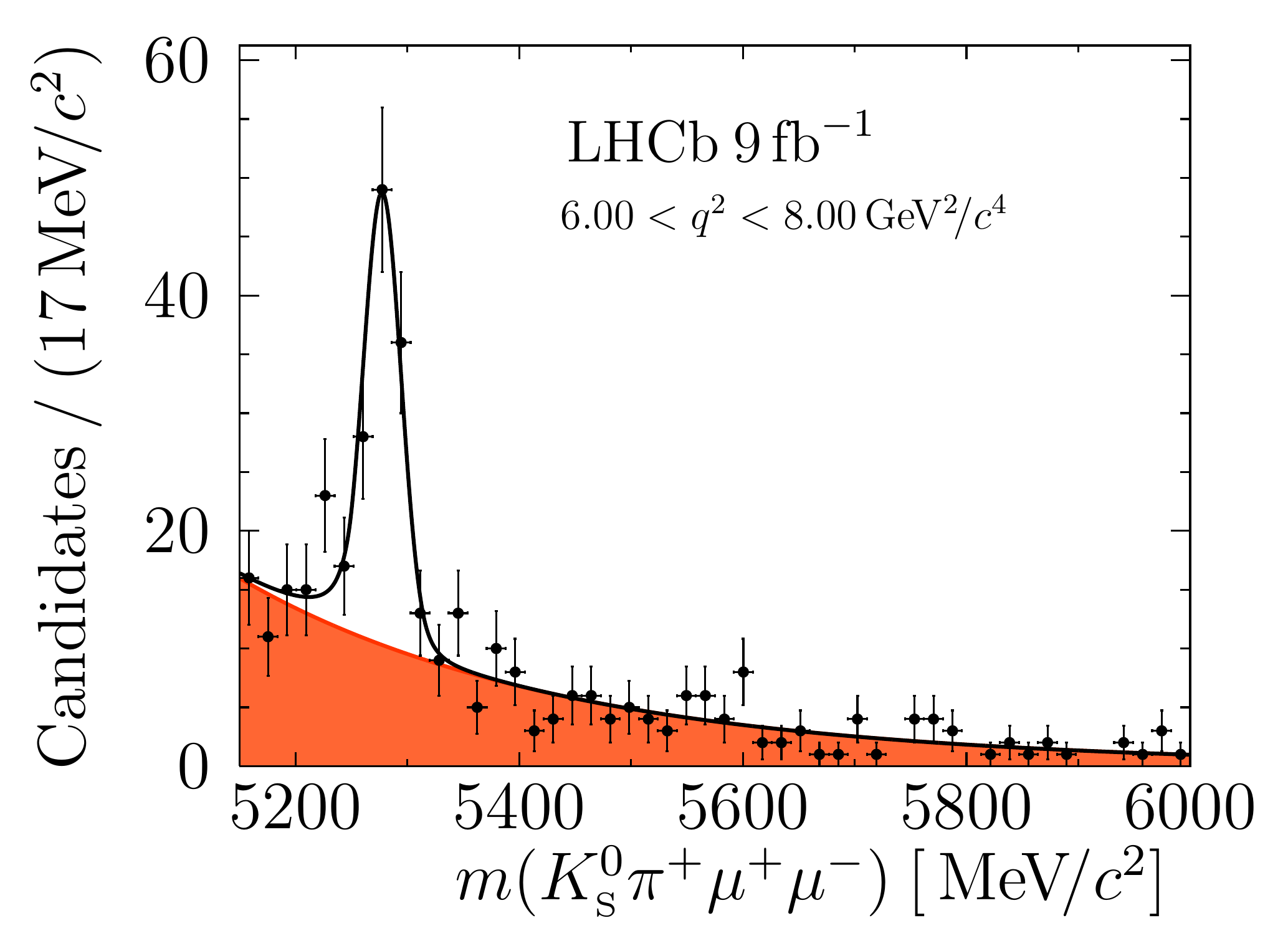}
    \includegraphics[width=0.3\linewidth]{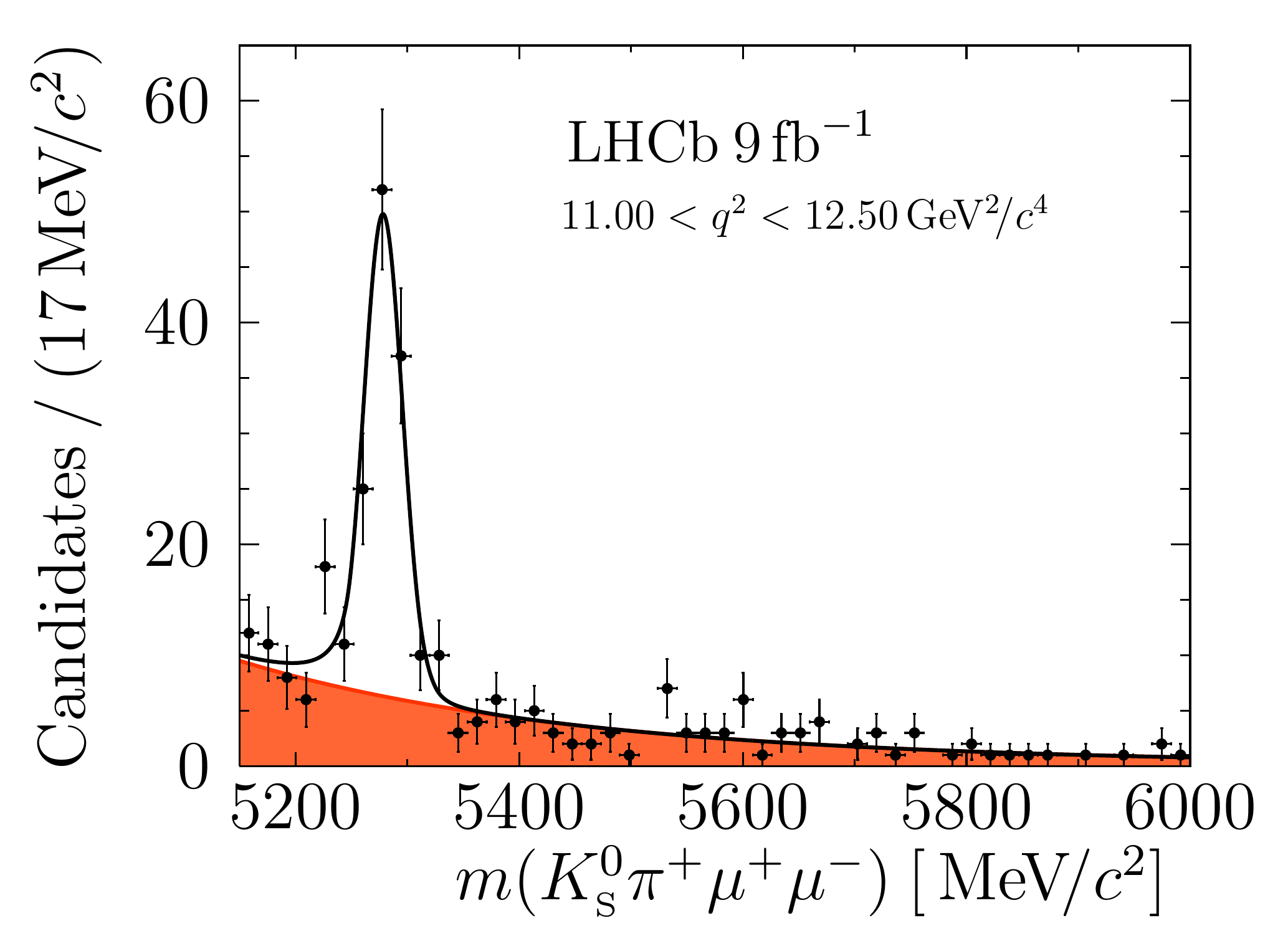}\\
    \vspace*{0.2cm}
    \includegraphics[width=0.3\linewidth]{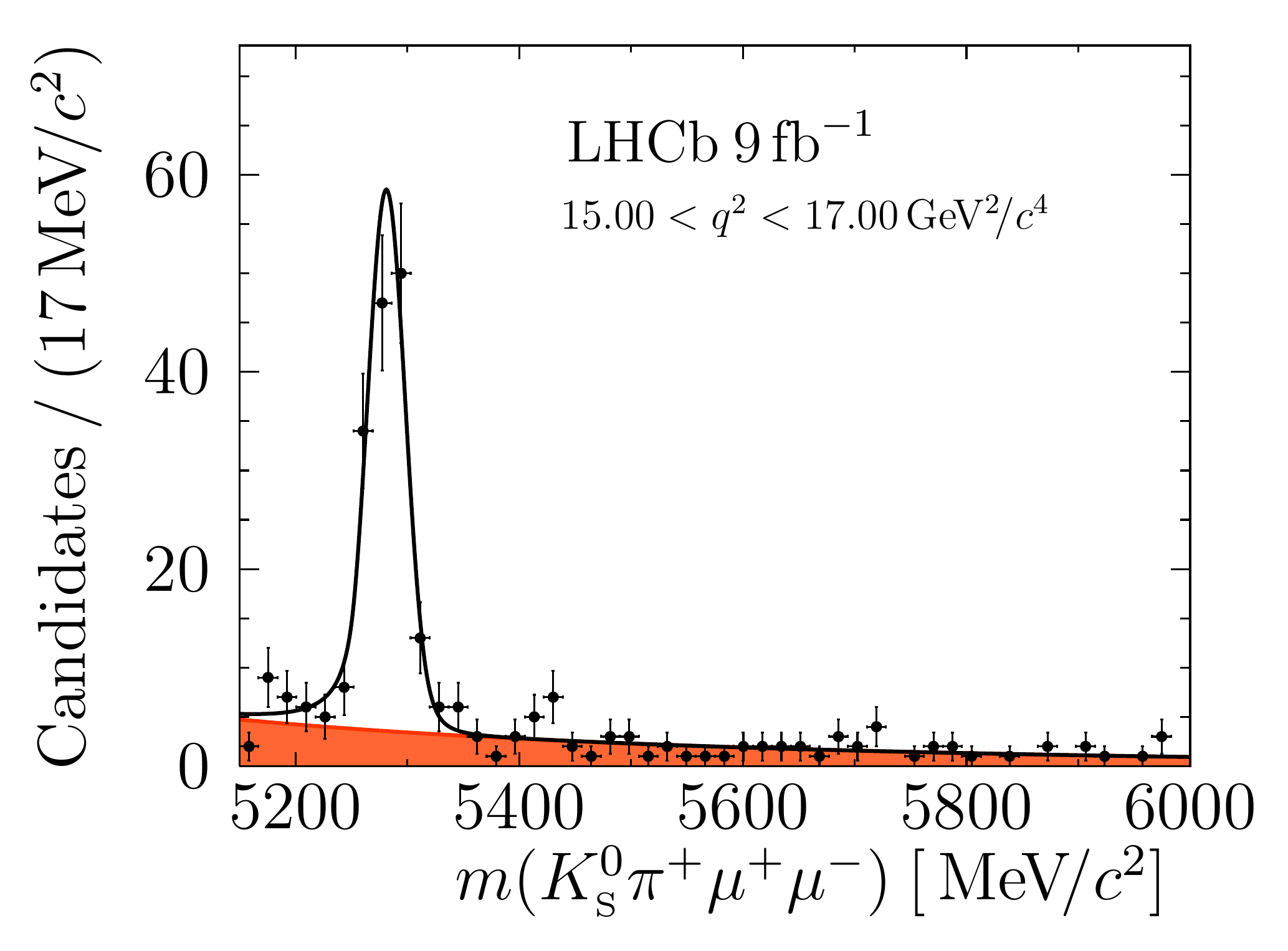}
    \includegraphics[width=0.3\linewidth]{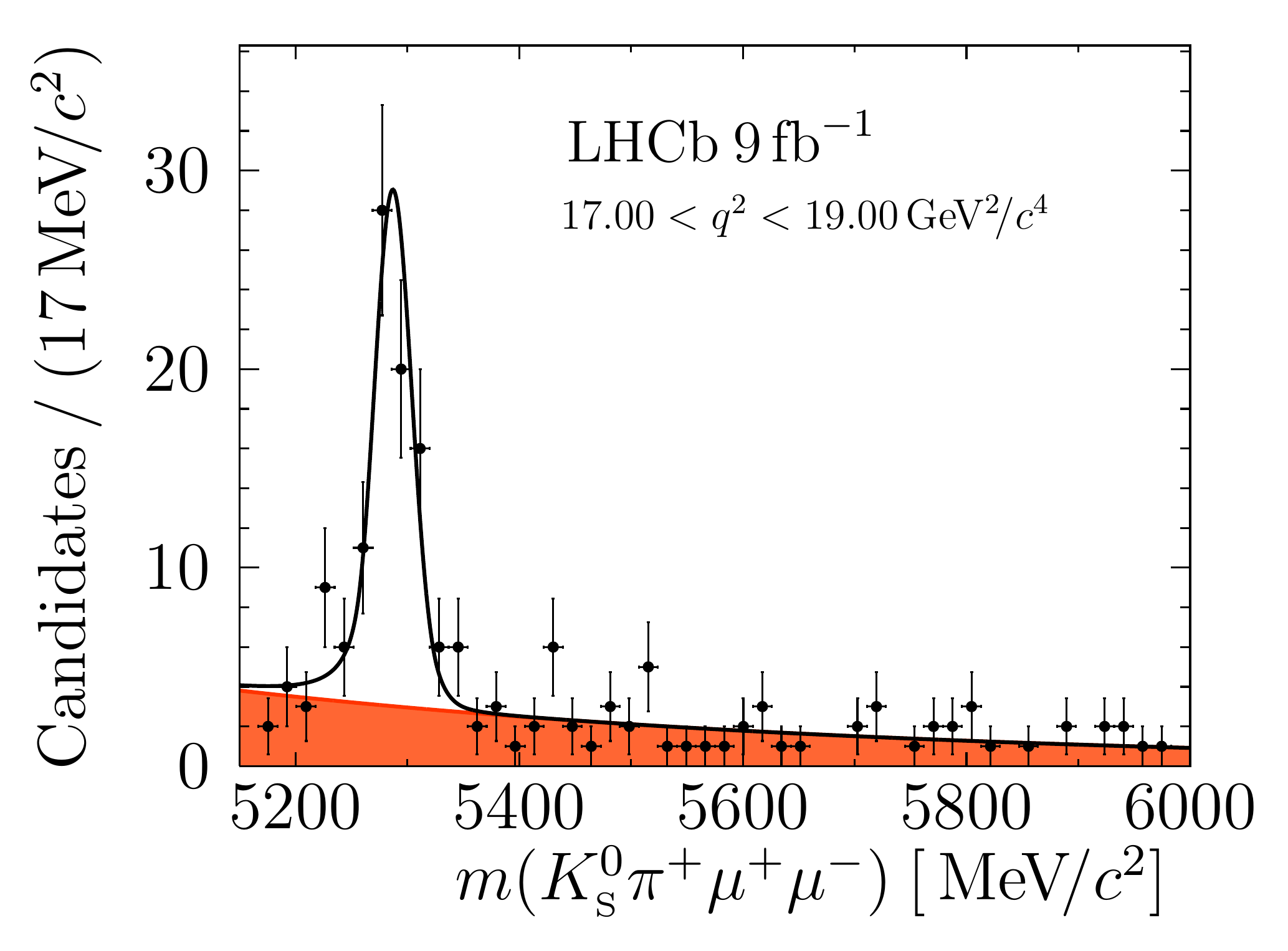}\\
    \vspace*{0.2cm}
    \includegraphics[width=0.3\linewidth]{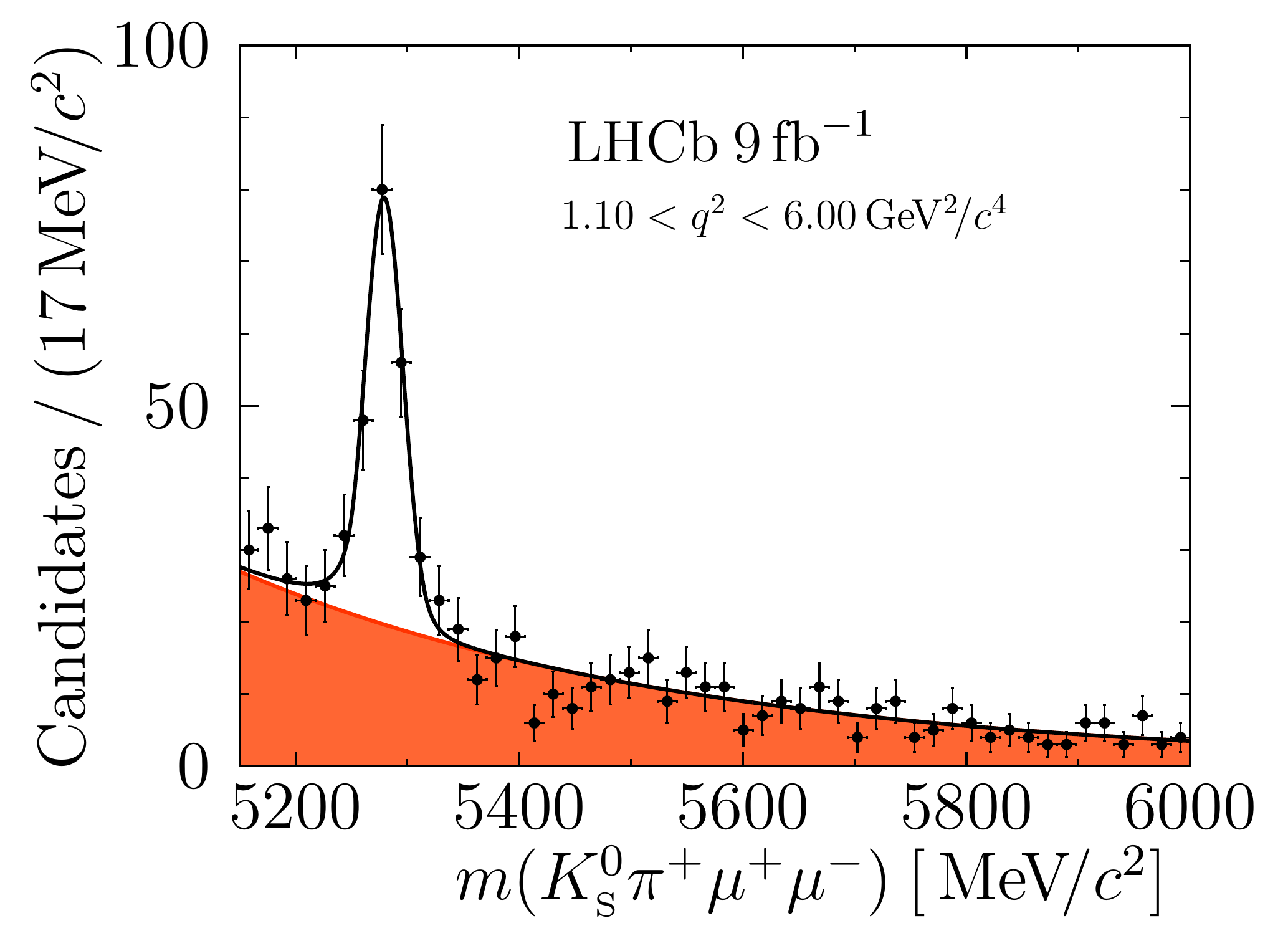}
    \includegraphics[width=0.3\linewidth]{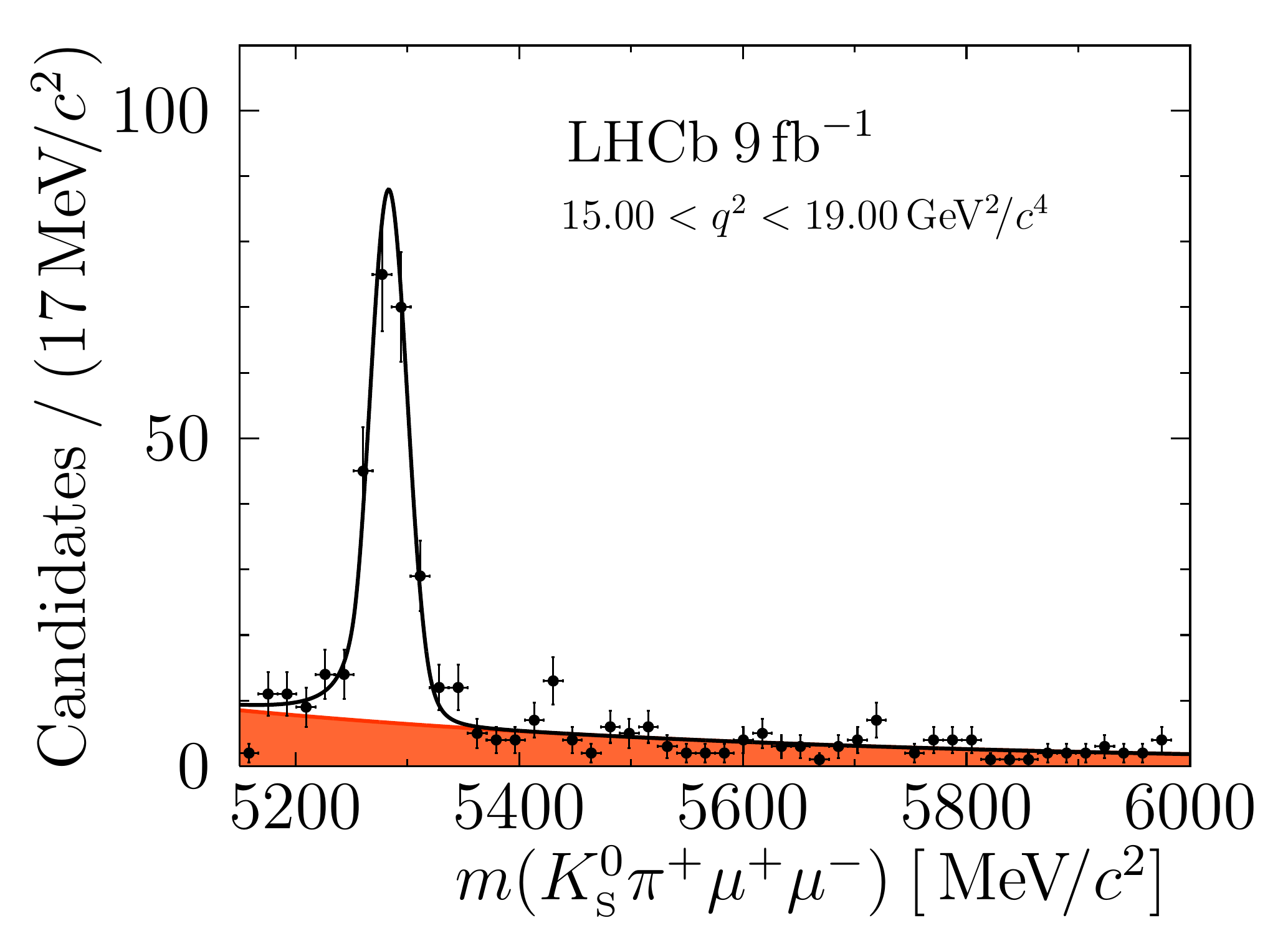}
    \vspace*{-0.6cm}
  \end{center}
  \caption{Projections for the invariant mass $m(\KS\pip\mumu)$ in the ten \qsq intervals. The black points represent the data, while the solid curve shows  the fit  result. The  background  component  is  represented by the orange shaded area.}
  \label{fig:proj_m}
\end{figure}

\begin{figure}[htb]
  \begin{center}
    \includegraphics[width=0.3\linewidth]{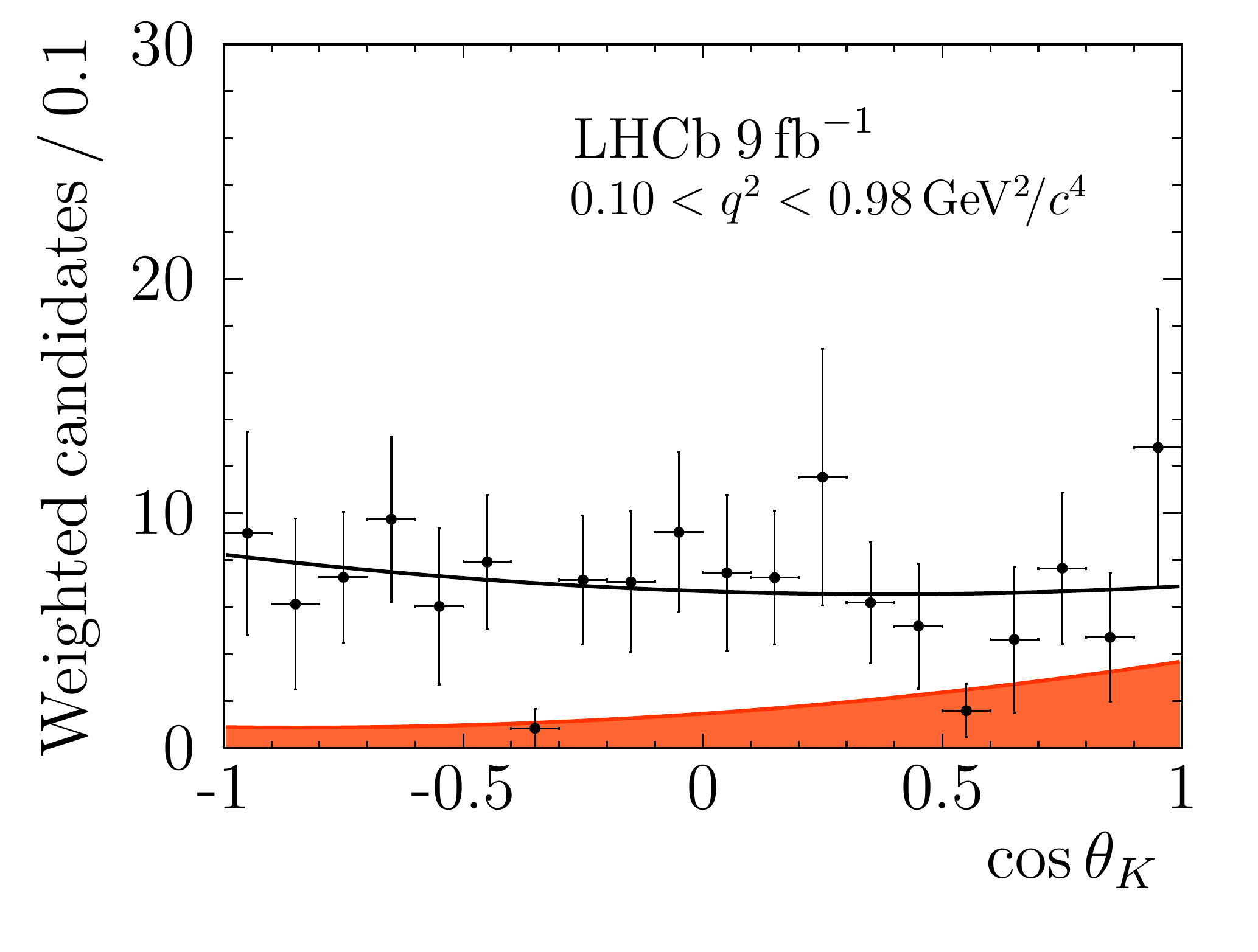}
    \includegraphics[width=0.3\linewidth]{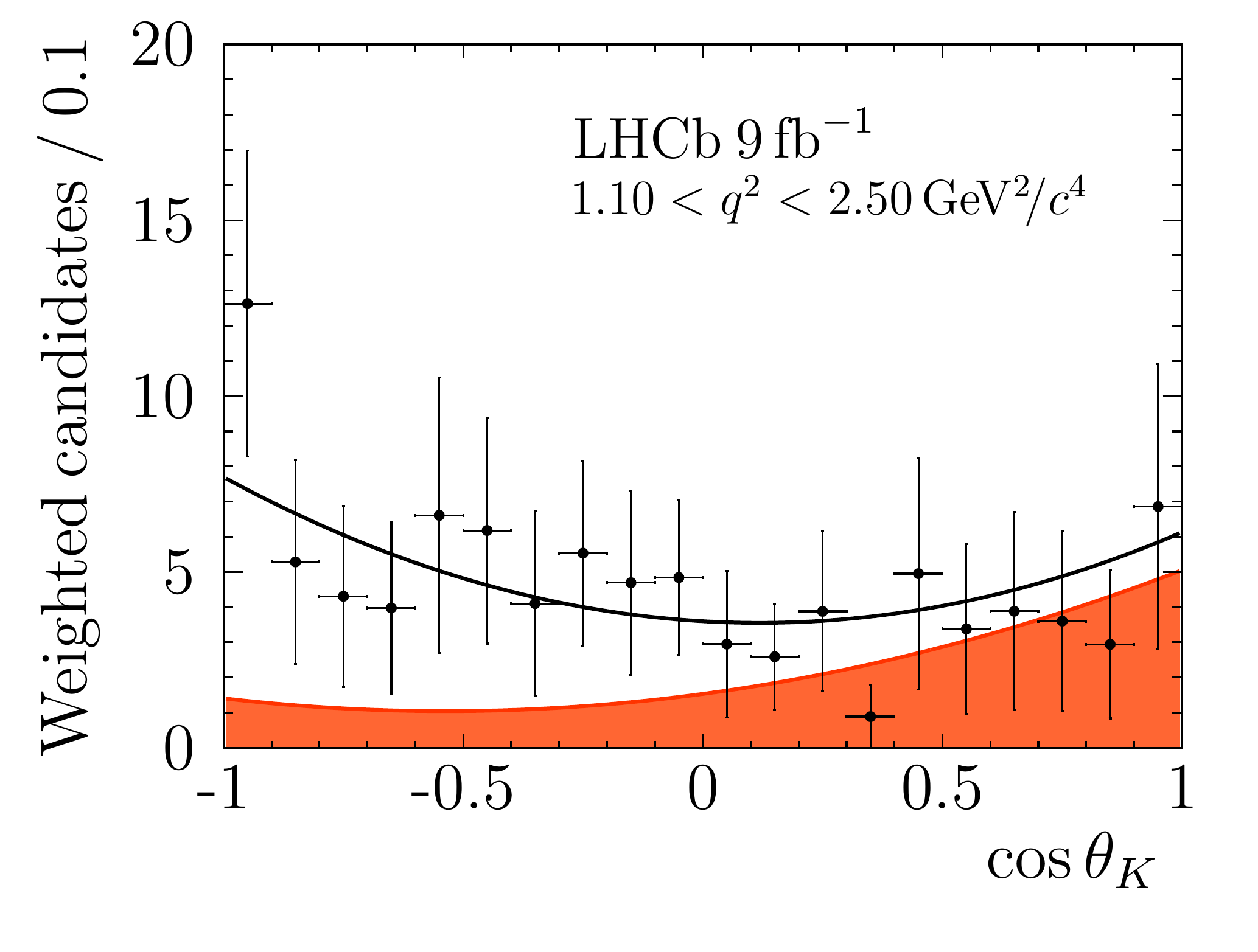}
    \includegraphics[width=0.3\linewidth]{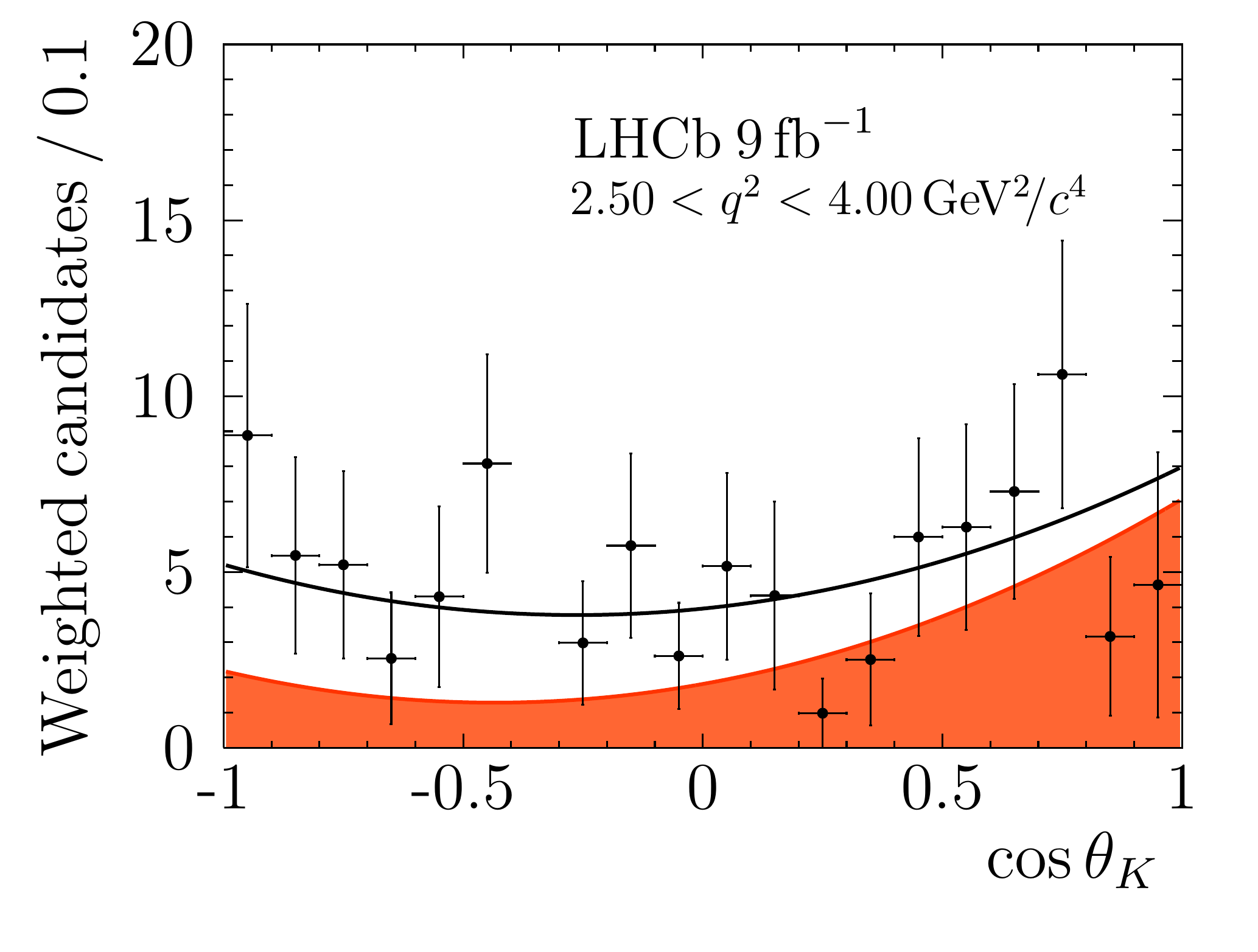}\\
    \includegraphics[width=0.3\linewidth]{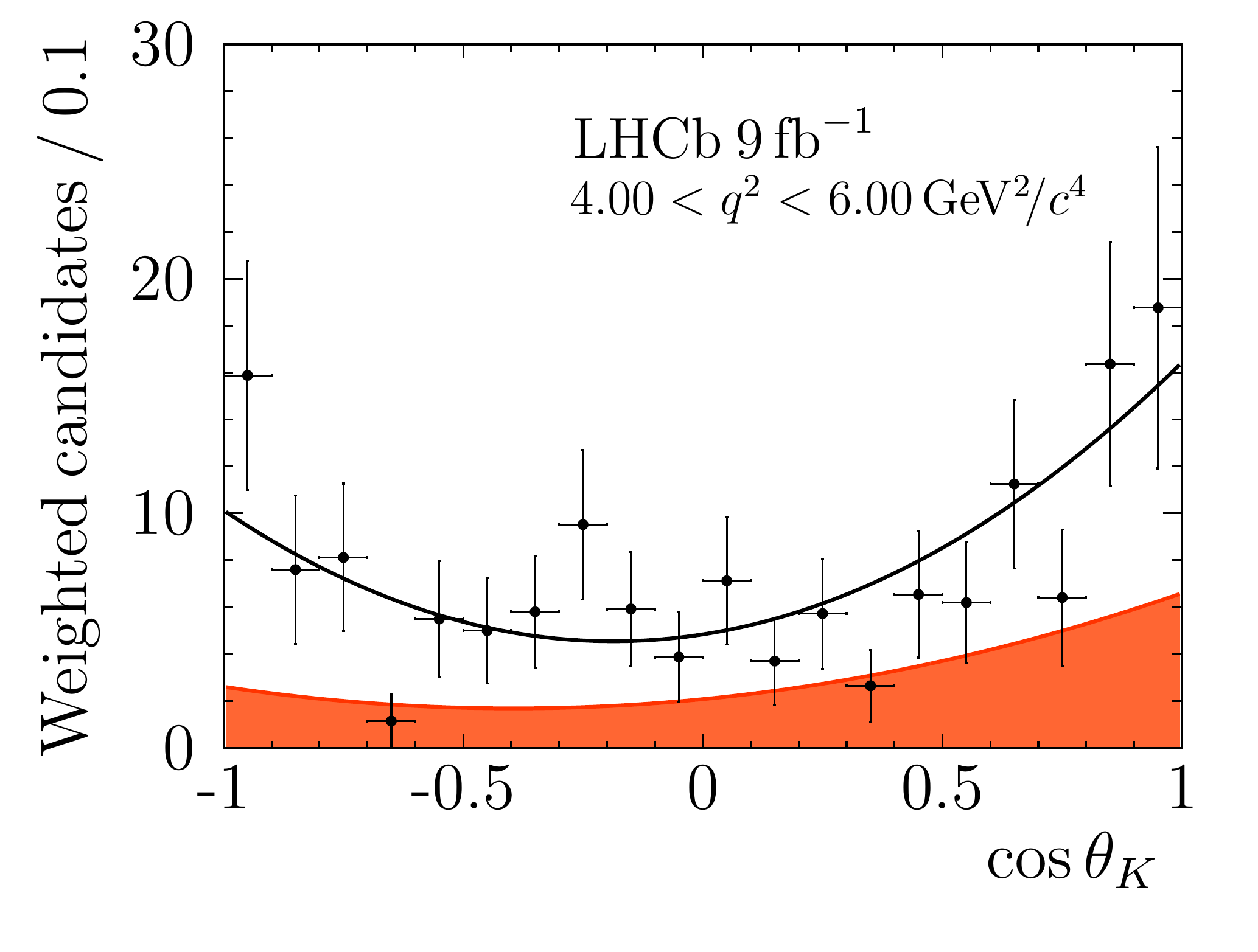}
    \includegraphics[width=0.3\linewidth]{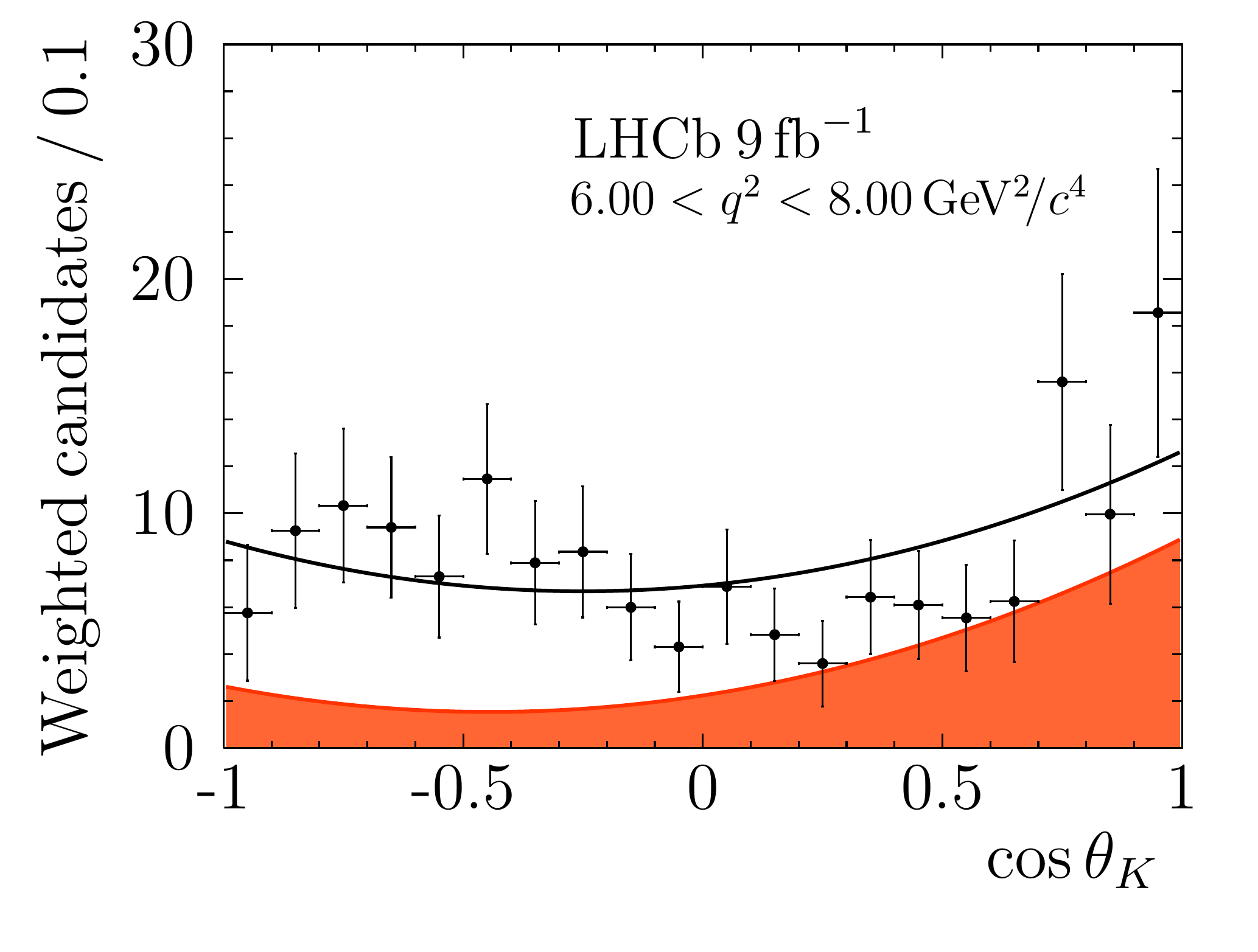}
    \includegraphics[width=0.3\linewidth]{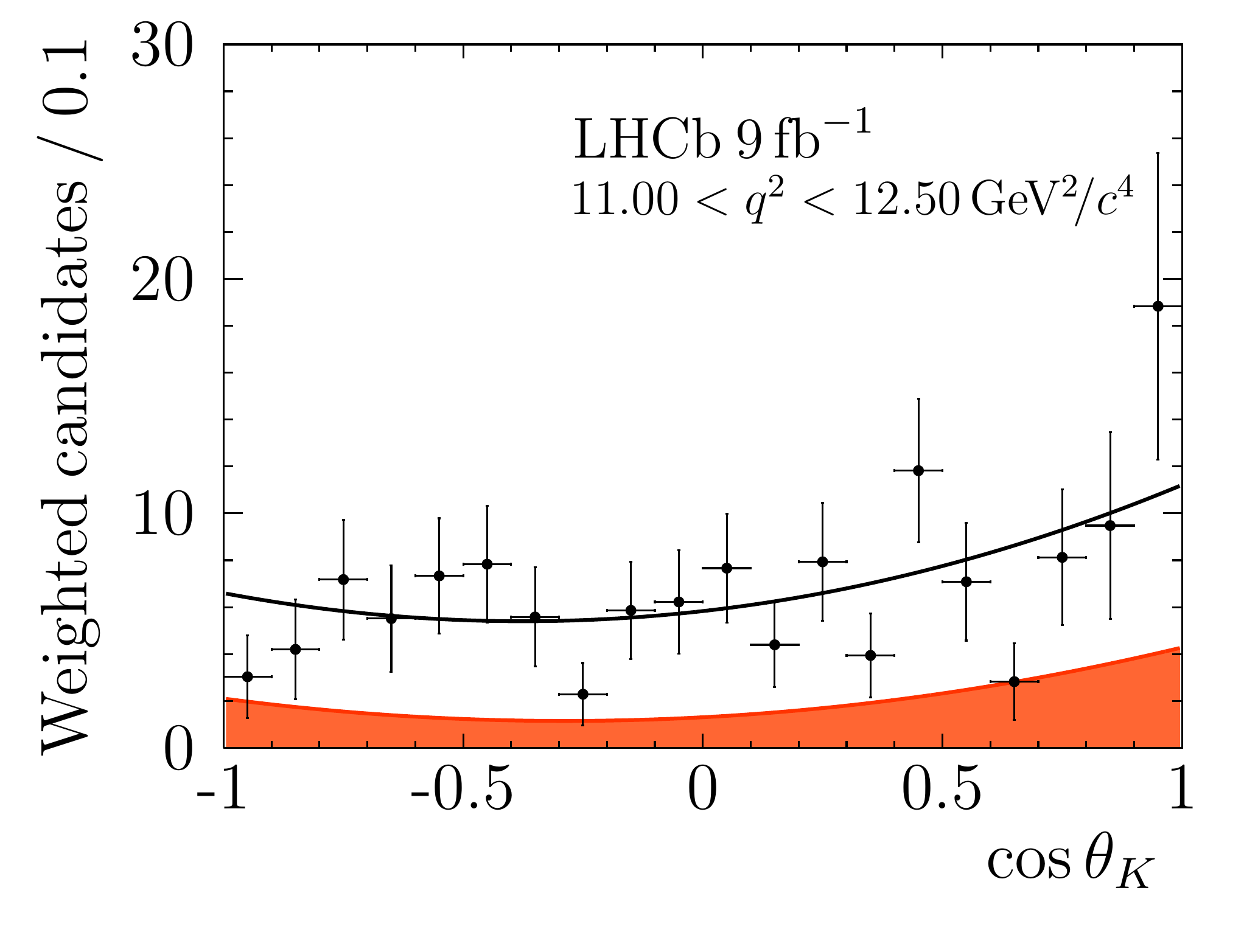}\\
    \includegraphics[width=0.3\linewidth]{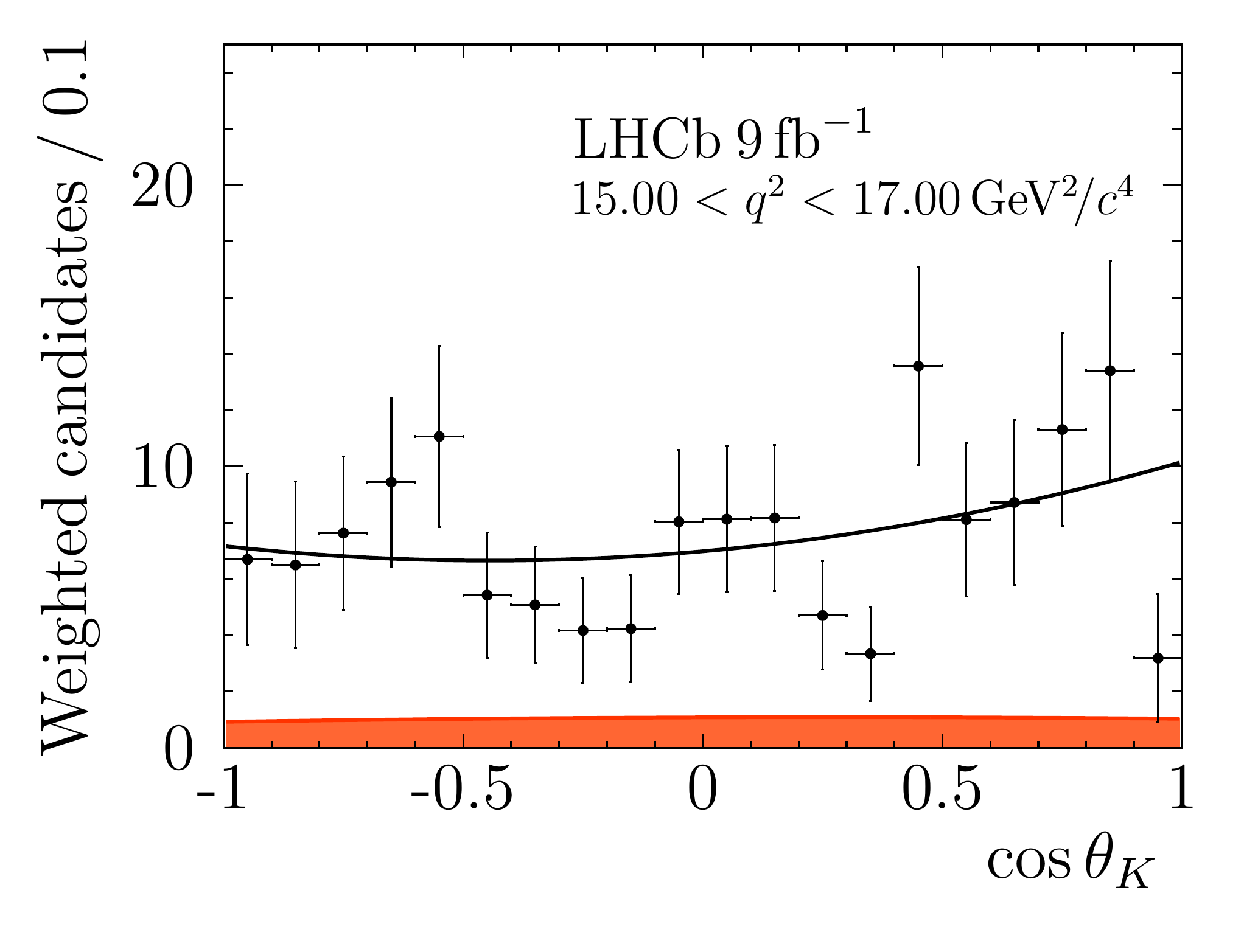}
    \includegraphics[width=0.3\linewidth]{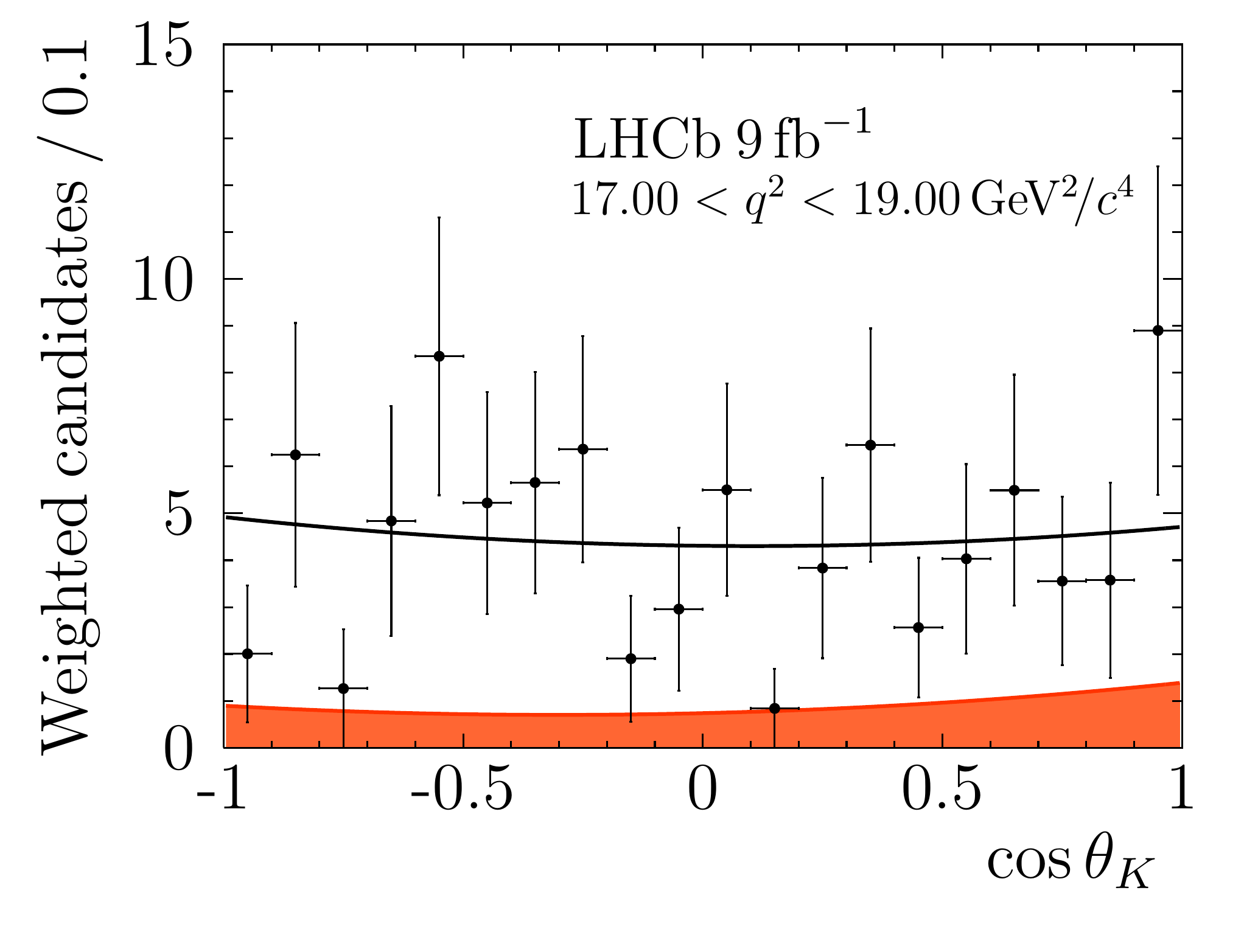}\\
    \includegraphics[width=0.3\linewidth]{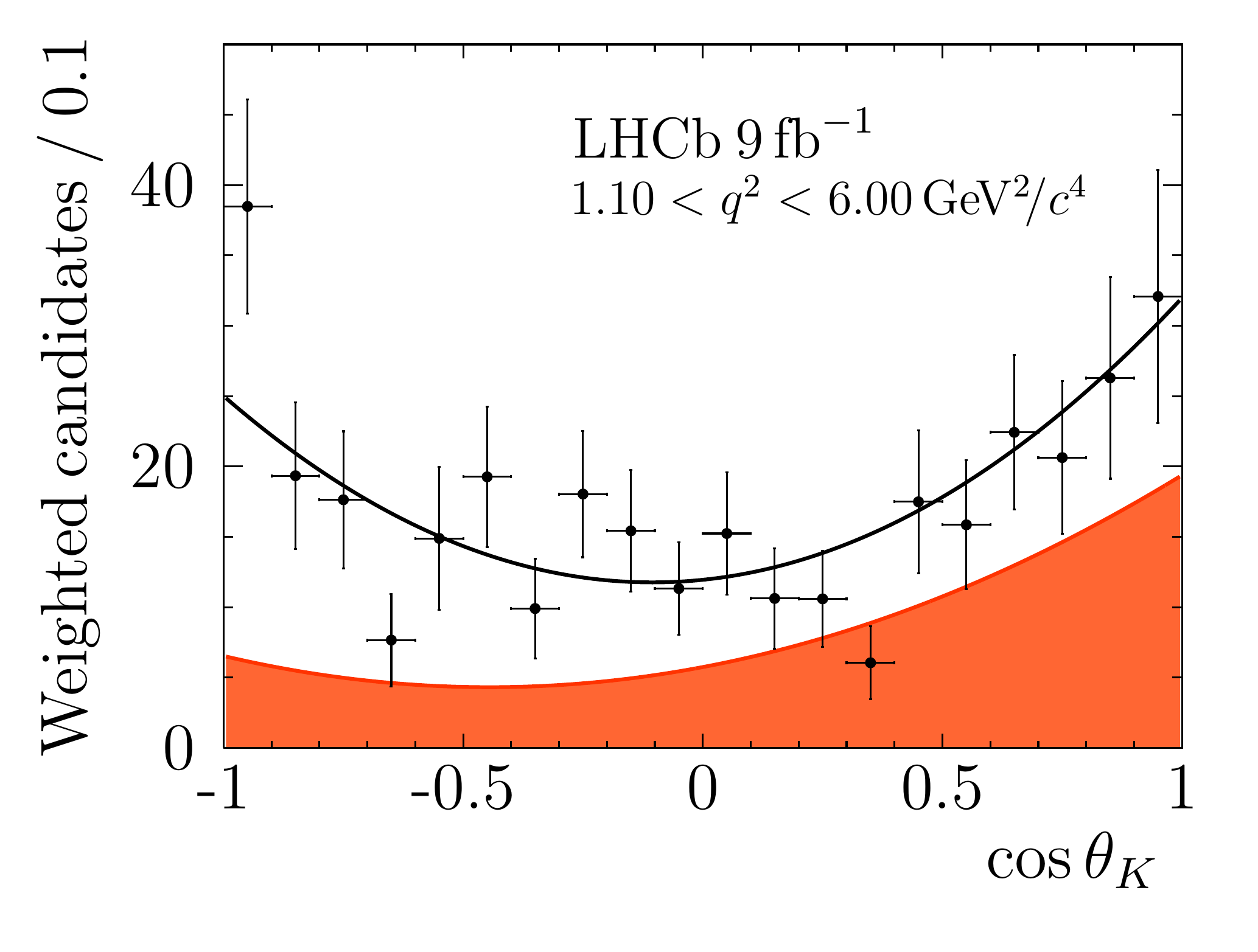}
    \includegraphics[width=0.3\linewidth]{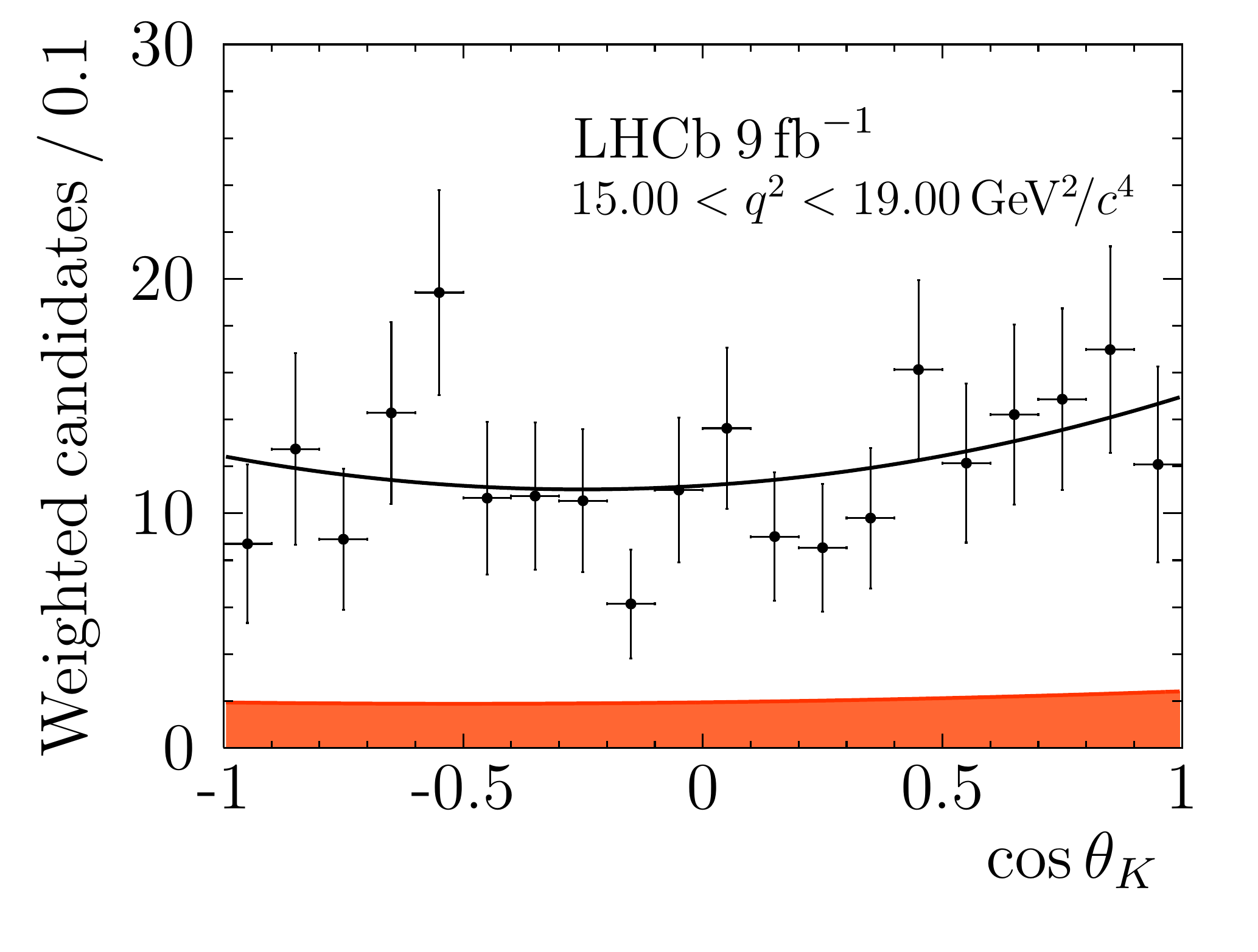}
    \vspace*{-0.6cm}
  \end{center}
  \caption{Projections for the angle \ctk in the ten \qsq intervals. The black points represent the data, while the solid curve shows  the fit  result. The  background  component  is  represented by the orange shaded area. The invariant mass $m(\KS\pip\mumu)$ is required to be within $50\mevcc$ of the measured \Bu meson mass.}
  \label{fig:proj_ctk}
\end{figure}

\begin{figure}[htb]
  \begin{center}
    \includegraphics[width=0.3\linewidth]{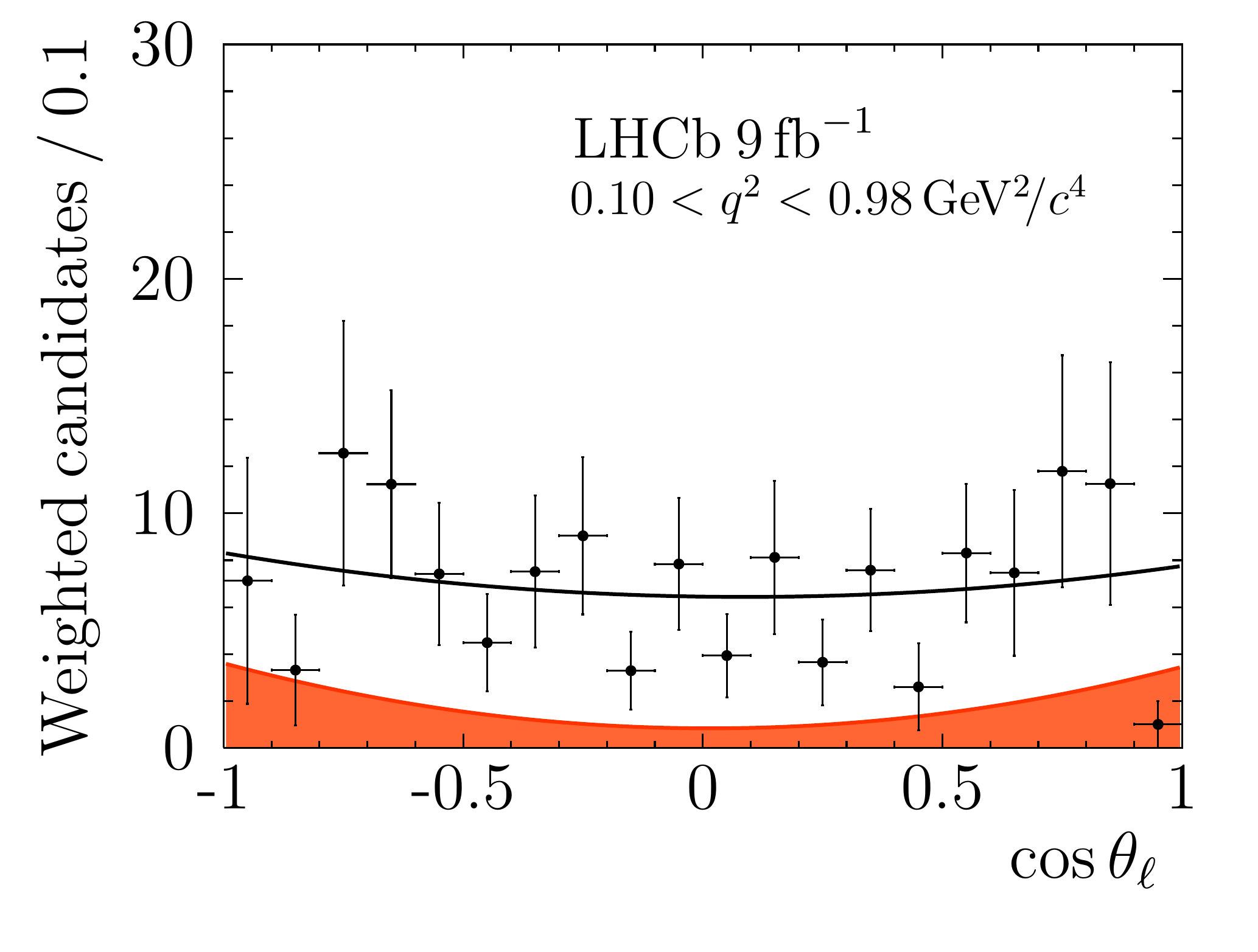}
    \includegraphics[width=0.3\linewidth]{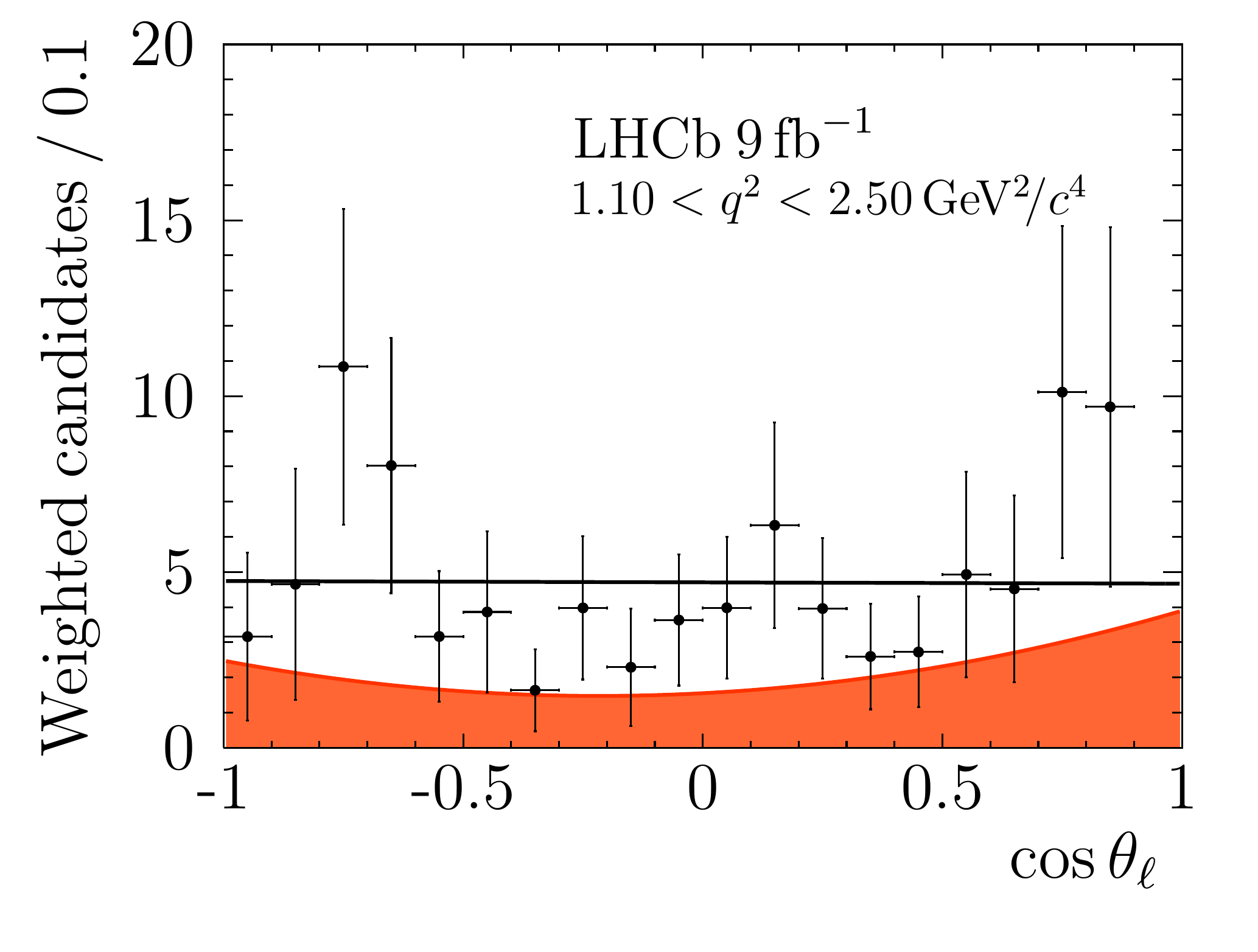}
    \includegraphics[width=0.3\linewidth]{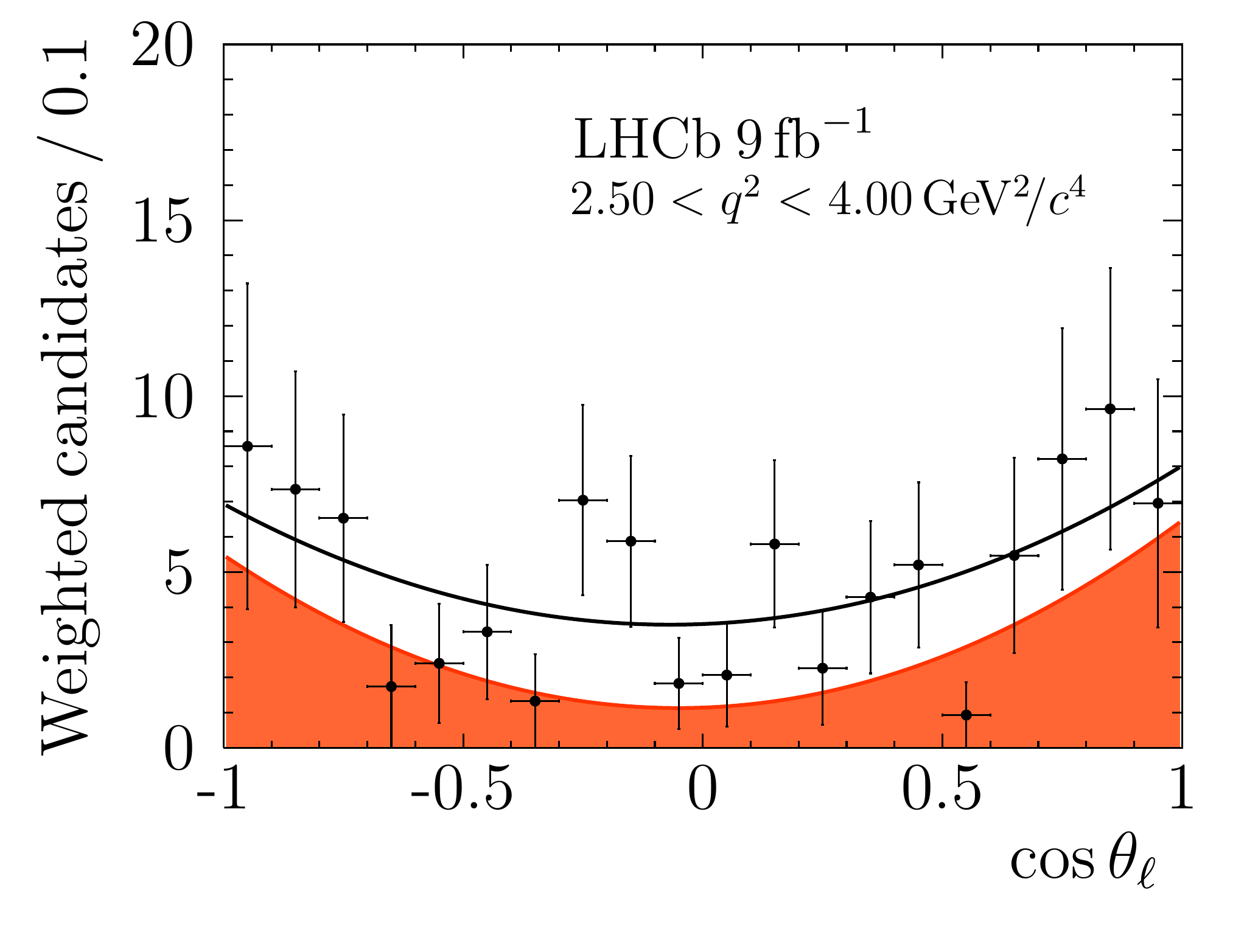}\\
    \includegraphics[width=0.3\linewidth]{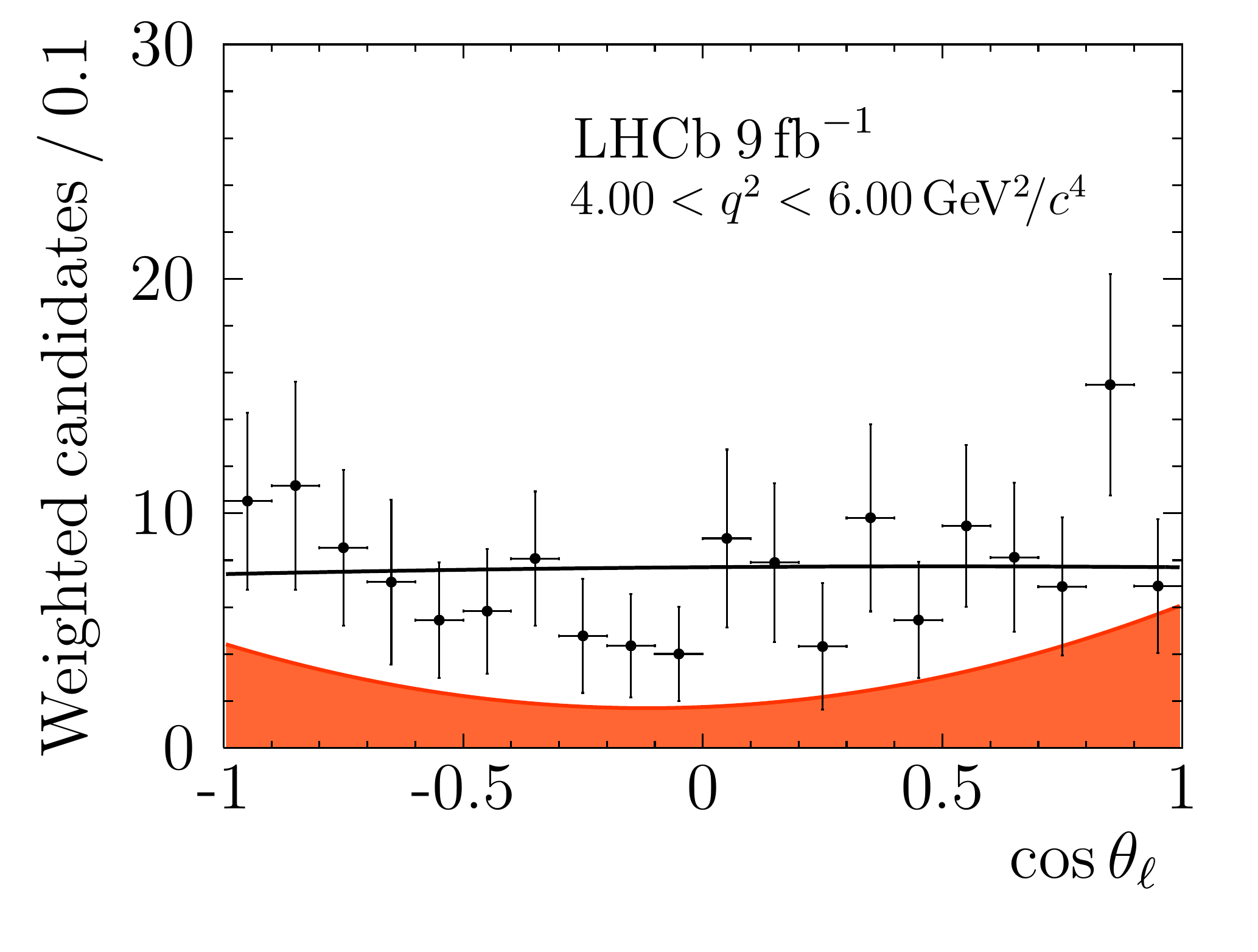}
    \includegraphics[width=0.3\linewidth]{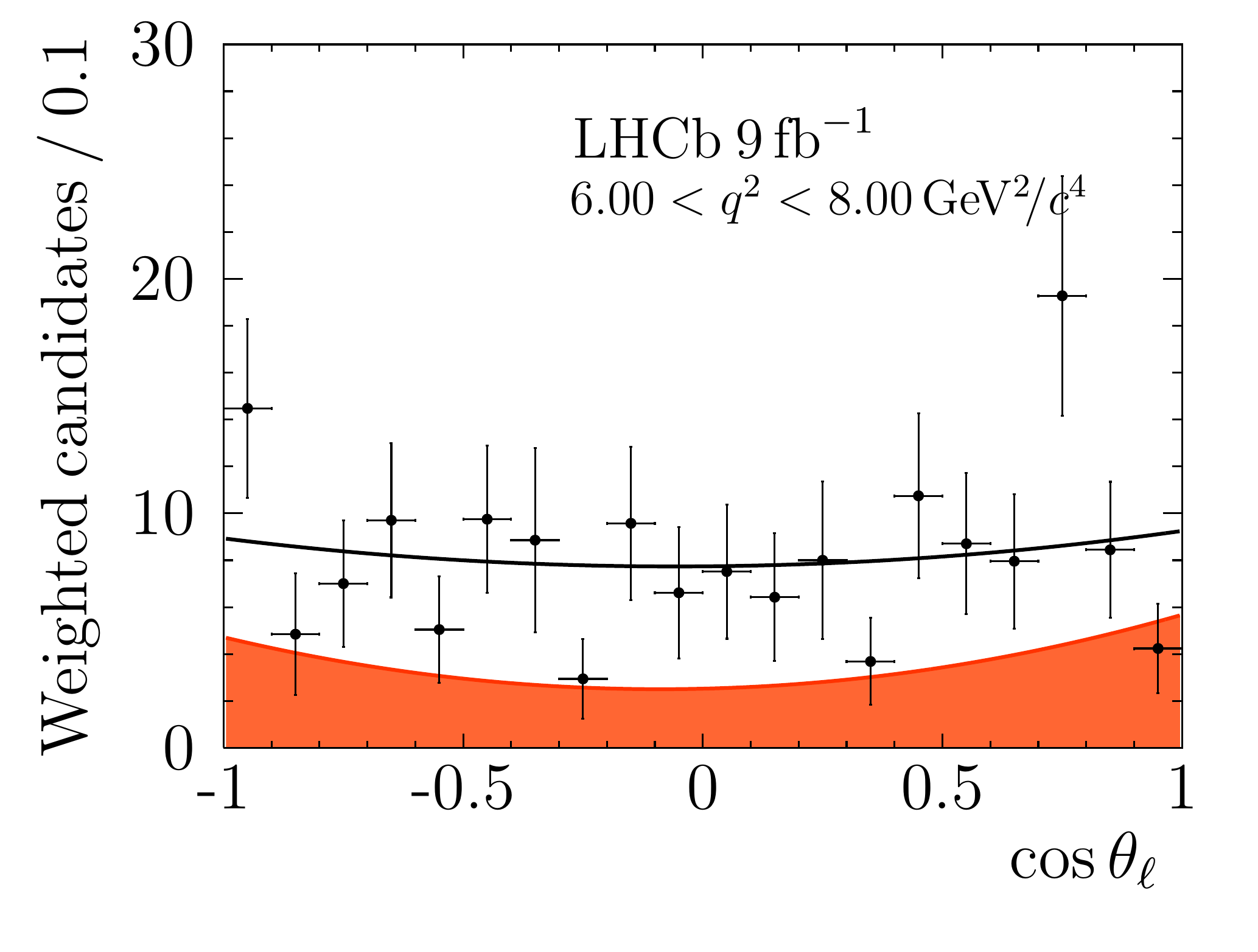}
    \includegraphics[width=0.3\linewidth]{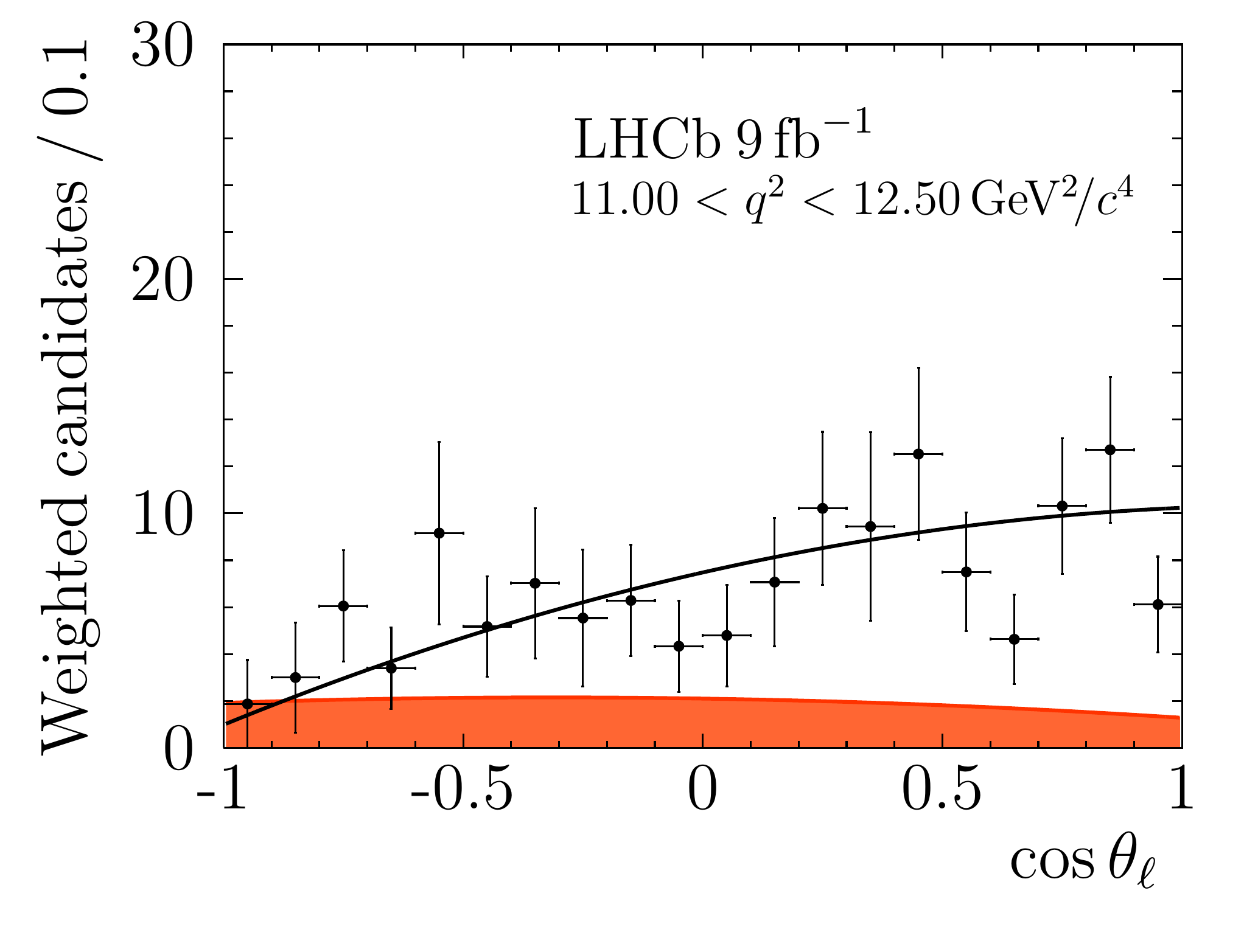}\\
    \includegraphics[width=0.3\linewidth]{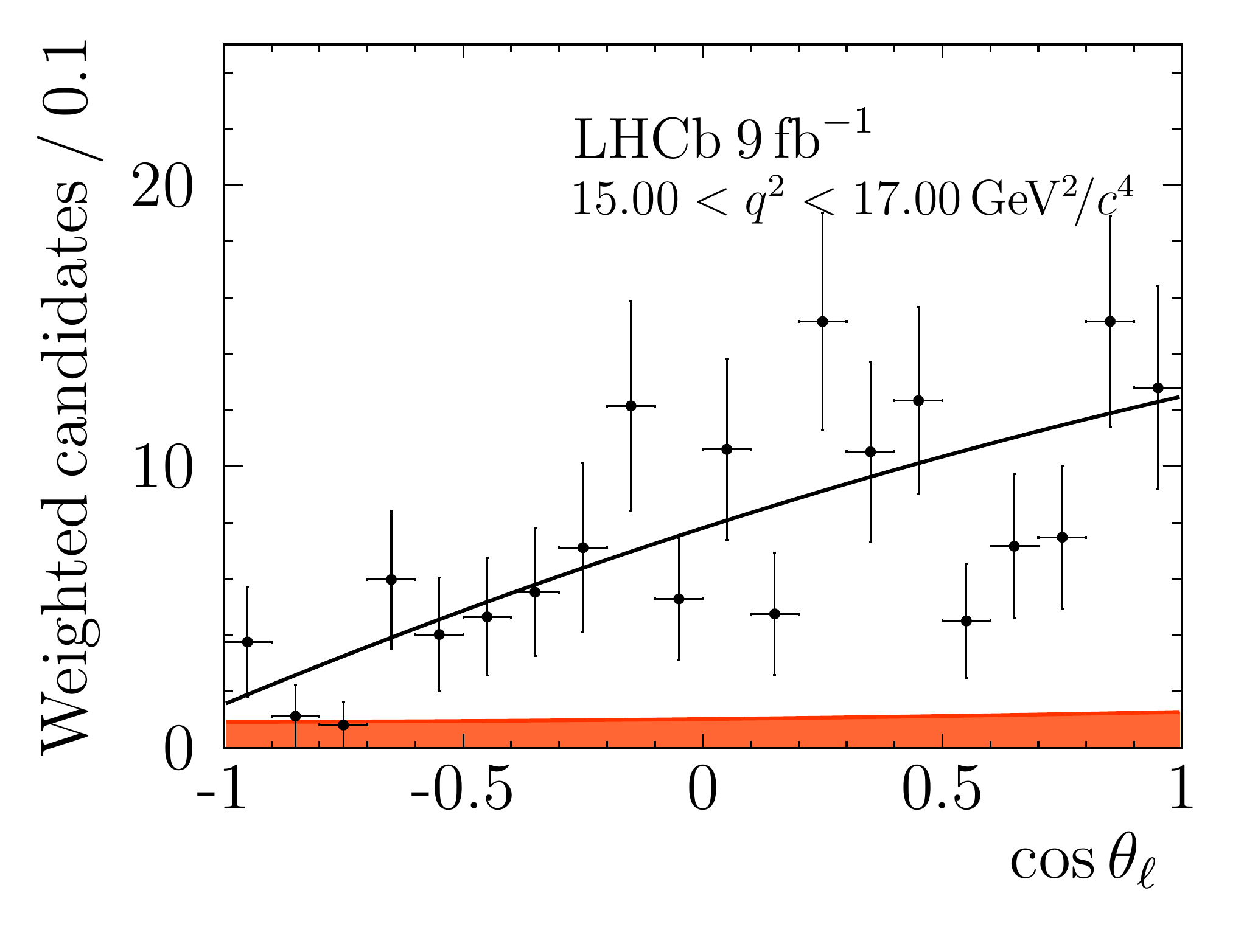}
    \includegraphics[width=0.3\linewidth]{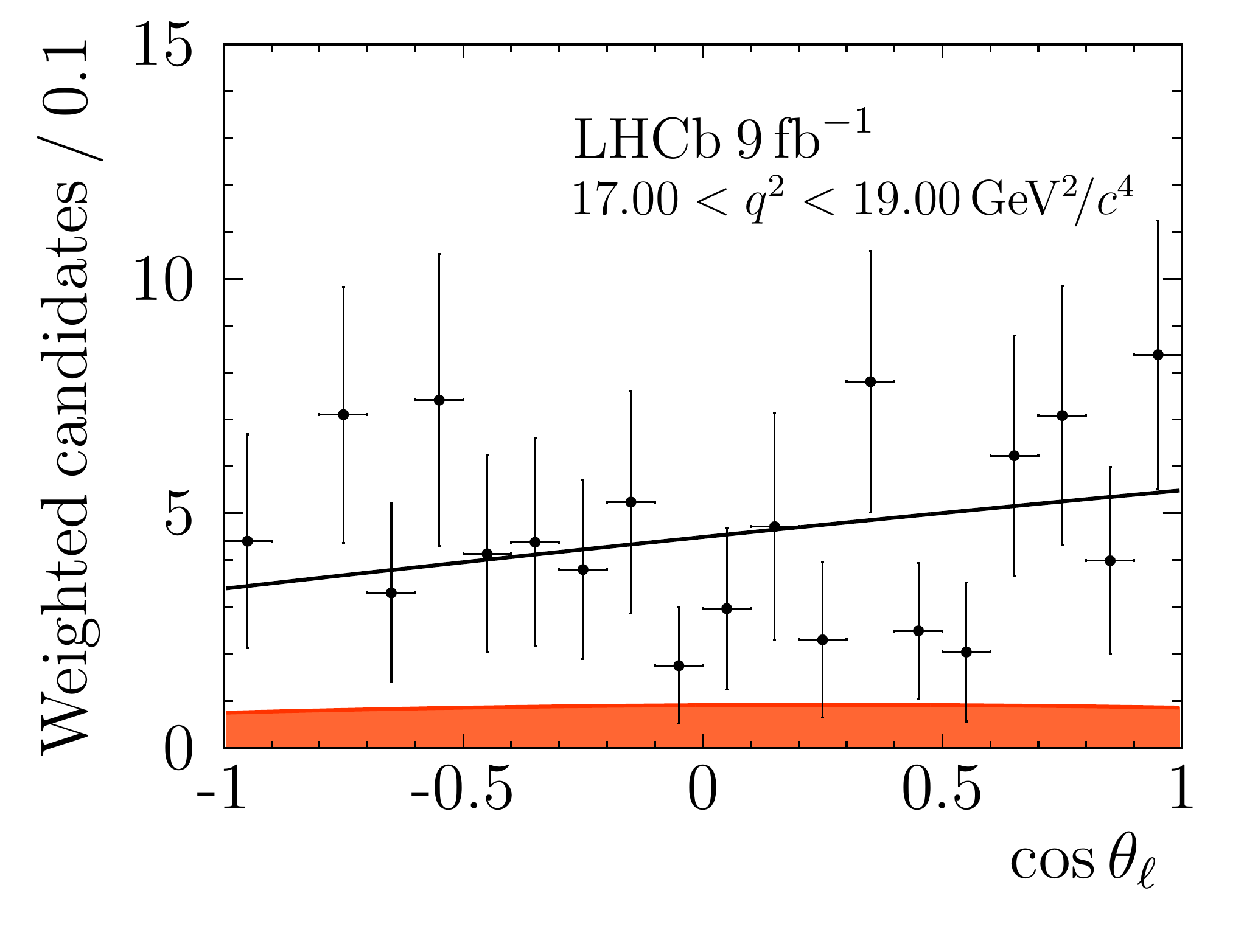}\\
    \includegraphics[width=0.3\linewidth]{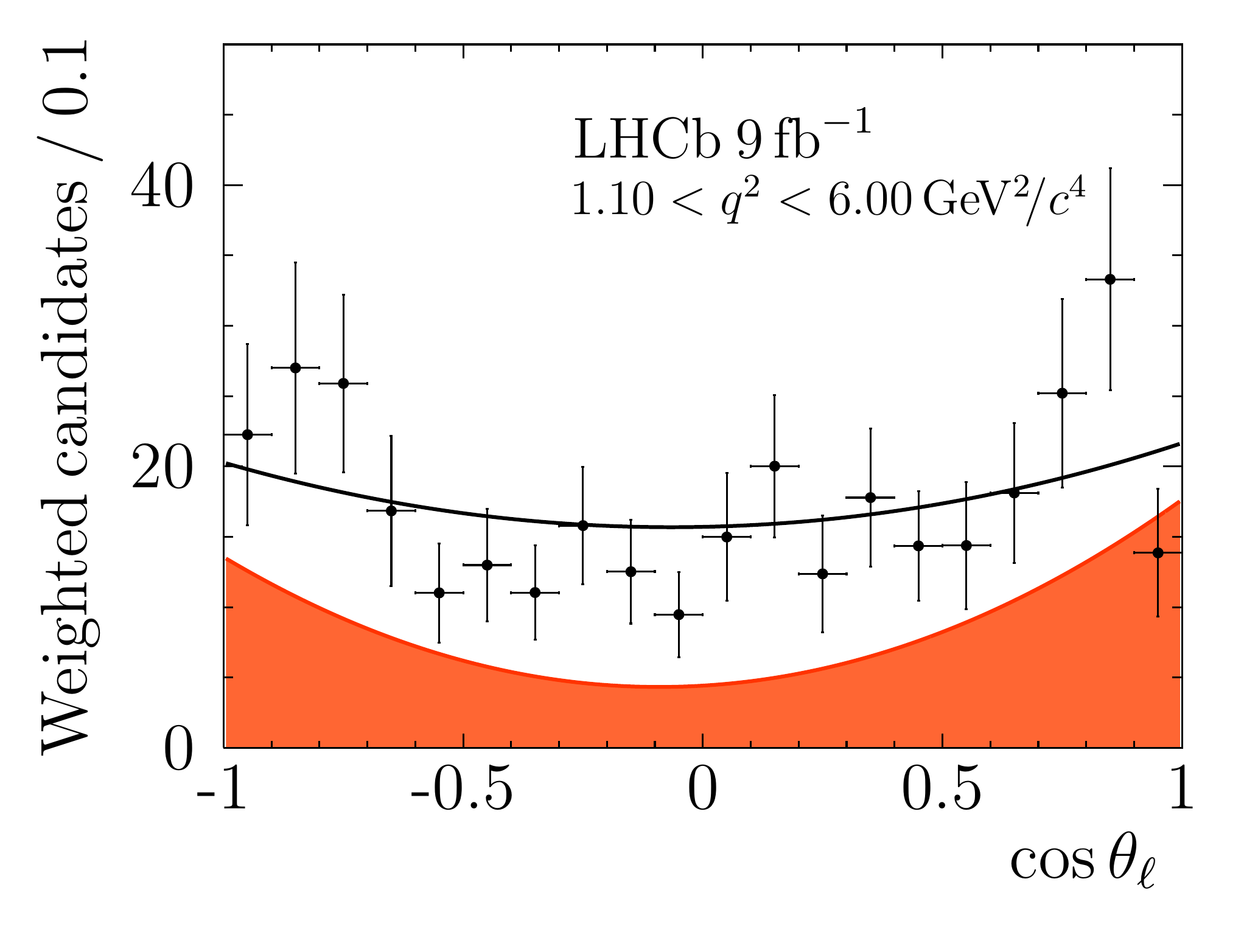}
    \includegraphics[width=0.3\linewidth]{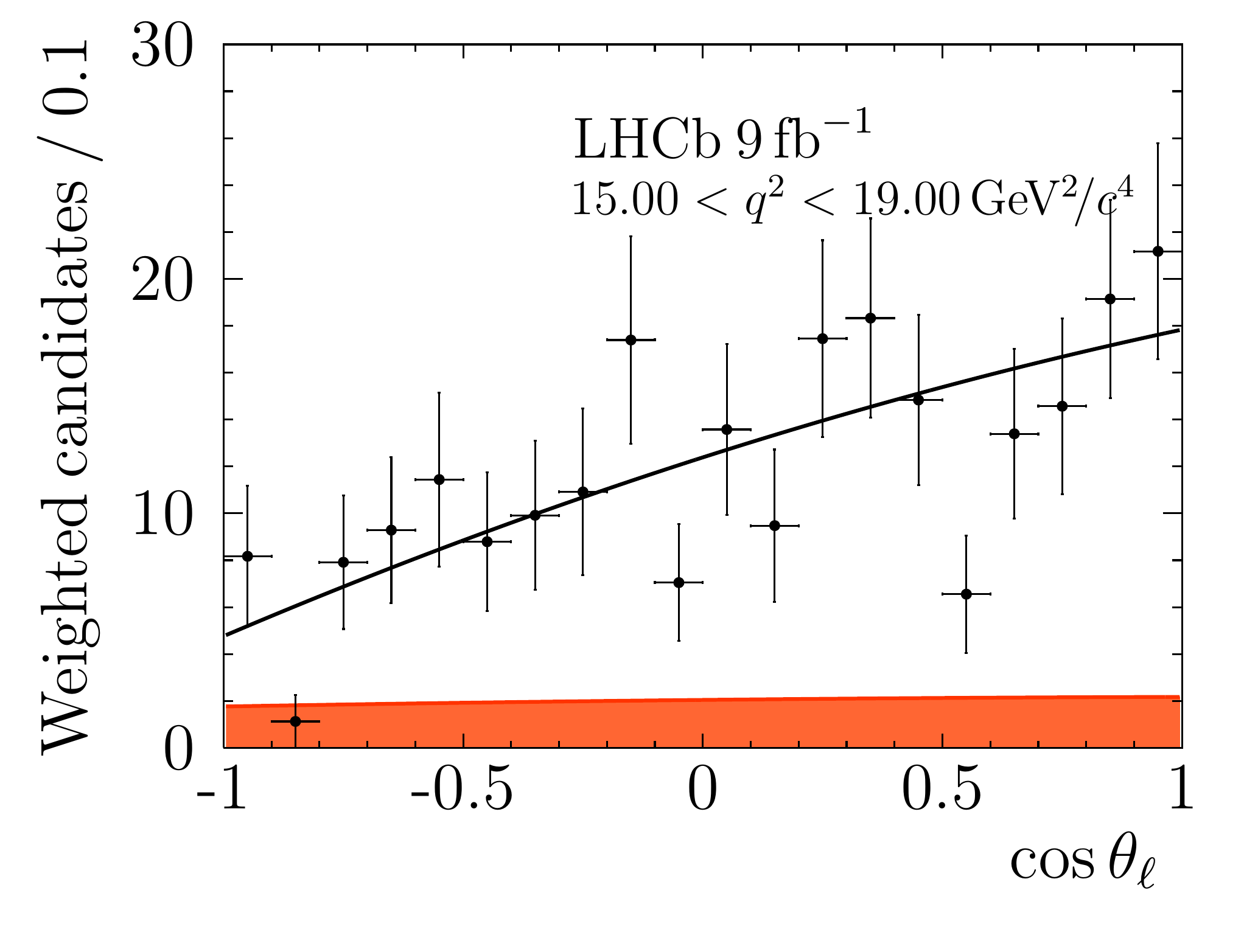}
    \vspace*{-0.6cm}
  \end{center}
  \caption{Projections for the angle \ctl in the ten \qsq intervals. The black points represent the data, while the solid curve shows  the fit  result. The  background  component  is  represented by the orange shaded area. The invariant mass $m(\KS\pip\mumu)$ is required to be within $50\mevcc$ of the measured \Bu meson mass.}
  \label{fig:proj_ctl}
\end{figure}

\begin{figure}[htb]
  \begin{center}
    \includegraphics[width=0.3\linewidth]{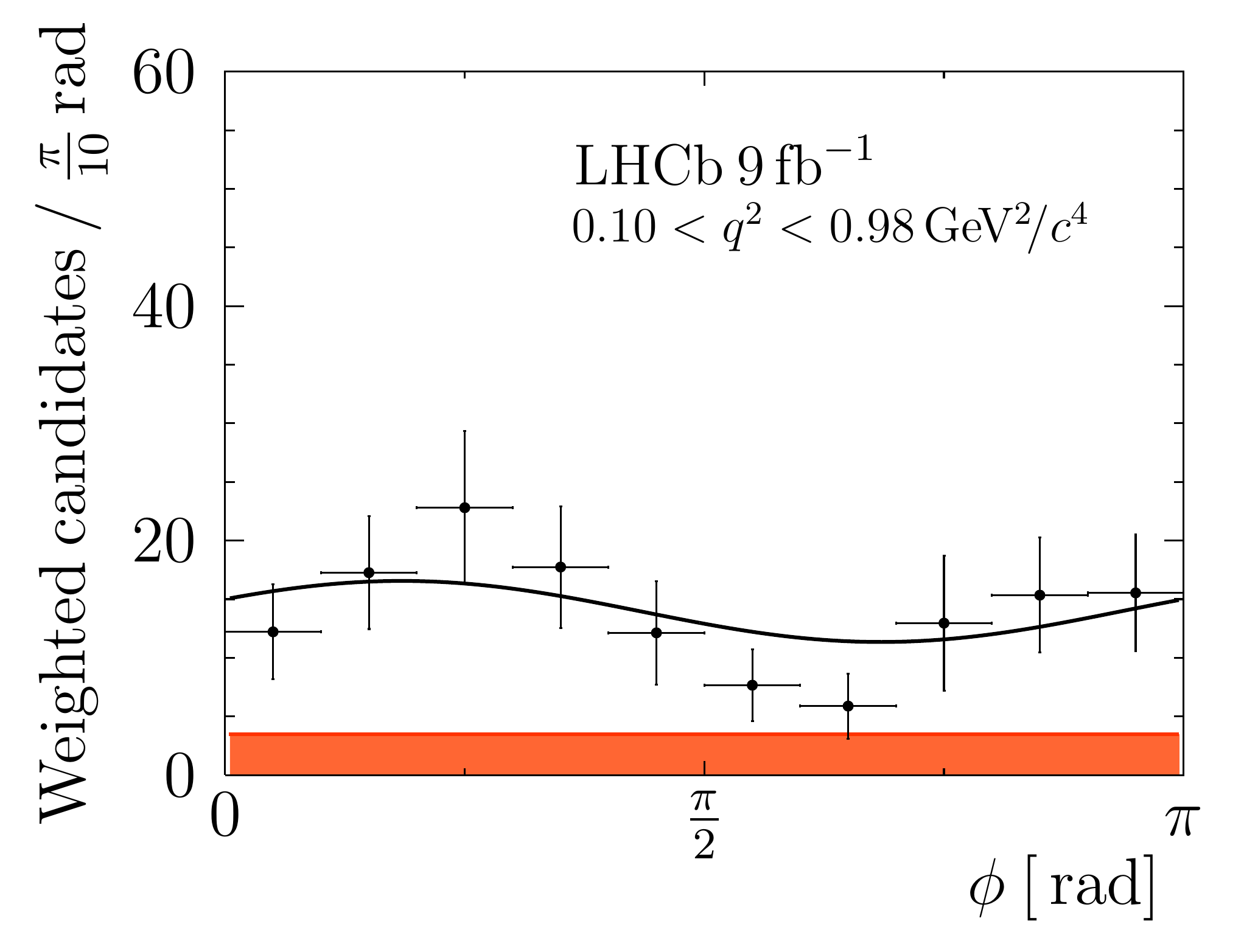}
    \includegraphics[width=0.3\linewidth]{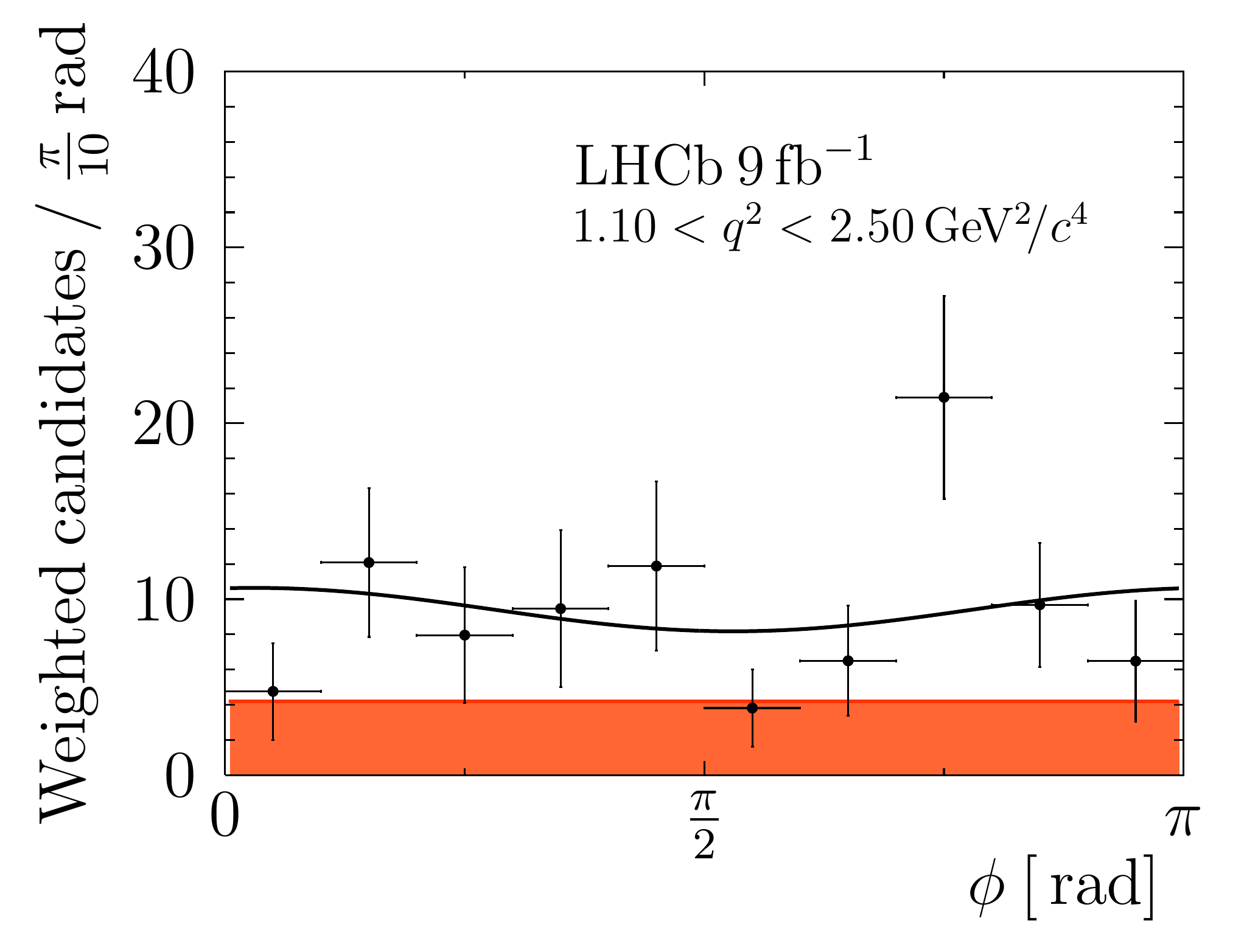}
    \includegraphics[width=0.3\linewidth]{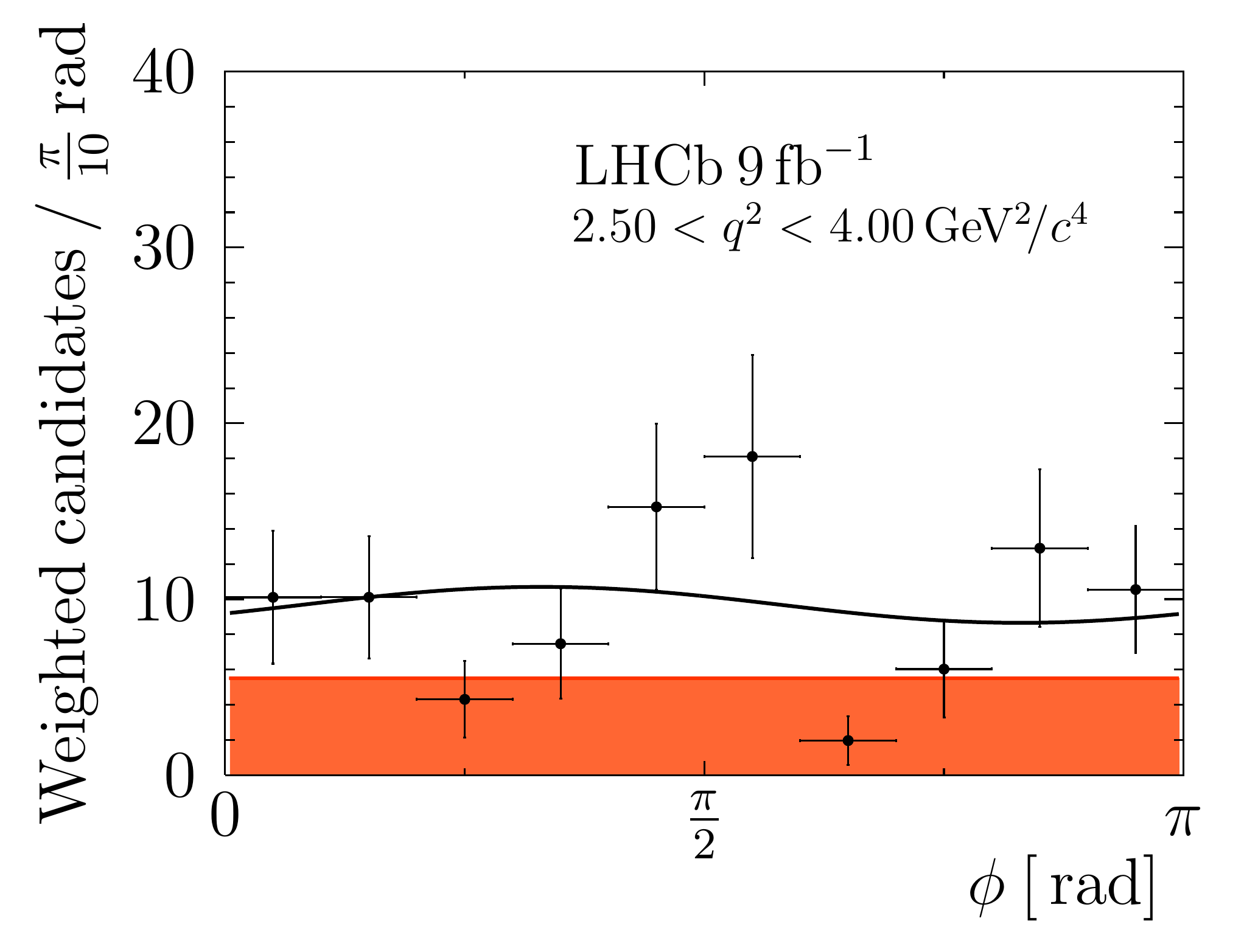}\\
    \includegraphics[width=0.3\linewidth]{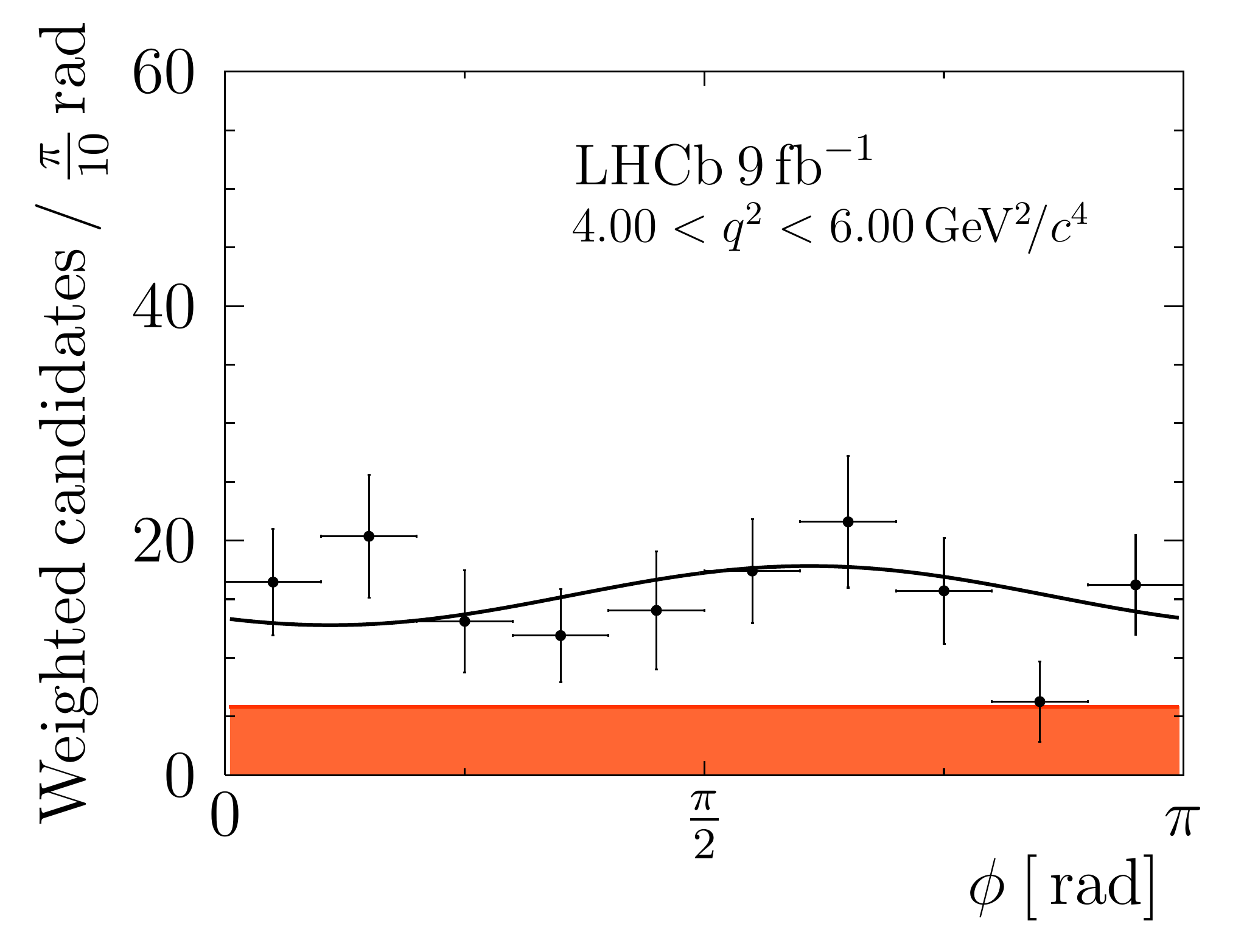}
    \includegraphics[width=0.3\linewidth]{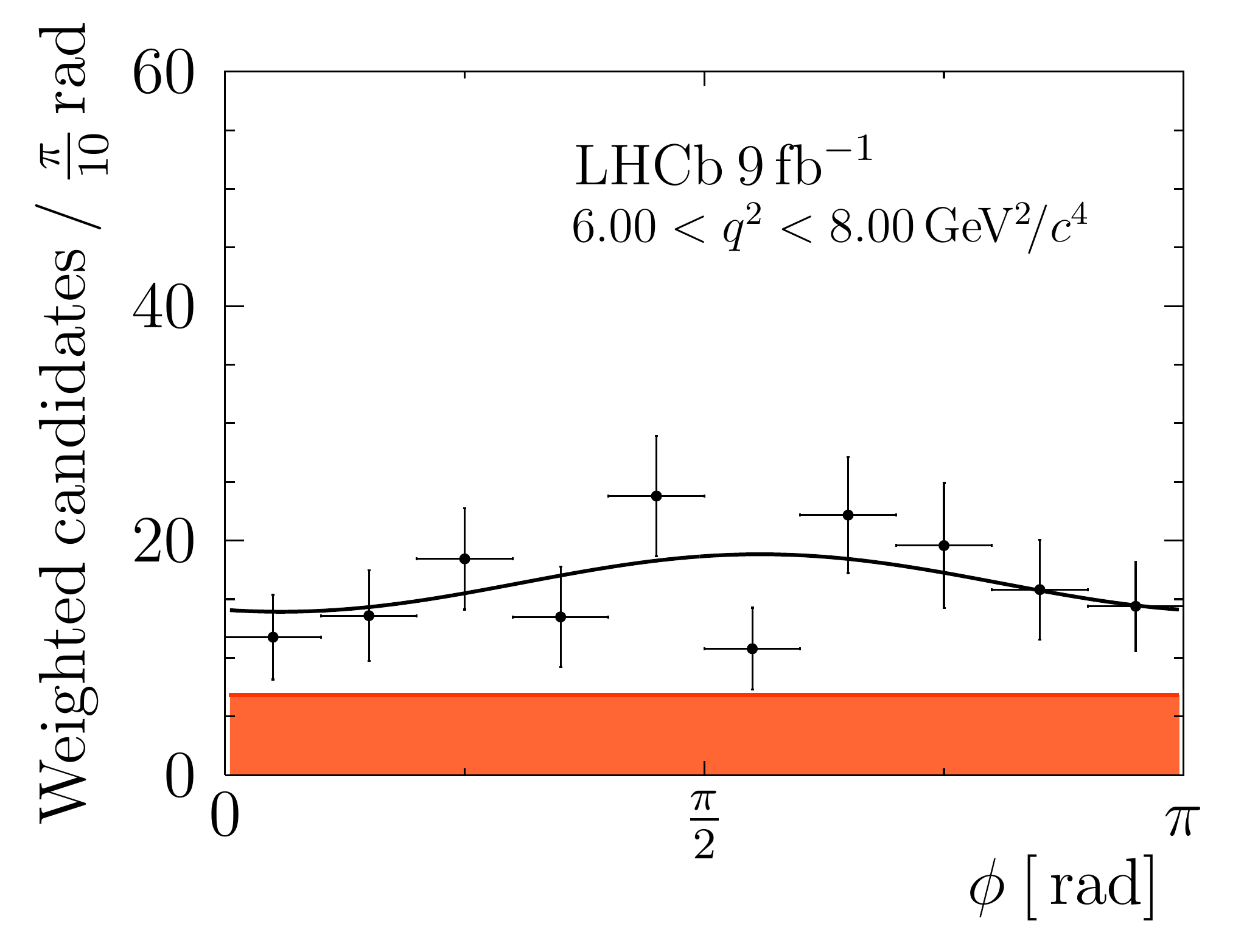}
    \includegraphics[width=0.3\linewidth]{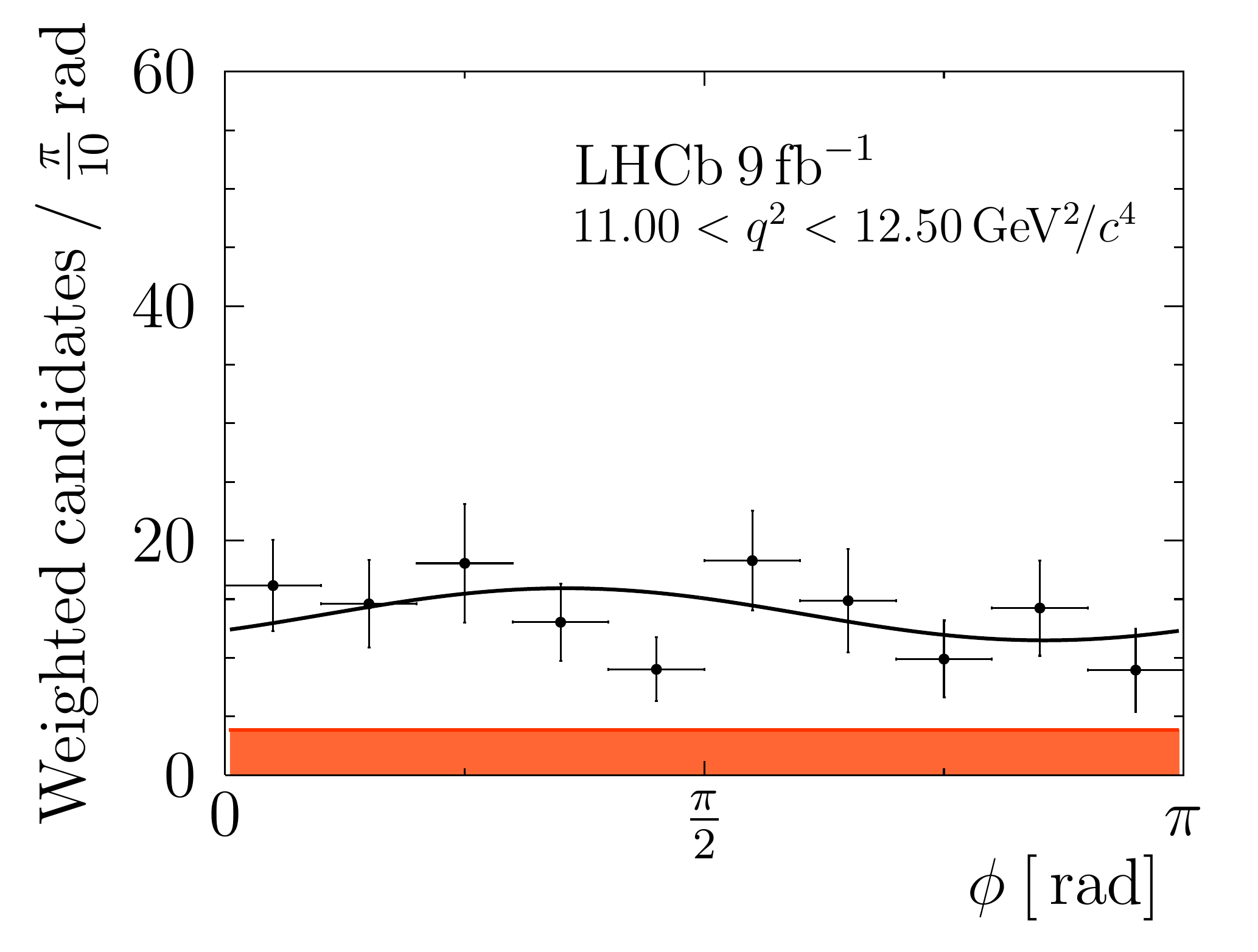}\\
    \includegraphics[width=0.3\linewidth]{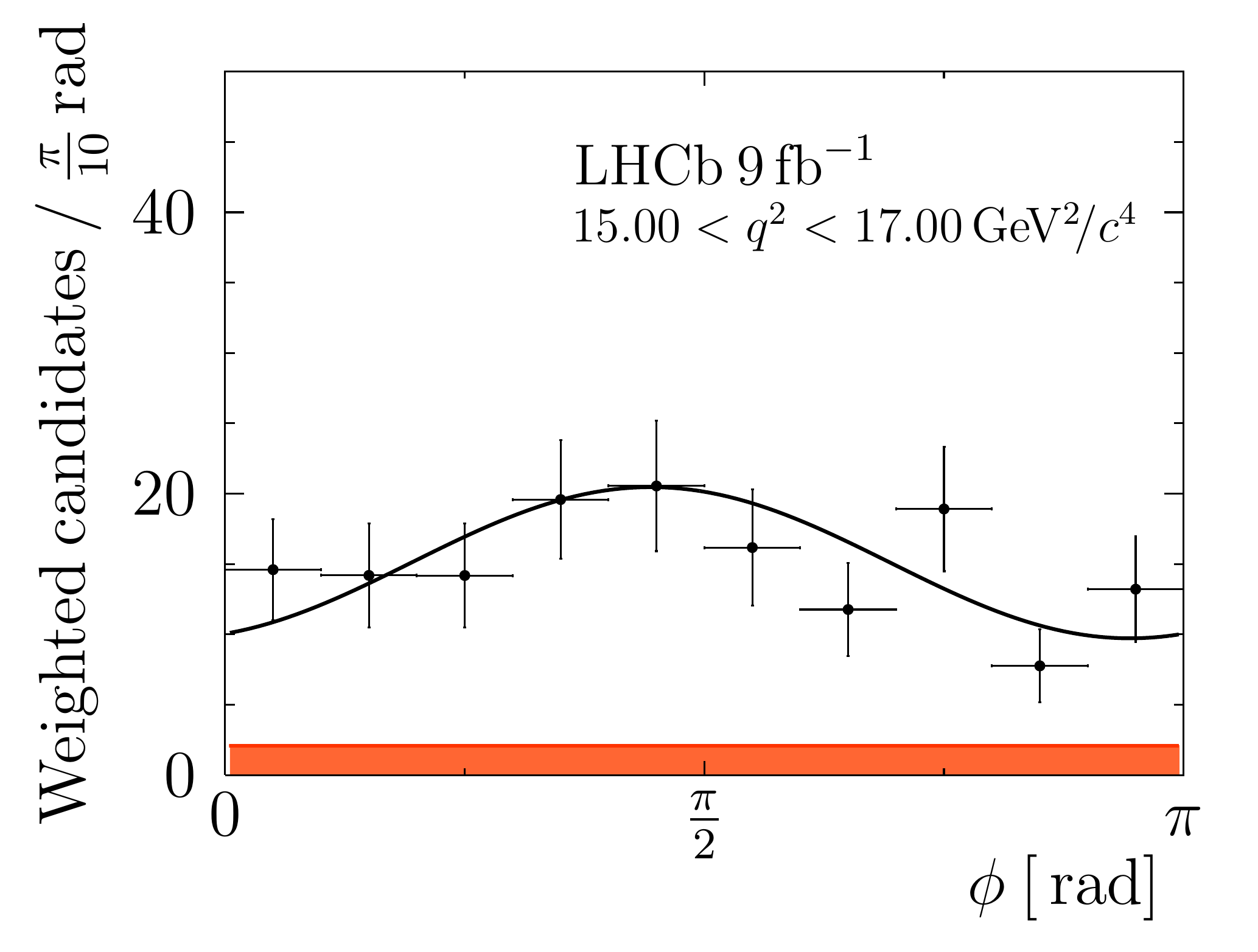}
    \includegraphics[width=0.3\linewidth]{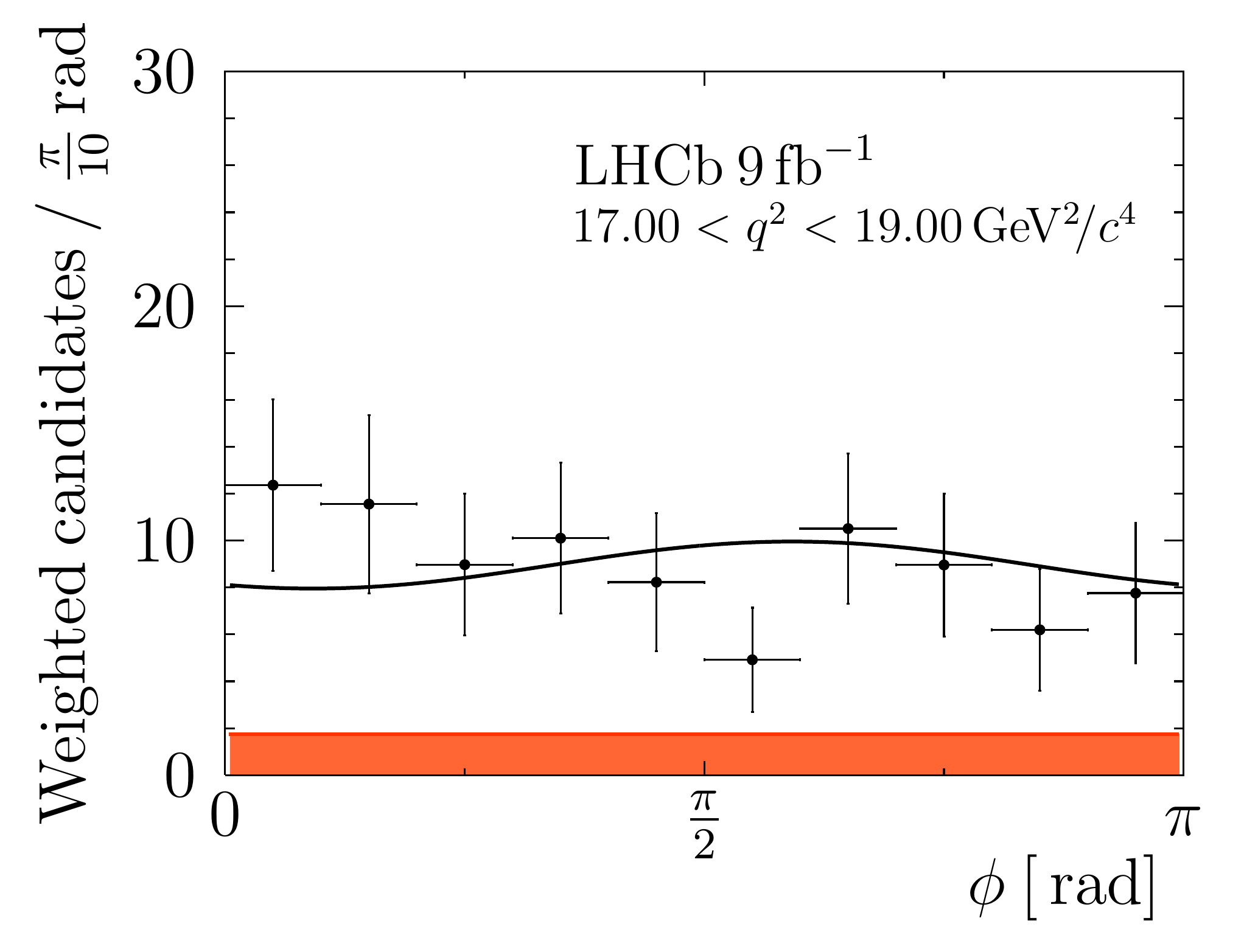}\\
    \includegraphics[width=0.3\linewidth]{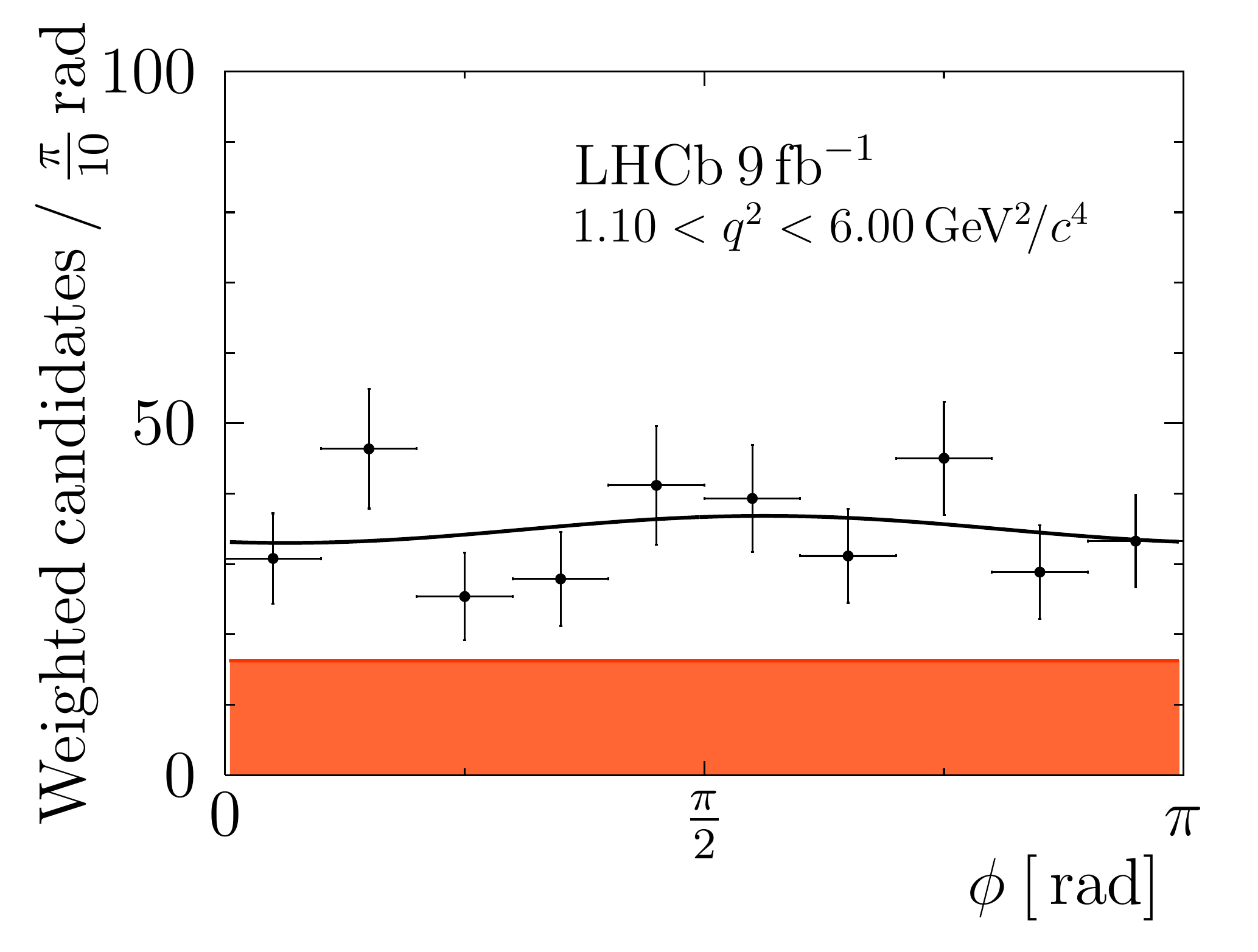}
    \includegraphics[width=0.3\linewidth]{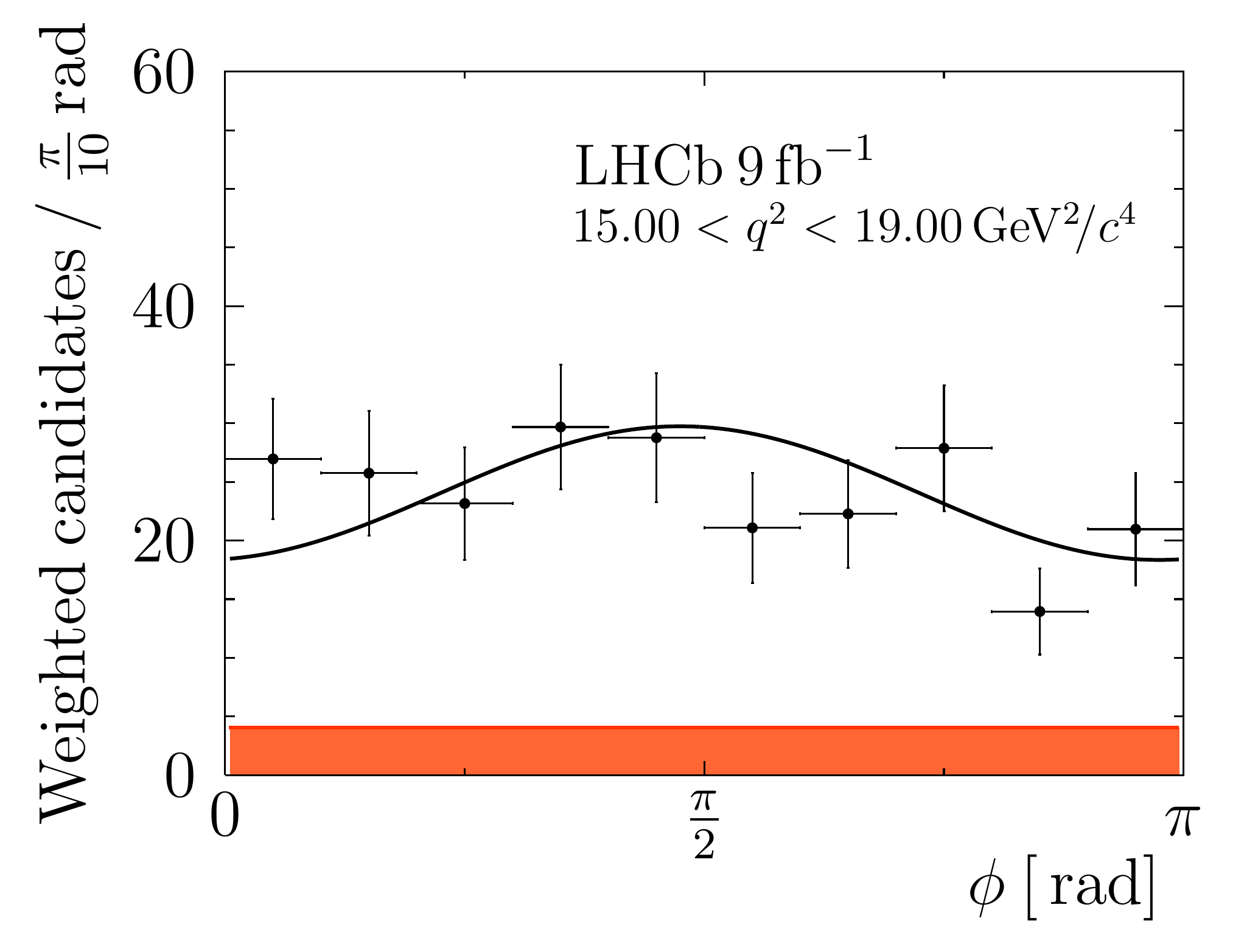}
    \vspace*{-0.6cm}
  \end{center}
  \caption{Projections for the angle $\phi$ in the ten \qsq intervals. The black points represent the data, while the solid curve shows  the fit  result. The  background  component  is  represented by the orange shaded area. The invariant mass $m(\KS\pip\mumu)$ is required to be within $50\mevcc$ of the measured \Bu meson mass.}
  \label{fig:proj_phi}
\end{figure}

The full set of results for both sets of angular observables is presented in graphical form in Sec.~\ref{sec:SandPPlots} and in tabular form in Sec.~\ref{sec:SandPTables}. The correlations between the angular observables are given in Sec.~\ref{sec:correlationsS} and Sec.~\ref{sec:correlationsP} for $S_{i}$ and $P_{i}^{(\prime)}$ observables, respectively. A summary of the systematic uncertainties is given in Sec.~\ref{sec:systematics}.
The signal yields in each \qsq interval are given in Table~\ref{tab:binyields}.
The projections of the data in $m(\KS\pip\mumu)$, \ctk, \ctl and $\phi$ using the angular fold with the transformation $\phi\to\phi+\pi$, for $\phi< 0$, are given in Figs.~\ref{fig:proj_m} - \ref{fig:proj_phi} along with the fitted probability density functions.

\section{Graphical results for the \boldmath{$S_{i}$} and \boldmath{$P_{i}^{(\prime)}$} observables}\label{sec:SandPPlots}
\begin{figure}[htb]
  \begin{center}
    \includegraphics[width=0.49\linewidth]{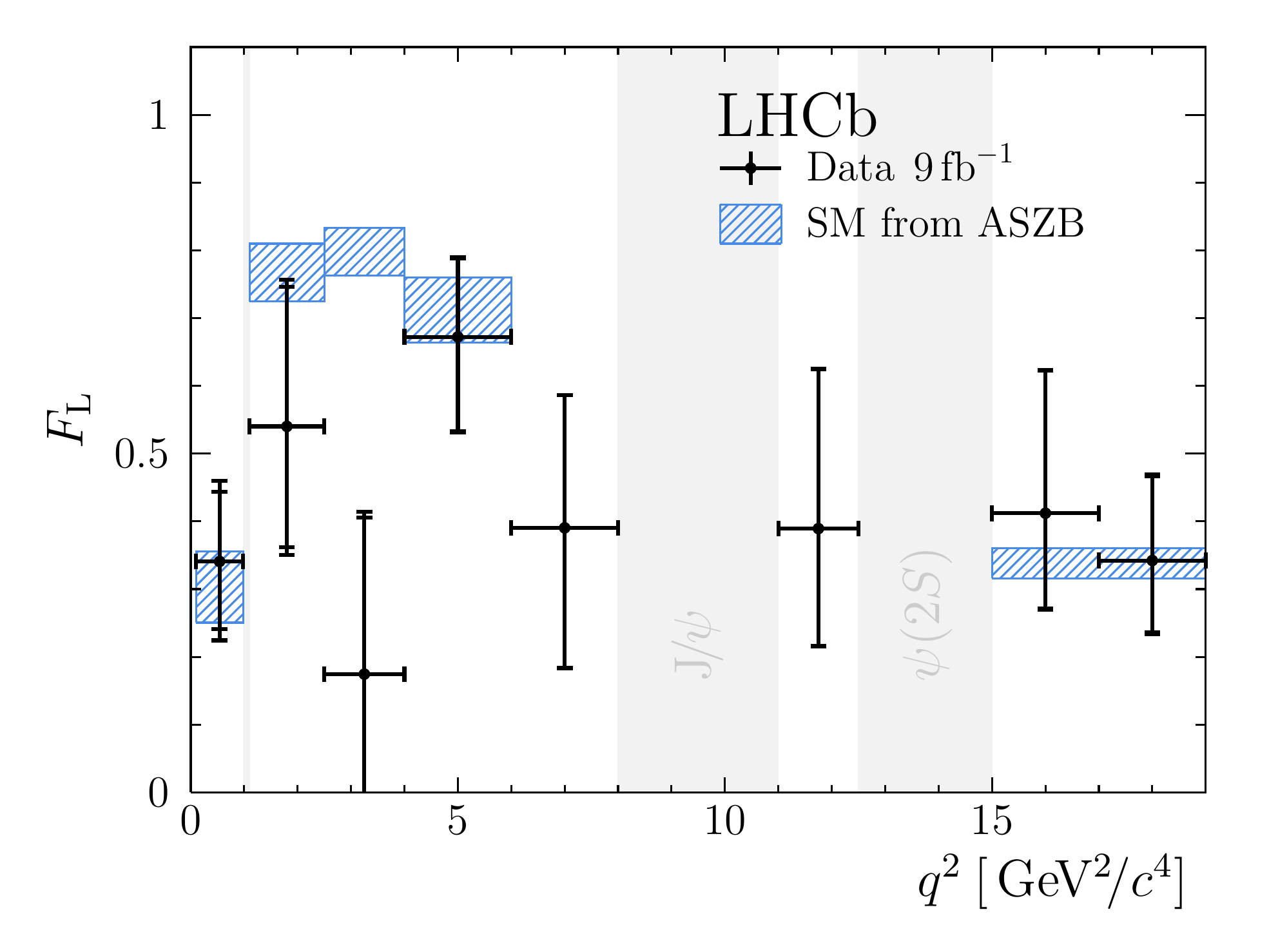}
    \includegraphics[width=0.49\linewidth]{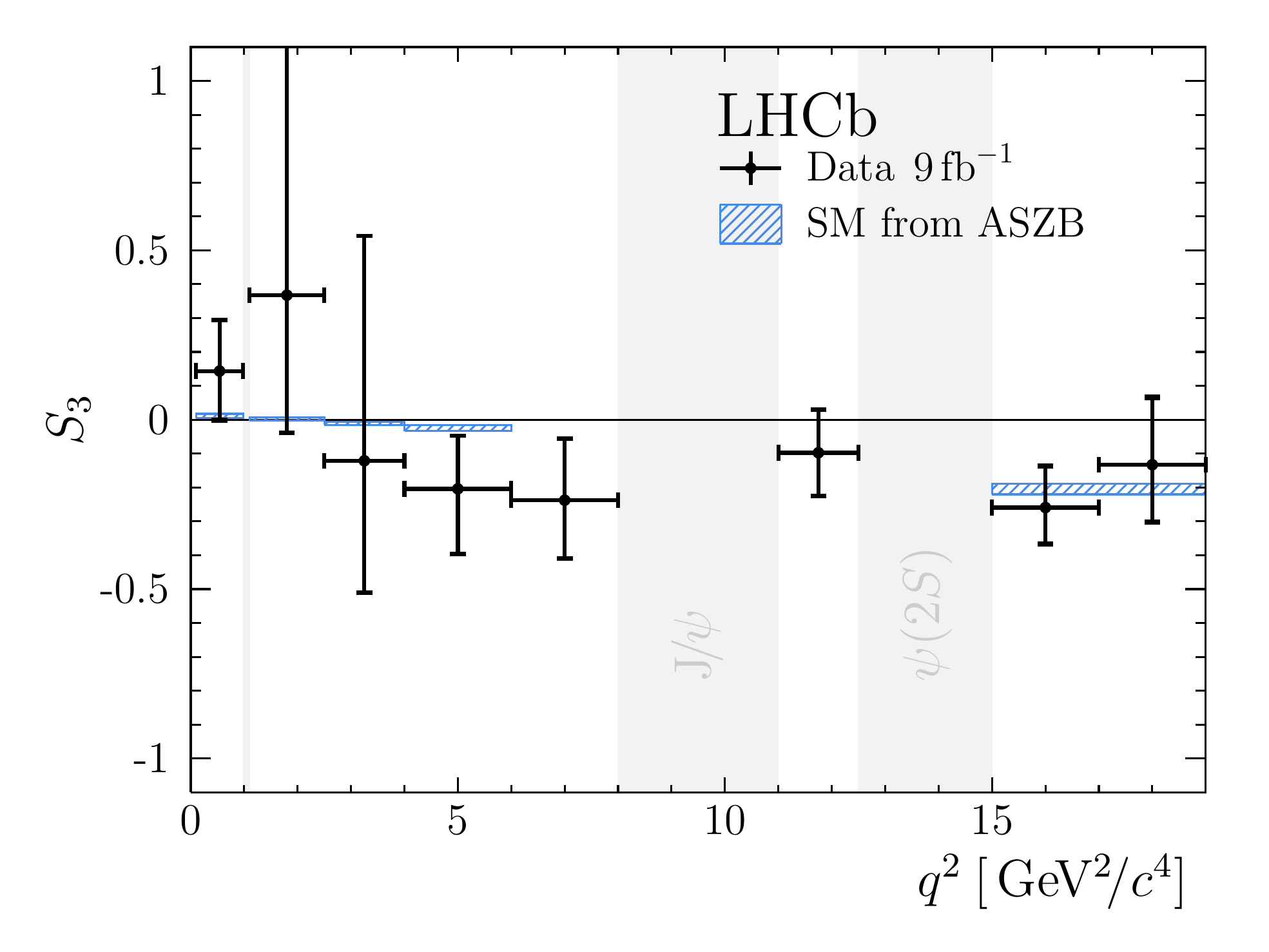}\\
    \vspace*{-0.2cm}
    \includegraphics[width=0.49\linewidth]{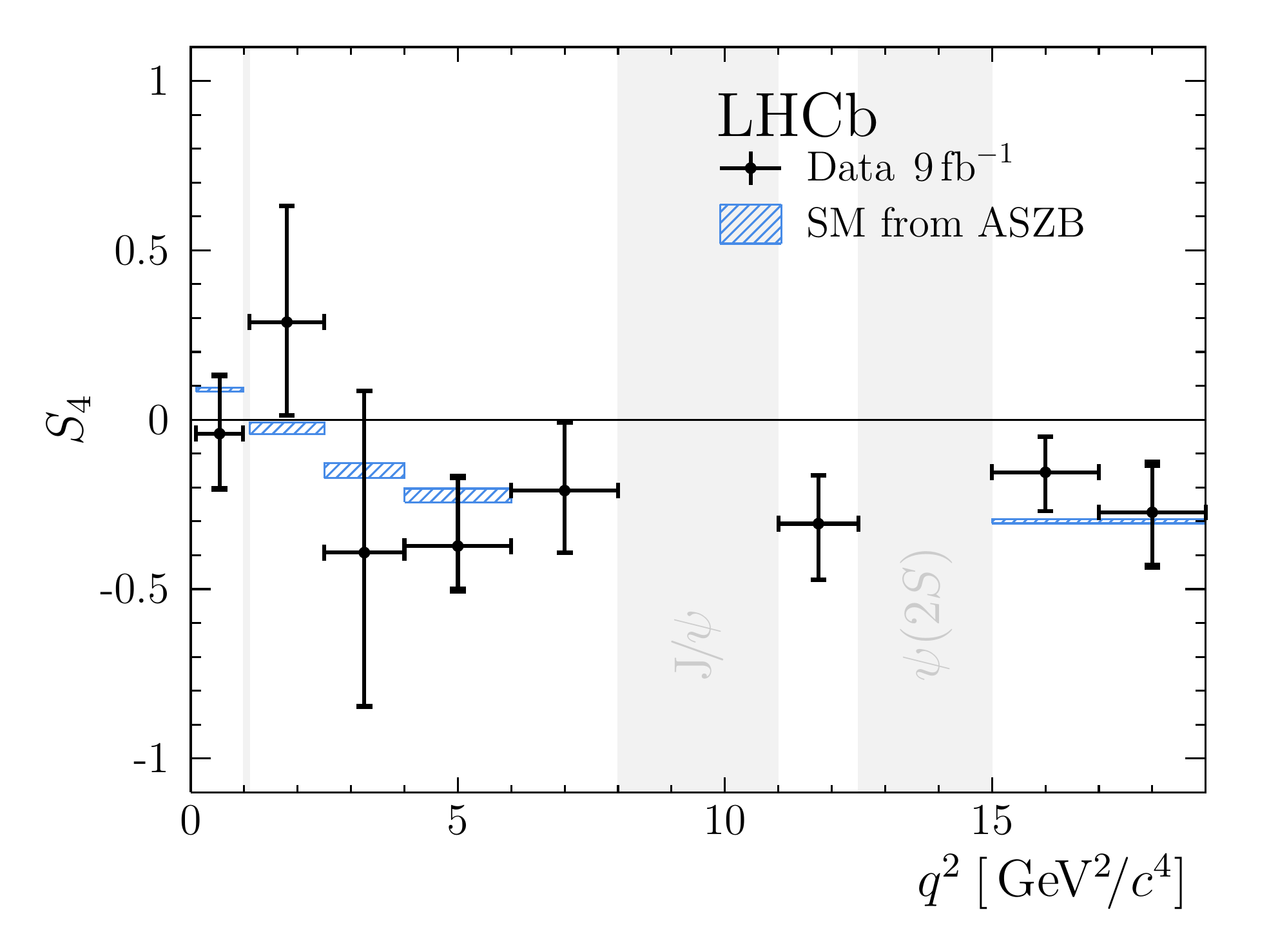}
    \includegraphics[width=0.49\linewidth]{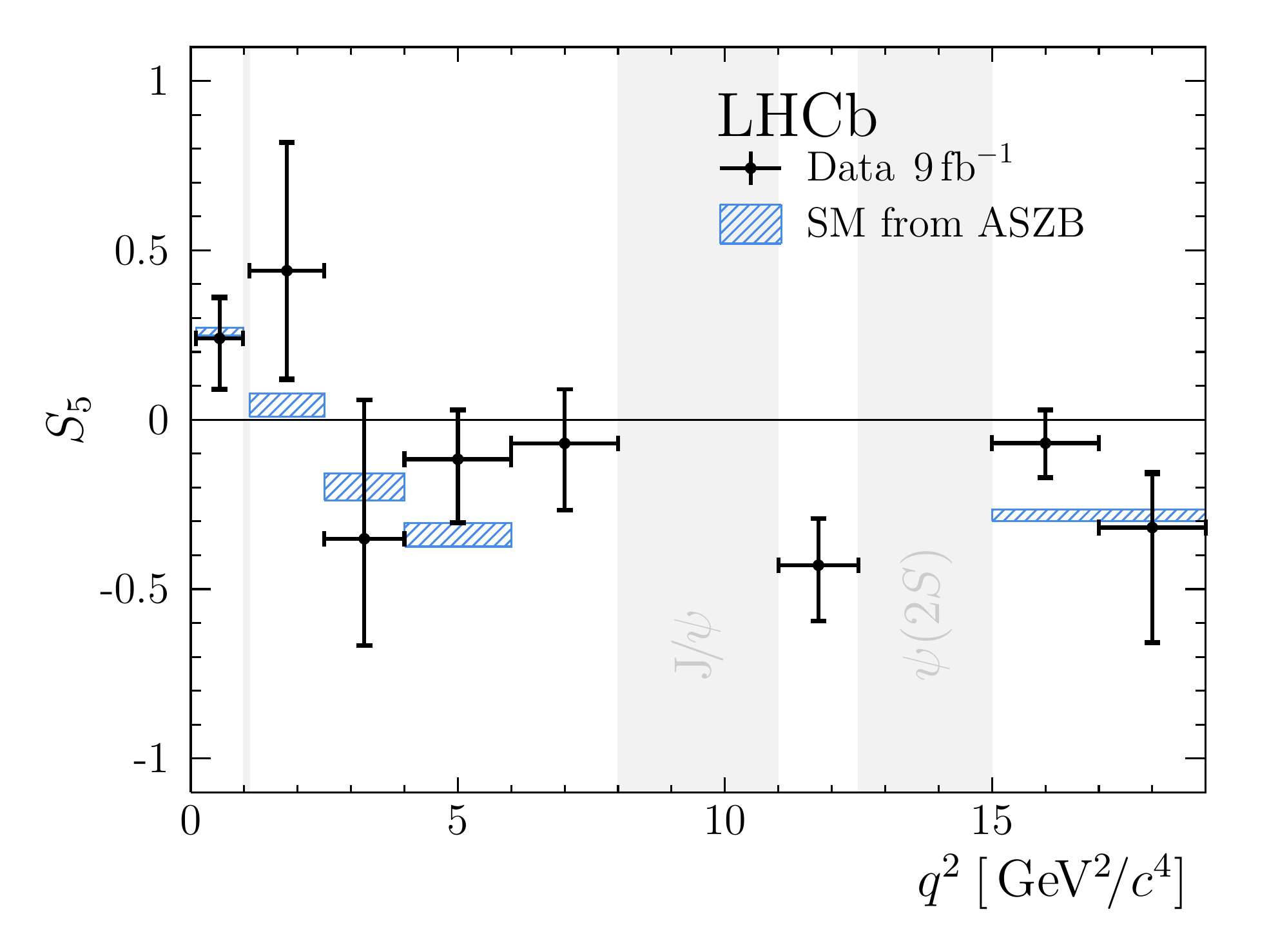}\\
    \vspace*{-0.2cm}
    \includegraphics[width=0.49\linewidth]{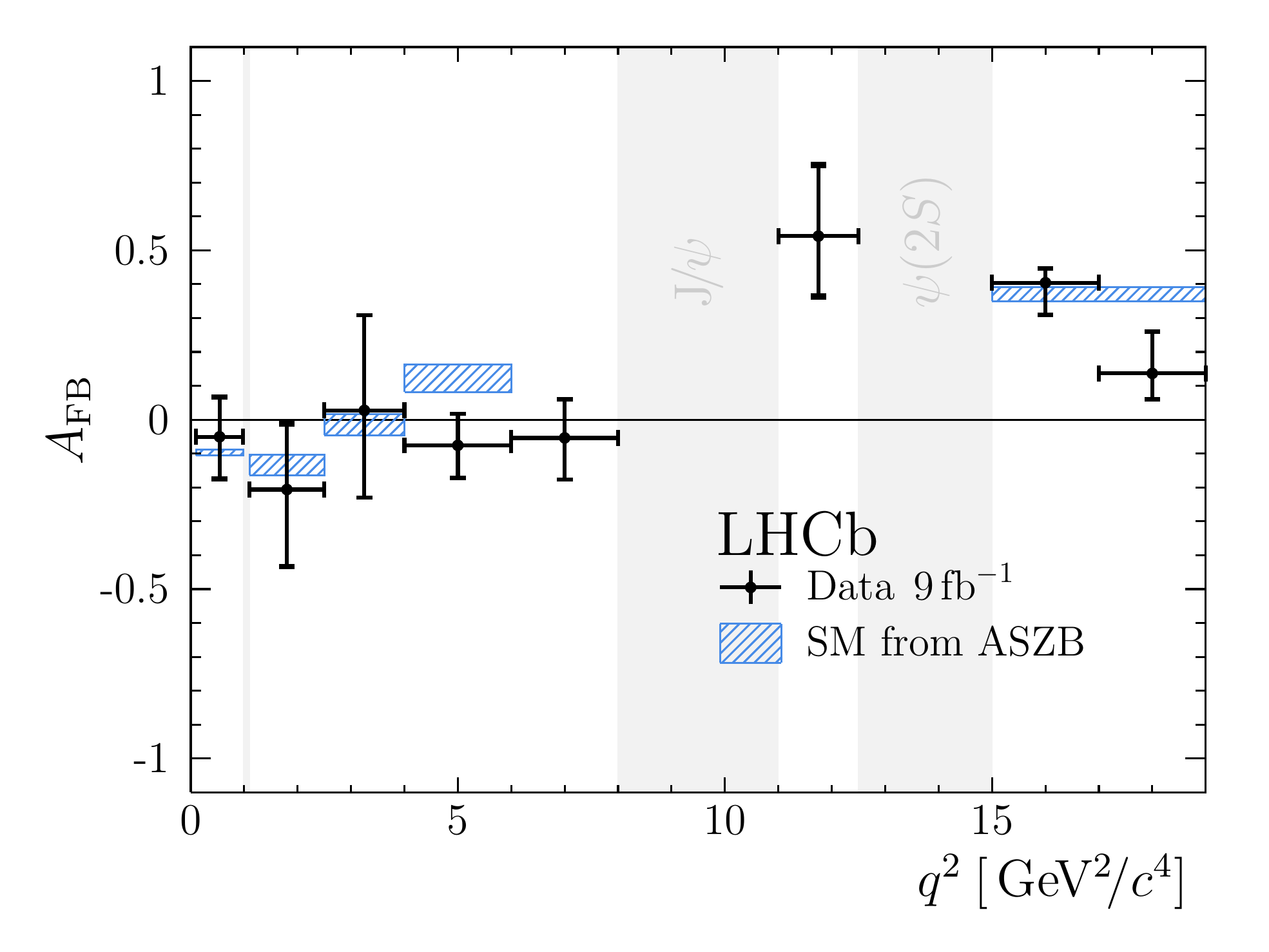}
    \includegraphics[width=0.49\linewidth]{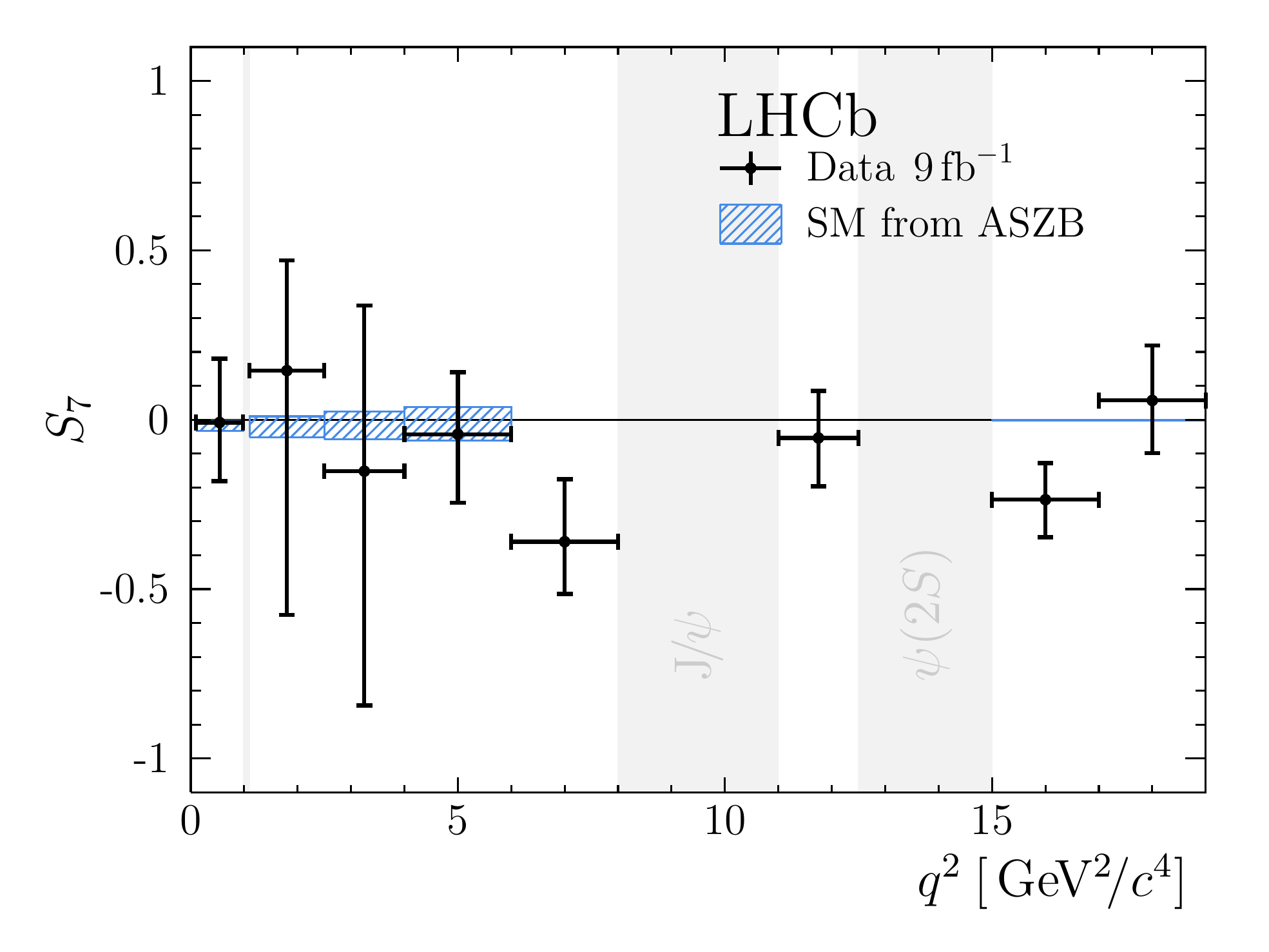}\\
    \vspace*{-0.2cm}
    \includegraphics[width=0.49\linewidth]{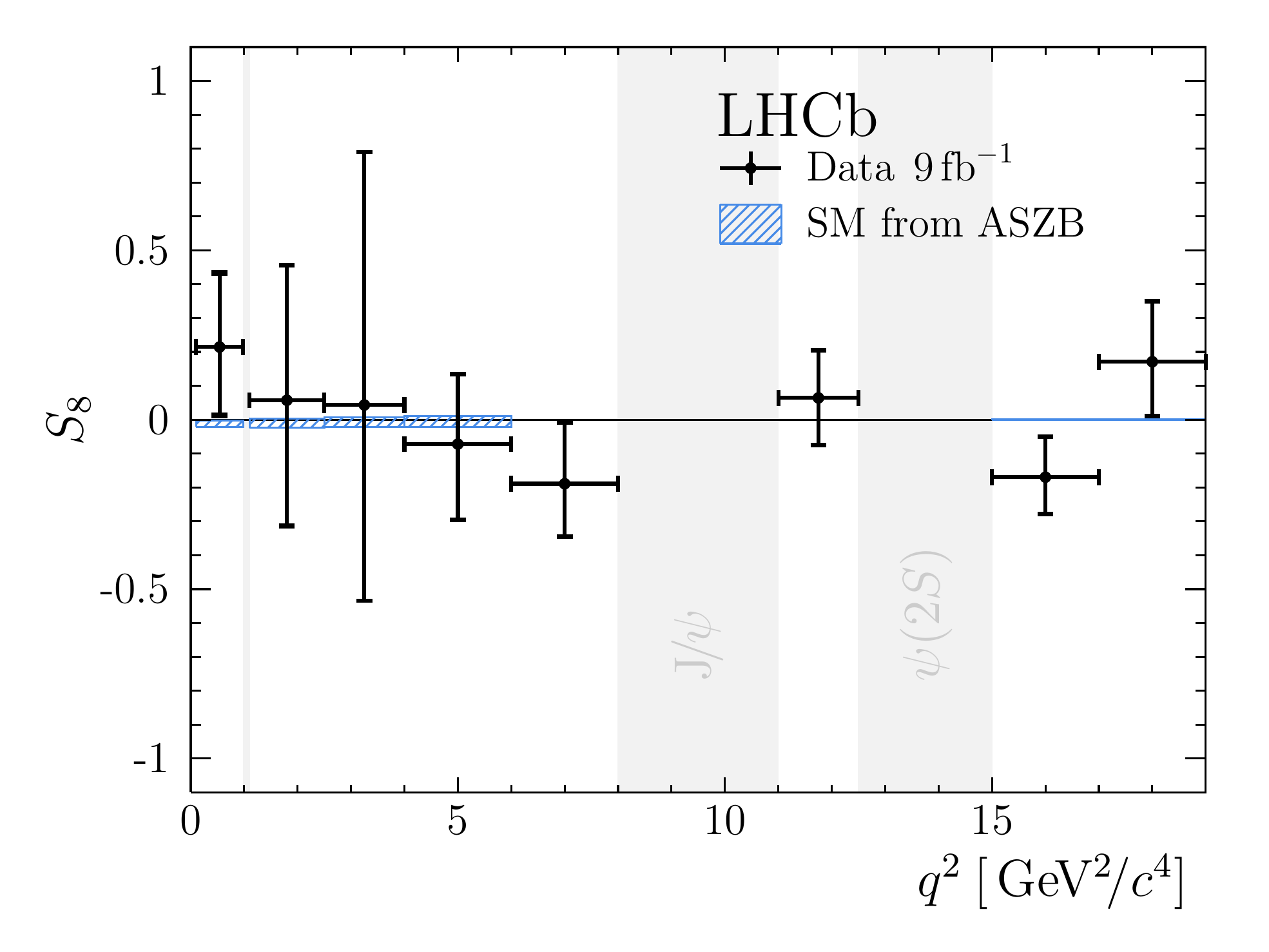}
    \includegraphics[width=0.49\linewidth]{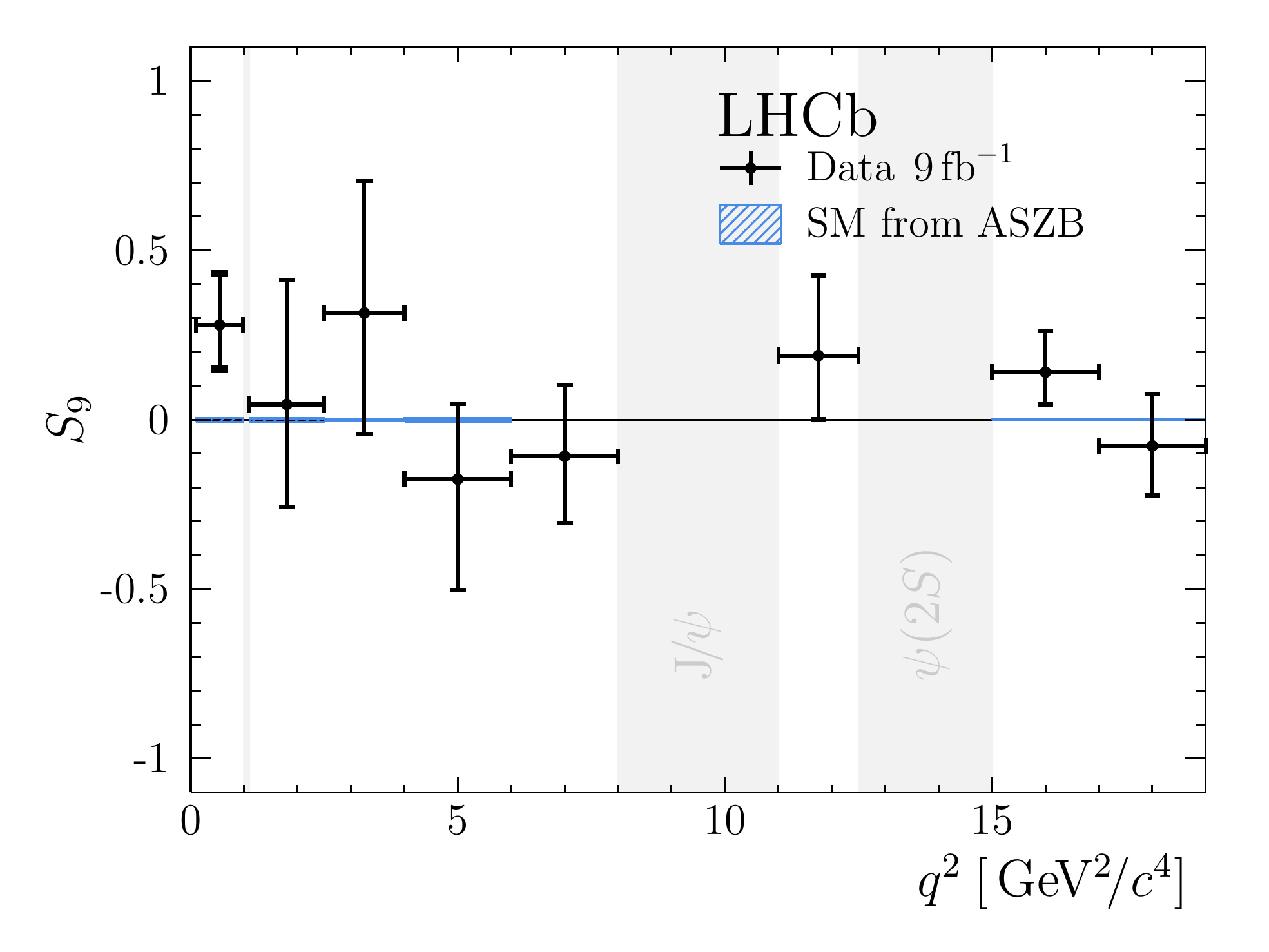}
    \vspace*{-0.6cm}
  \end{center}
  \caption{ The \CP-averaged observables \FL, \AFB and $S_{3} \text{--} S_{9}$ versus \qsq. The first (second) error bars represent the statistical (total) uncertainties. The theoretical predictions are based on Refs.~\cite{Altmannshofer:2014rta,Straub:2015ica,Horgan:2013hoa,Horgan:2015vla}. The grey bands indicate regions of excluded resonances.}
  \label{fig:Sobservables}
\end{figure}

\begin{figure}[htb]
  \begin{center}
    \includegraphics[width=0.49\linewidth]{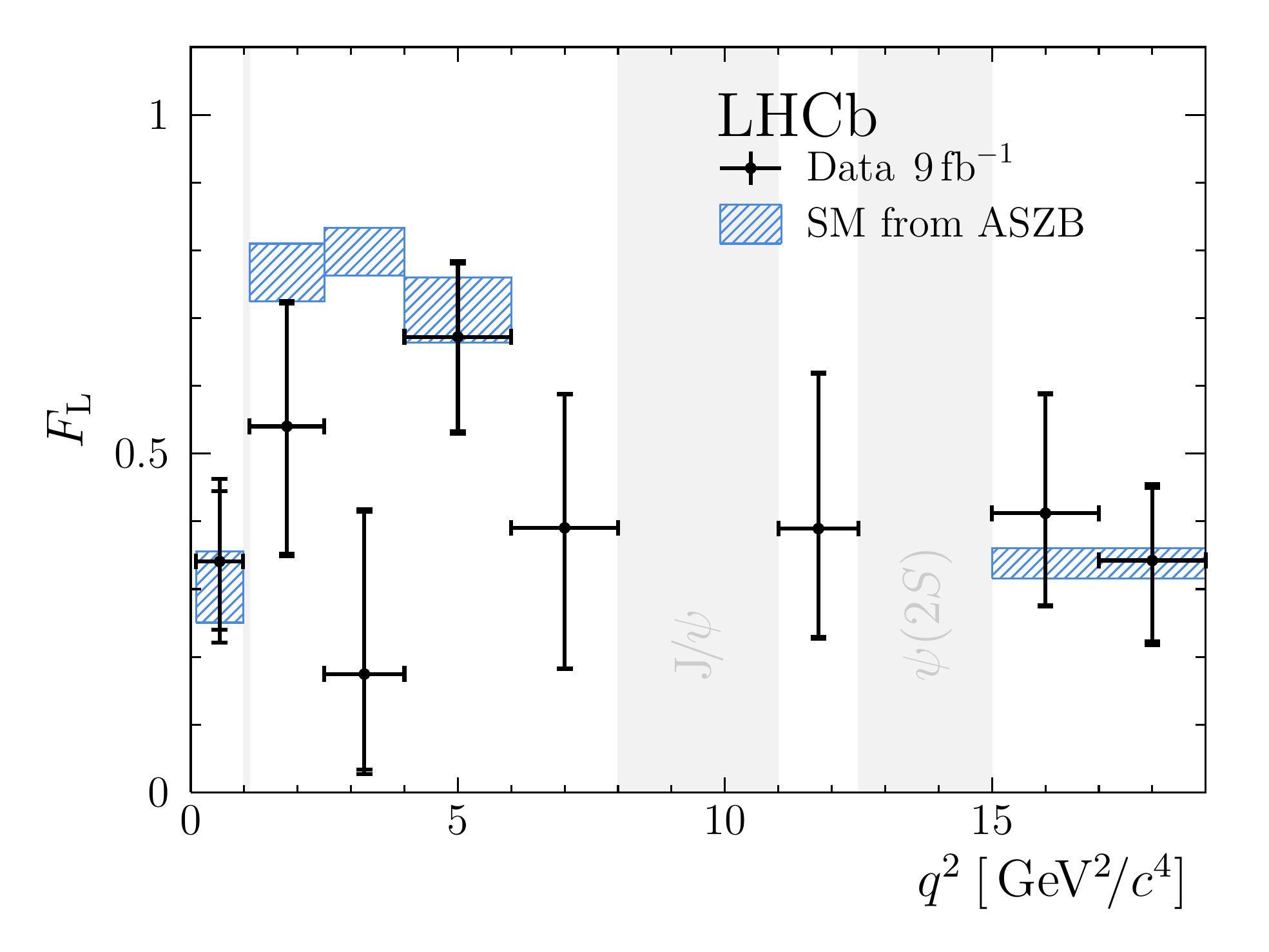}
    \includegraphics[width=0.49\linewidth]{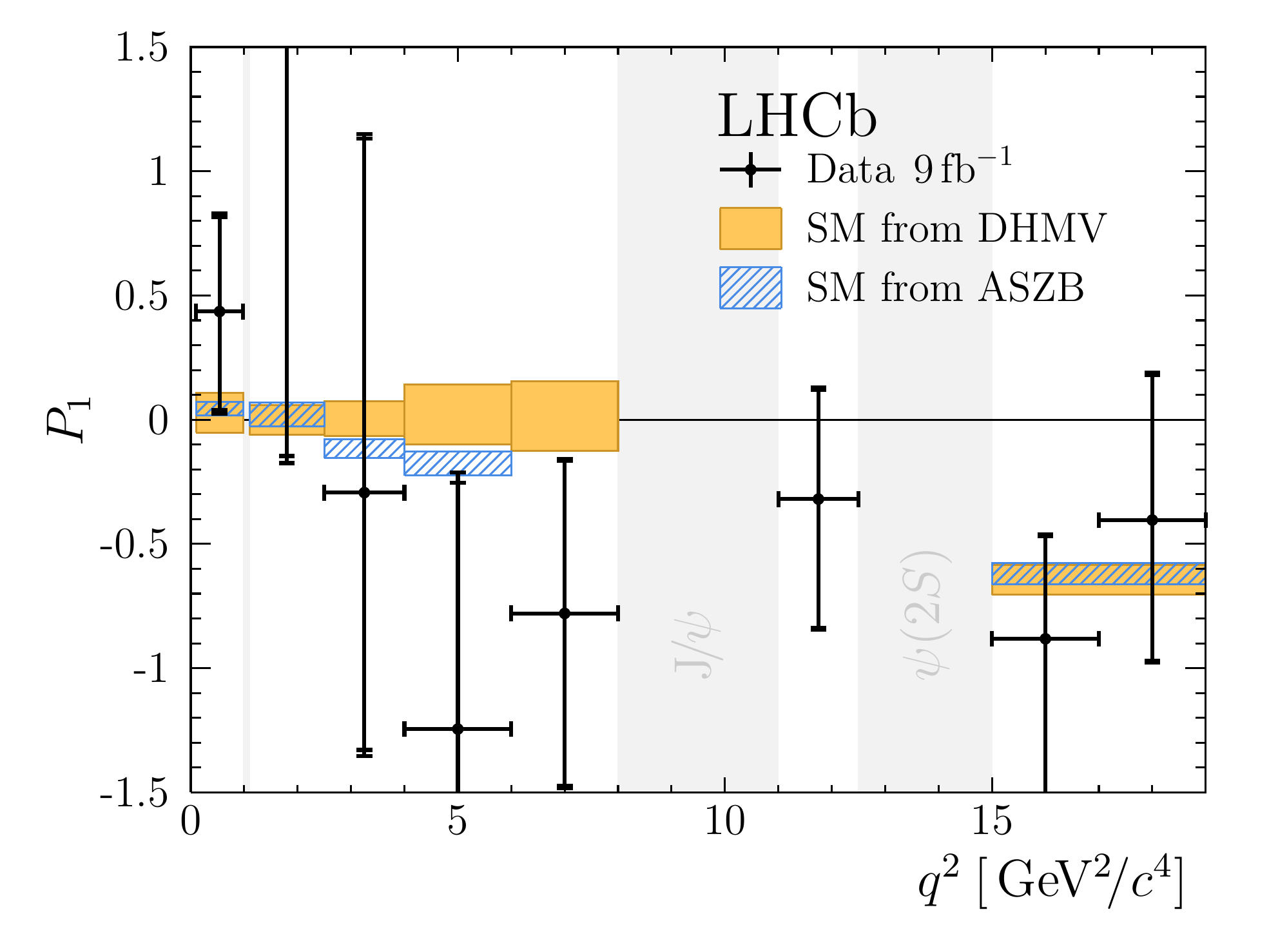}\\
    \vspace*{-0.2cm}
    \includegraphics[width=0.49\linewidth]{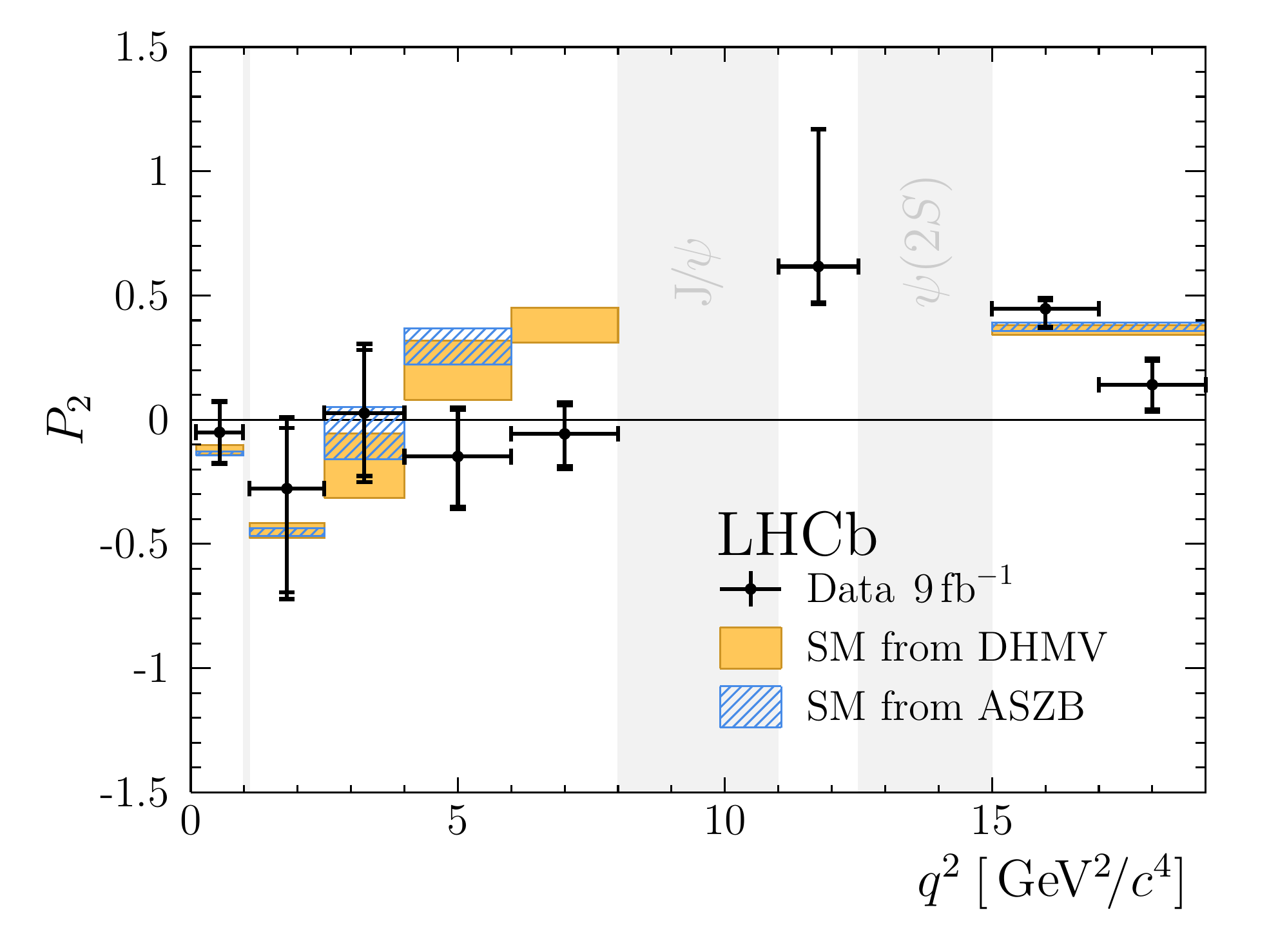}
    \includegraphics[width=0.49\linewidth]{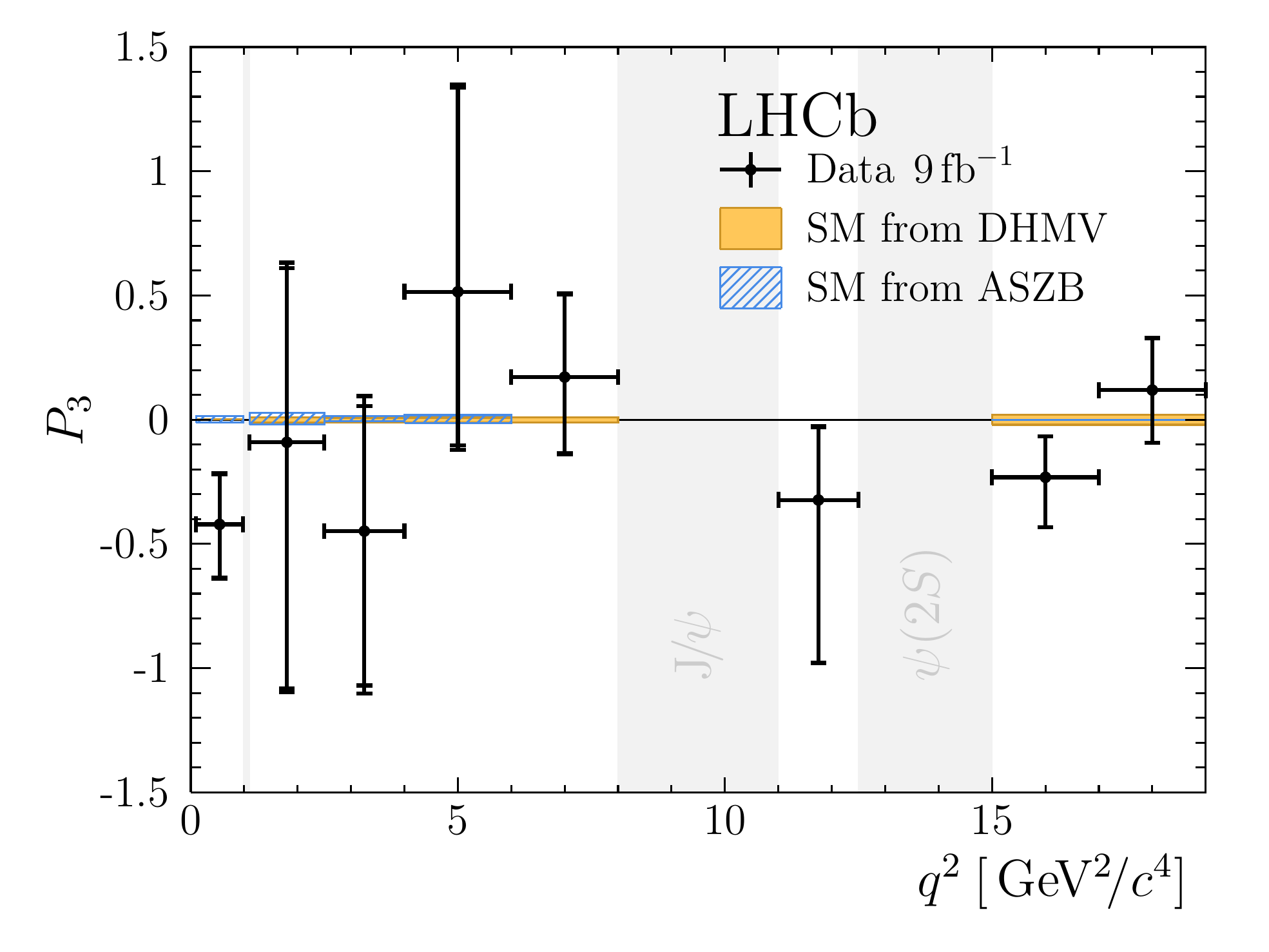}\\
    \vspace*{-0.2cm}
    \includegraphics[width=0.49\linewidth]{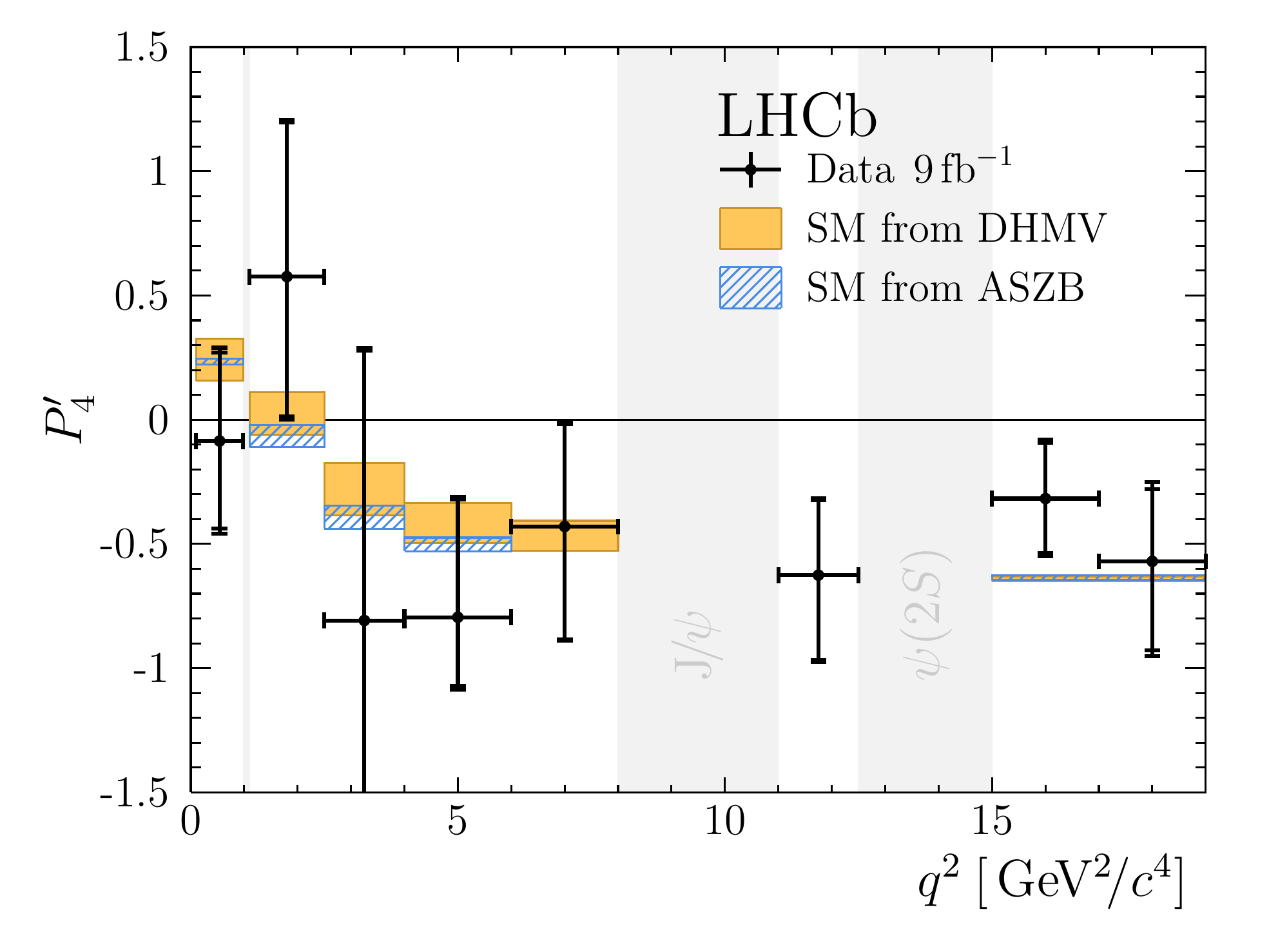}
    \includegraphics[width=0.49\linewidth]{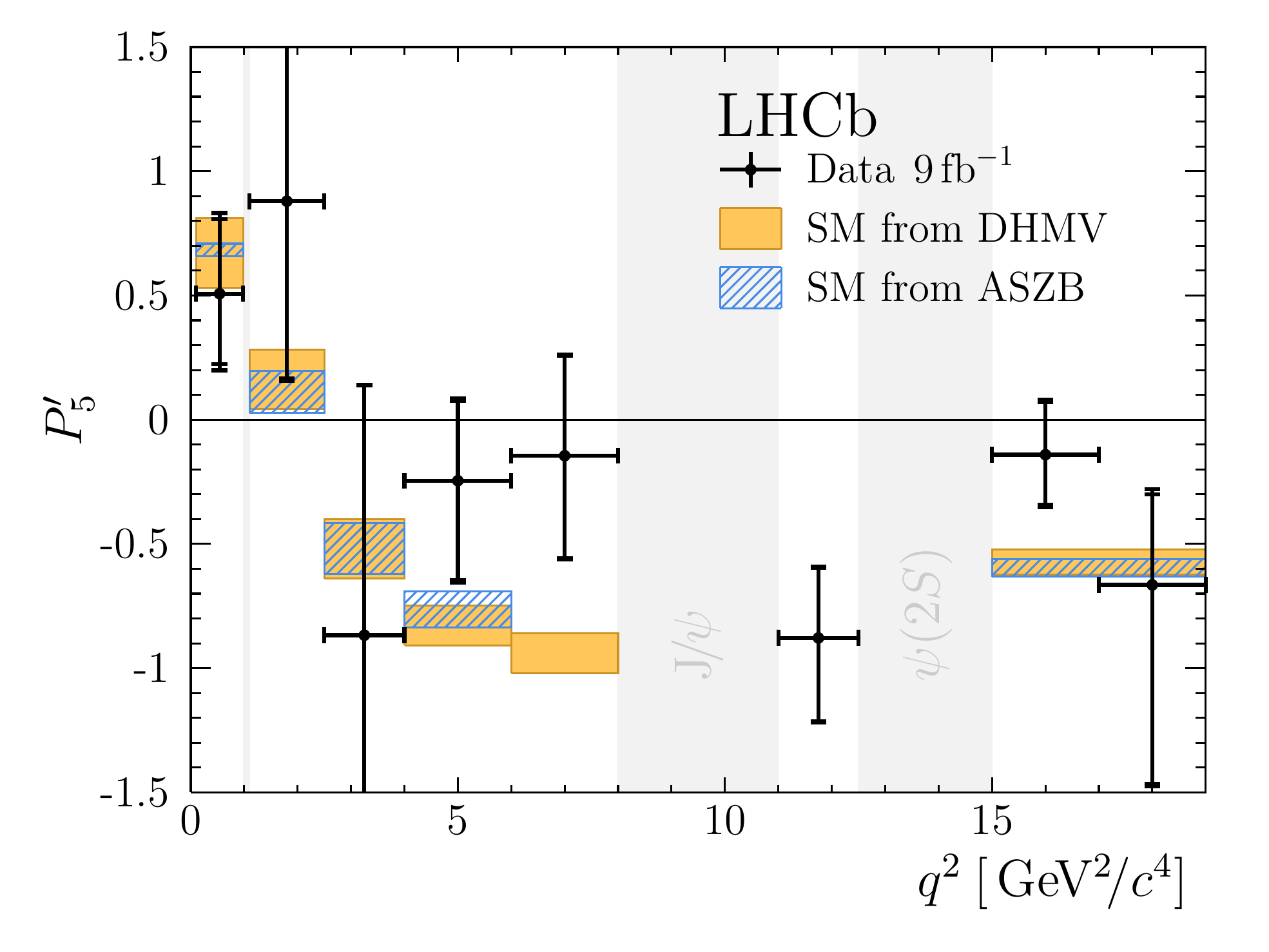}\\
    \vspace*{-0.2cm}
    \includegraphics[width=0.49\linewidth]{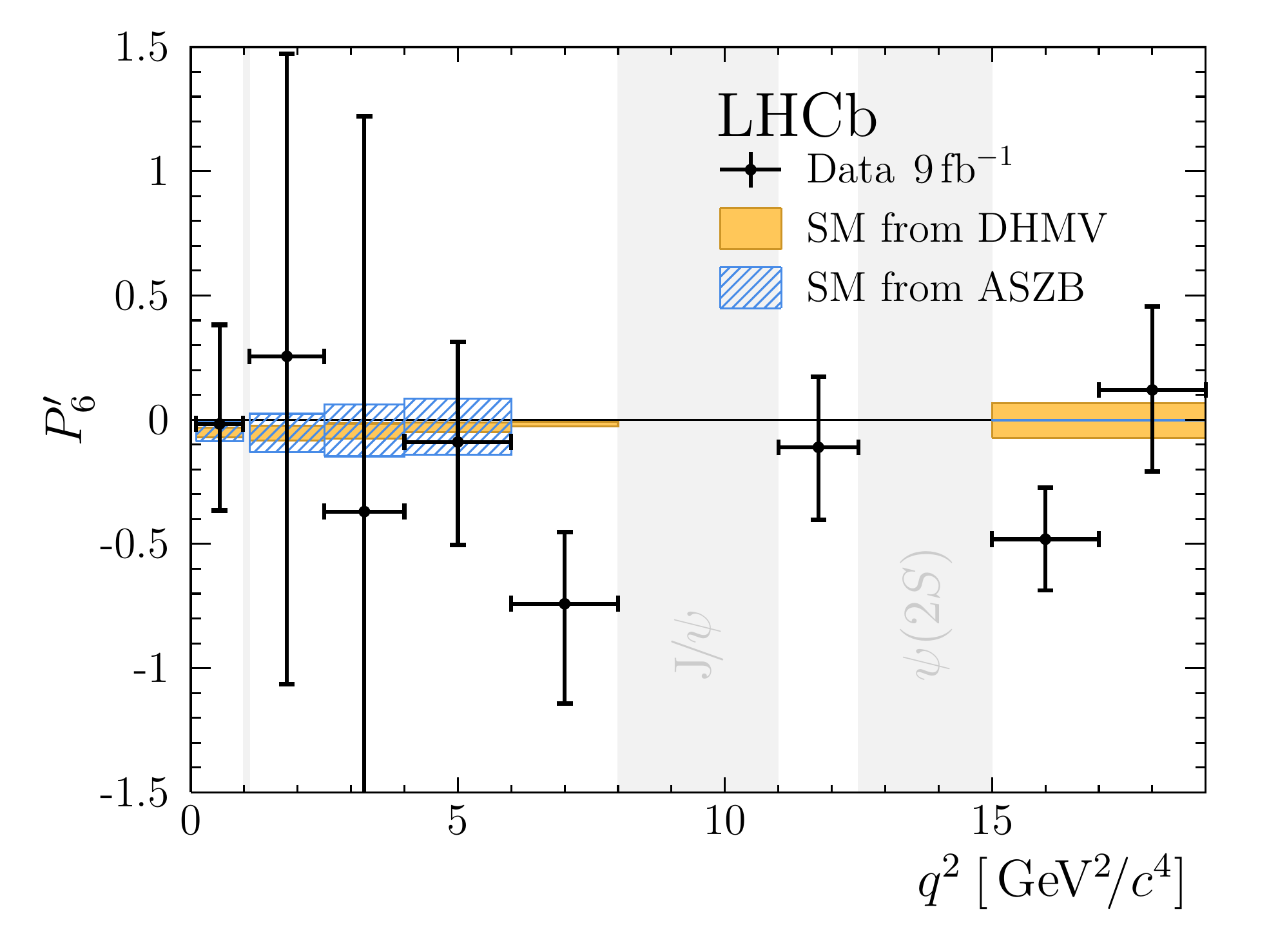}
    \includegraphics[width=0.49\linewidth]{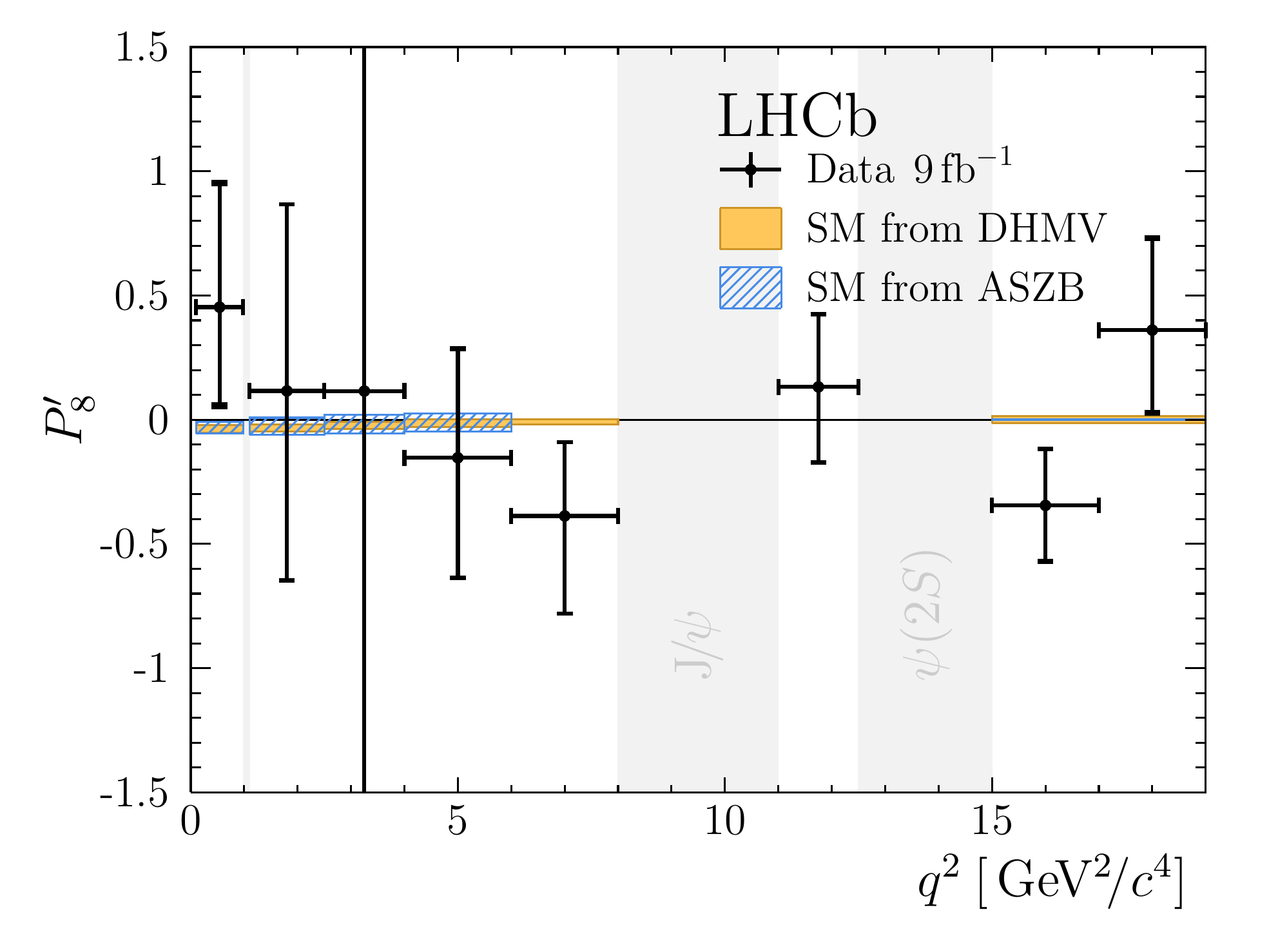}
    \vspace*{-0.6cm}
  \end{center}
  \caption{ The optimised observables $P_1$ to $P^{\prime}_8$ versus \qsq. The first (second) error bars represent the statistical (total) uncertainties. The theoretical predictions are based on Refs.~\cite{Descotes-Genon2016,Capdevila2018,Khodjamirian:2010vf} (orange) and on Refs.~\cite{Altmannshofer:2014rta,Straub:2015ica,Horgan:2013hoa,Horgan:2015vla} (blue). The grey bands indicate regions of excluded resonances.}
  \label{fig:Pobservables}
\end{figure}
\clearpage

\section{Tabular results for the \boldmath{$S_{i}$} and \boldmath{$P_{i}^{(\prime)}$} observables}\label{sec:SandPTables}

\begin{table}[h]
\caption{Results for the \CP-averaged observables \FL, \AFB and $S_{3} \text{--} S_{9}$. The first uncertainties are statistical and the second systematic.\label{tab:Sobservables}}

\begin{center}\footnotesize
\newcommand\xrowht[2][0]{\addstackgap[.5\dimexpr#2\relax]{\vphantom{#1}}}
\begin{tabular}{l|rrrr}
\qsq [\gevgevcccc]	&\multicolumn{1}{c}{\FL}	& \multicolumn{1}{c}{$S_{3}$}	& \multicolumn{1}{c}{$S_{4}$}	& \multicolumn{1}{c}{$S_{5}$}\\
\hline
\xrowht[()]{8pt}
$[0.10, 0.98]$  &$\phantom{-}0.34\:^{+0.10}_{-0.10} \pm 0.06$      &$0.14\:^{+0.15}_{-0.14} \pm 0.02$      &$-0.04\:^{+0.17}_{-0.16} \pm 0.04$     &$0.24\:^{+0.12}_{-0.15} \pm 0.04$\\
\xrowht[()]{8pt}
$[1.1, 2.5]$    &$0.54\:^{+0.21}_{-0.18} \pm 0.06$      &$0.37\:^{+0.97}_{-0.41} \pm 0.03$      &$0.29\:^{+0.34}_{-0.27} \pm 0.03$      &$0.44\:^{+0.38}_{-0.32} \pm 0.05$\\
\xrowht[()]{8pt}
$[2.5, 4.0]$    &$0.17\:^{+0.23}_{-0.32} \pm 0.06$      &$-0.12\:^{+0.66}_{-0.39} \pm 0.02$     &$-0.39\:^{+0.48}_{-0.45} \pm 0.04$     &$-0.35\:^{+0.41}_{-0.31} \pm 0.02$\\
\xrowht[()]{8pt}
$[4.0, 6.0]$    &$0.67\:^{+0.12}_{-0.14} \pm 0.02$      &$-0.20\:^{+0.16}_{-0.19} \pm 0.01$     &$-0.37\:^{+0.20}_{-0.13} \pm 0.05$     &$-0.12\:^{+0.14}_{-0.19} \pm 0.03$\\
\xrowht[()]{8pt}
$[6.0, 8.0]$    &$0.39\:^{+0.20}_{-0.21} \pm 0.01$      &$-0.24\:^{+0.18}_{-0.17} \pm 0.02$     &$-0.21\:^{+0.20}_{-0.18} \pm 0.02$     &$-0.07\:^{+0.16}_{-0.20} \pm 0.02$\\
\xrowht[()]{8pt}
$[11.0, 12.5]$  &$0.39\:^{+0.24}_{-0.17} \pm 0.02$      &$-0.10\:^{+0.13}_{-0.13} \pm 0.02$     &$-0.31\:^{+0.14}_{-0.17} \pm 0.02$     &$-0.43\:^{+0.14}_{-0.16} \pm 0.02$\\
\xrowht[()]{8pt}
$[15.0, 17.0]$  &$0.41\:^{+0.21}_{-0.14} \pm 0.02$      &$-0.26\:^{+0.12}_{-0.11} \pm 0.03$     &$-0.16\:^{+0.10}_{-0.11} \pm 0.02$     &$-0.07\:^{+0.10}_{-0.10} \pm 0.03$\\
\xrowht[()]{8pt}
$[17.0, 19.0]$  &$0.34\:^{+0.12}_{-0.11} \pm 0.03$      &$-0.13\:^{+0.20}_{-0.17} \pm 0.04$     &$-0.27\:^{+0.14}_{-0.15} \pm 0.06$     &$-0.32\:^{+0.16}_{-0.34} \pm 0.04$\\
\hline
\xrowht[()]{8pt}
$[1.1, 6.0]$    &$0.59\:^{+0.09}_{-0.09} \pm 0.03$      &$-0.10\:^{+0.11}_{-0.11} \pm 0.01$     &$-0.20\:^{+0.13}_{-0.14} \pm 0.03$     &$-0.04\:^{+0.12}_{-0.12} \pm 0.02$\\
\xrowht[()]{8pt}
$[15.0, 19.0]$  &$0.40\:^{+0.13}_{-0.11} \pm 0.03$      &$-0.21\:^{+0.09}_{-0.09} \pm 0.03$     &$-0.19\:^{+0.10}_{-0.13} \pm 0.06$     &$-0.12\:^{+0.07}_{-0.07} \pm 0.02$\\
\end{tabular}\\
\vspace*{1.0cm}
\begin{tabular}{l|rrrr}
\qsq [\gevgevcccc]	& \multicolumn{1}{c}{\AFB}	& \multicolumn{1}{c}{$S_{7}$}	& \multicolumn{1}{c}{$S_{8}$}	& \multicolumn{1}{c}{$S_{9}$}\\
\hline
\xrowht[()]{8pt}
$[0.10, 0.98]$  &$-0.05\:^{+0.12}_{-0.12} \pm 0.03$     &$-0.01\:^{+0.19}_{-0.17} \pm 0.01$     &$0.21\:^{+0.22}_{-0.20} \pm 0.05$      &$0.28\:^{+0.15}_{-0.12} \pm 0.06$\\
\xrowht[()]{8pt}
$[1.1, 2.5]$    &$-0.21\:^{+0.19}_{-0.23} \pm 0.04$     &$0.15\:^{+0.32}_{-0.72} \pm 0.02$      &$0.06\:^{+0.40}_{-0.37} \pm 0.04$      &$0.05\:^{+0.37}_{-0.30} \pm 0.02$\\
\xrowht[()]{8pt}
$[2.5, 4.0]$    &$0.03\:^{+0.28}_{-0.26} \pm 0.01$      &$-0.15\:^{+0.49}_{-0.69} \pm 0.03$     &$0.04\:^{+0.75}_{-0.58} \pm 0.03$      &$0.31\:^{+0.39}_{-0.36} \pm 0.02$\\
\xrowht[()]{8pt}
$[4.0, 6.0]$    &$-0.08\:^{+0.09}_{-0.10} \pm 0.01$     &$-0.04\:^{+0.18}_{-0.20} \pm 0.01$     &$-0.07\:^{+0.21}_{-0.22} \pm 0.03$     &$-0.18\:^{+0.22}_{-0.33} \pm 0.02$\\
\xrowht[()]{8pt}
$[6.0, 8.0]$    &$-0.05\:^{+0.11}_{-0.12} \pm 0.01$     &$-0.36\:^{+0.18}_{-0.15} \pm 0.02$     &$-0.19\:^{+0.18}_{-0.16} \pm 0.02$     &$-0.11\:^{+0.21}_{-0.20} \pm 0.02$\\
\xrowht[()]{8pt}
$[11.0, 12.5]$  &$0.54\:^{+0.21}_{-0.18} \pm 0.05$      &$-0.05\:^{+0.14}_{-0.14} \pm 0.01$     &$0.06\:^{+0.14}_{-0.14} \pm 0.01$      &$0.19\:^{+0.24}_{-0.19} \pm 0.03$\\
\xrowht[()]{8pt}
$[15.0, 17.0]$  &$0.40\:^{+0.04}_{-0.09} \pm 0.01$      &$-0.24\:^{+0.11}_{-0.11} \pm 0.02$     &$-0.17\:^{+0.12}_{-0.11} \pm 0.02$     &$0.14\:^{+0.12}_{-0.09} \pm 0.02$\\
\xrowht[()]{8pt}
$[17.0, 19.0]$  &$0.14\:^{+0.12}_{-0.07} \pm 0.02$      &$0.06\:^{+0.16}_{-0.16} \pm 0.01$      &$0.17\:^{+0.18}_{-0.16} \pm 0.02$      &$-0.08\:^{+0.15}_{-0.15} \pm 0.02$\\
\hline
\xrowht[()]{8pt}
$[1.1, 6.0]$    &$-0.08\:^{+0.07}_{-0.08} \pm 0.02$     &$-0.10\:^{+0.11}_{-0.13} \pm 0.01$     &$0.02\:^{+0.13}_{-0.14} \pm 0.02$      &$-0.05\:^{+0.11}_{-0.12} \pm 0.01$\\
\xrowht[()]{8pt}
$[15.0, 19.0]$  &$0.31\:^{+0.06}_{-0.06} \pm 0.04$      &$-0.14\:^{+0.08}_{-0.09} \pm 0.01$     &$-0.06\:^{+0.09}_{-0.09} \pm 0.01$     &$0.04\:^{+0.08}_{-0.06} \pm 0.02$\\
\end{tabular}
\end{center}
\end{table}

\begin{table}
\caption{Results for the optimised observables \FL and $P_1\text{--}P_8^{\prime}$. The first uncertainties are statistical and the second systematic.\label{tab:Pobservables}}
\begin{center}\footnotesize
\newcommand\xrowht[2][0]{\addstackgap[.5\dimexpr#2\relax]{\vphantom{#1}}}
\begin{tabular}{l|rrrr}
\qsq [\gevgevcccc]	& \multicolumn{1}{c}{\FL}   & \multicolumn{1}{c}{$P_{1}$}	& \multicolumn{1}{c}{$P_{2}$}	& \multicolumn{1}{c}{$P_{3}$}\\
\hline
\xrowht[()]{8pt}
$[0.10, 0.98]$  &$\phantom{-}0.34\:^{+0.10}_{-0.10} \pm 0.06$      &$0.44\:^{+0.38}_{-0.40} \pm 0.11$      &$-0.05\:^{+0.12}_{-0.12} \pm 0.03$     &$-0.42\:^{+0.20}_{-0.21} \pm 0.05$\\
\xrowht[()]{8pt}
$[1.1, 2.5]$    &$0.54\:^{+0.18}_{-0.19} \pm 0.03$      &$1.60\:^{+4.92}_{-1.75} \pm 0.32$      &$-0.28\:^{+0.24}_{-0.42} \pm 0.15$     &$-0.09\:^{+0.70}_{-0.99} \pm 0.18$\\
\xrowht[()]{8pt}
$[2.5, 4.0]$    &$0.17\:^{+0.24}_{-0.14} \pm 0.04$      &$-0.29\:^{+1.43}_{-1.04} \pm 0.22$     &$0.03\:^{+0.26}_{-0.25} \pm 0.11$      &$-0.45\:^{+0.50}_{-0.62} \pm 0.20$\\
\xrowht[()]{8pt}
$[4.0, 6.0]$    &$0.67\:^{+0.11}_{-0.14} \pm 0.03$      &$-1.24\:^{+0.99}_{-1.17} \pm 0.29$     &$-0.15\:^{+0.19}_{-0.20} \pm 0.06$     &$0.52\:^{+0.82}_{-0.62} \pm 0.15$\\
\xrowht[()]{8pt}
$[6.0, 8.0]$    &$0.39\:^{+0.20}_{-0.21} \pm 0.02$      &$-0.78\:^{+0.61}_{-0.69} \pm 0.10$     &$-0.06\:^{+0.12}_{-0.13} \pm 0.05$     &$0.17\:^{+0.33}_{-0.31} \pm 0.06$\\
\xrowht[()]{8pt}
$[11.0, 12.5]$  &$0.39\:^{+0.23}_{-0.16} \pm 0.03$      &$-0.32\:^{+0.44}_{-0.52} \pm 0.09$     &$0.62\:^{+0.55}_{-0.14} \pm 0.04$      &$-0.32\:^{+0.29}_{-0.65} \pm 0.05$\\
\xrowht[()]{8pt}
$[15.0, 17.0]$  &$0.41\:^{+0.18}_{-0.14} \pm 0.02$      &$-0.88\:^{+0.41}_{-0.67} \pm 0.07$     &$0.45\:^{+0.03}_{-0.07} \pm 0.03$      &$-0.23\:^{+0.16}_{-0.20} \pm 0.02$\\
\xrowht[()]{8pt}
$[17.0, 19.0]$  &$0.34\:^{+0.11}_{-0.12} \pm 0.04$      &$-0.40\:^{+0.58}_{-0.57} \pm 0.09$     &$0.14\:^{+0.10}_{-0.10} \pm 0.04$      &$0.12\:^{+0.21}_{-0.21} \pm 0.02$\\
\hline
\xrowht[()]{8pt}
$[1.1, 6.0]$    &$0.59\:^{+0.10}_{-0.10} \pm 0.03$      &$-0.51\:^{+0.56}_{-0.54} \pm 0.08$     &$-0.13\:^{+0.13}_{-0.13} \pm 0.05$     &$0.12\:^{+0.27}_{-0.28} \pm 0.04$\\
\xrowht[()]{8pt}
$[15.0, 19.0]$  &$0.40\:^{+0.13}_{-0.11} \pm 0.02$      &$-0.70\:^{+0.35}_{-0.43} \pm 0.07$     &$0.34\:^{+0.09}_{-0.07} \pm 0.04$      &$-0.07\:^{+0.12}_{-0.13} \pm 0.03$\\
\end{tabular}\\
\vspace*{1.0cm}
\begin{tabular}{l|rrrr}
\qsq [\gevgevcccc]	& \multicolumn{1}{c}{$P_{4}^{\prime}$}	& \multicolumn{1}{c}{$P_{5}^{\prime}$}	& \multicolumn{1}{c}{$P_{6}^{\prime}$}	& \multicolumn{1}{c}{$P_{8}^{\prime}$}\\
\hline
\xrowht[()]{8pt}
$[0.10, 0.98]$  &$-0.09\:^{+0.36}_{-0.35} \pm 0.12$     &$0.51\:^{+0.30}_{-0.28} \pm 0.12$      &$-0.02\:^{+0.40}_{-0.34} \pm 0.06$     &$0.45\:^{+0.50}_{-0.39} \pm 0.09$\\
\xrowht[()]{8pt}
$[1.1, 2.5]$    &$0.58\:^{+0.62}_{-0.56} \pm 0.11$      &$0.88\:^{+0.70}_{-0.71} \pm 0.10$      &$0.25\:^{+1.22}_{-1.32} \pm 0.08$      &$0.12\:^{+0.75}_{-0.76} \pm 0.05$\\
\xrowht[()]{8pt}
$[2.5, 4.0]$    &$-0.81\:^{+1.09}_{-0.84} \pm 0.14$     &$-0.87\:^{+1.00}_{-1.68} \pm 0.09$     &$-0.37\:^{+1.59}_{-3.91} \pm 0.05$     &$0.12\:^{+7.89}_{-4.95} \pm 0.07$\\
\xrowht[()]{8pt}
$[4.0, 6.0]$    &$-0.79\:^{+0.47}_{-0.28} \pm 0.09$     &$-0.25\:^{+0.32}_{-0.40} \pm 0.09$     &$-0.09\:^{+0.40}_{-0.41} \pm 0.05$     &$-0.15\:^{+0.44}_{-0.48} \pm 0.05$\\
\xrowht[()]{8pt}
$[6.0, 8.0]$    &$-0.43\:^{+0.41}_{-0.45} \pm 0.06$     &$-0.15\:^{+0.40}_{-0.41} \pm 0.06$     &$-0.74\:^{+0.29}_{-0.40} \pm 0.03$     &$-0.39\:^{+0.30}_{-0.39} \pm 0.02$\\
\xrowht[()]{8pt}
$[11.0, 12.5]$  &$-0.63\:^{+0.30}_{-0.34} \pm 0.07$     &$-0.88\:^{+0.28}_{-0.34} \pm 0.05$     &$-0.11\:^{+0.28}_{-0.29} \pm 0.03$     &$0.13\:^{+0.29}_{-0.30} \pm 0.04$\\
\xrowht[()]{8pt}
$[15.0, 17.0]$  &$-0.32\:^{+0.23}_{-0.22} \pm 0.08$     &$-0.14\:^{+0.21}_{-0.20} \pm 0.06$     &$-0.48\:^{+0.21}_{-0.21} \pm 0.02$     &$-0.34\:^{+0.23}_{-0.22} \pm 0.04$\\
\xrowht[()]{8pt}
$[17.0, 19.0]$  &$-0.57\:^{+0.29}_{-0.36} \pm 0.13$     &$-0.66\:^{+0.36}_{-0.80} \pm 0.13$     &$0.12\:^{+0.33}_{-0.33} \pm 0.04$      &$0.36\:^{+0.37}_{-0.33} \pm 0.07$\\
\hline
\xrowht[()]{8pt}
$[1.1, 6.0]$    &$-0.41\:^{+0.28}_{-0.28} \pm 0.07$     &$-0.07\:^{+0.25}_{-0.25} \pm 0.04$     &$-0.21\:^{+0.23}_{-0.23} \pm 0.04$     &$0.03\:^{+0.26}_{-0.28} \pm 0.06$\\
\xrowht[()]{8pt}
$[15.0, 19.0]$  &$-0.39\:^{+0.18}_{-0.21} \pm 0.10$     &$-0.24\:^{+0.16}_{-0.16} \pm 0.05$     &$-0.28\:^{+0.19}_{-0.14} \pm 0.03$     &$-0.11\:^{+0.19}_{-0.18} \pm 0.03$\\
\end{tabular}\\
\end{center}
\end{table}
\clearpage

\clearpage
\section{Correlation matrices for the \boldmath{$S_{i}$} observables}\label{sec:correlationsS}

Correlation matrices between the \CP-averaged observables \FL, \AFB and $S_{3} \text{--} S_{9}$ in the different \qsq intervals are provided in Tables~\ref{tab:correlationsSBin0}--\ref{tab:correlationsSBinB}. Correlations between observables measured with different folds are obtained using the bootstrapping technique~\cite{efron:1979}. The different $\qsq$ intervals are statistically independent.

\begin{table}[!ht]
\caption{
Correlation matrix for the \CP-averaged observables \FL, \AFB and $S_{3} \text{--} S_{9}$ from the maximum-likelihood fit in the interval $0.10<\qsq<0.98\gevgevcccc$.
\label{tab:correlationsSBin0}
}
\centering 
\begin{tabular}{l|rrrrrrrr} 
&\multicolumn{1}{c}{\FL}&\multicolumn{1}{c}{$S_{3}$}&\multicolumn{1}{c}{$S_{4}$}&\multicolumn{1}{c}{$S_{5}$}&\multicolumn{1}{c}{\AFB}&\multicolumn{1}{c}{$S_{7}$}&\multicolumn{1}{c}{$S_{8}$}&\multicolumn{1}{c}{$S_{9}$}\\
\hline
       \FL  &       $\phantom{-}1\phantom{.00}$&       $0.04$&       $-0.01$&       $0.03$&       $0.04$&       $0.12$&       $-0.00$&       $-0.11$\\
       $S_{3}$       &  &       $\phantom{-}1\phantom{.00}$&       $-0.02$&       $0.12$&       $-0.02$&       $0.02$&       $0.06$&       $0.02$\\
       $S_{4}$       &       &  &       $\phantom{-}1\phantom{.00}$&       $-0.27$&       $-0.09$&       $-0.25$&       $0.24$&       $-0.06$\\
       $S_{5}$       &       &       &  &       $\phantom{-}1\phantom{.00}$&       $0.10$&       $0.22$&       $-0.18$&       $0.06$\\
       \AFB       &       &       &       &  &       $\phantom{-}1\phantom{.00}$&       $0.19$&       $-0.27$&       $-0.06$\\
       $S_{7}$       &       &       &       &       &  &       $\phantom{-}1\phantom{.00}$&       $-0.35$&       $0.22$\\
       $S_{8}$       &       &       &       &       &       &  &       $\phantom{-}1\phantom{.00}$&       $-0.08$\\
       $S_{9}$       &       &       &       &       &       &       &  &       $\phantom{-}1\phantom{.00}$\\
\end{tabular}
\end{table}

\begin{table}[!ht]
\caption{
Correlation matrix for the \CP-averaged observables \FL, \AFB and $S_{3} \text{--} S_{9}$ from the maximum-likelihood fit in the interval $1.1<\qsq<2.5\gevgevcccc$. 
\label{tab:correlationsSBin1}
}
\centering 
\begin{tabular}{l|rrrrrrrr} 
&\multicolumn{1}{c}{\FL}&\multicolumn{1}{c}{$S_{3}$}&\multicolumn{1}{c}{$S_{4}$}&\multicolumn{1}{c}{$S_{5}$}&\multicolumn{1}{c}{\AFB}&\multicolumn{1}{c}{$S_{7}$}&\multicolumn{1}{c}{$S_{8}$}&\multicolumn{1}{c}{$S_{9}$}\\
\hline
       \FL  &       $\phantom{-}1\phantom{.00}$&       $0.16$&       $-0.05$&       $0.11$&       $0.11$&       $0.04$&       $-0.10$&       $-0.03$\\
       $S_{3}$       &  &       $\phantom{-}1\phantom{.00}$&       $0.06$&       $0.09$&       $-0.02$&       $0.13$&       $-0.01$&       $-0.12$\\
       $S_{4}$       &       &  &       $\phantom{-}1\phantom{.00}$&       $-0.02$&       $0.17$&       $0.05$&       $0.33$&       $0.09$\\
       $S_{5}$       &       &       &  &       $\phantom{-}1\phantom{.00}$&       $0.20$&       $0.22$&       $-0.06$&       $0.04$\\
       \AFB       &       &       &       &  &       $\phantom{-}1\phantom{.00}$&       $0.20$&       $0.11$&       $0.12$\\
       $S_{7}$       &       &       &       &       &  &       $\phantom{-}1\phantom{.00}$&       $0.06$&       $0.16$\\
       $S_{8}$       &       &       &       &       &       &  &       $\phantom{-}1\phantom{.00}$&       $0.22$\\
       $S_{9}$       &       &       &       &       &       &       &  &       $\phantom{-}1\phantom{.00}$\\
\end{tabular}
\end{table} 

\begin{table}[!ht]
\caption{
Correlation matrix for the \CP-averaged observables \FL, \AFB and $S_{3} \text{--} S_{9}$ from the maximum-likelihood fit in the interval $2.5<\qsq<4.0\gevgevcccc$.
\label{tab:correlationsSBin2}
}
\centering 
\begin{tabular}{l|rrrrrrrr} 
&\multicolumn{1}{c}{\FL}&\multicolumn{1}{c}{$S_{3}$}&\multicolumn{1}{c}{$S_{4}$}&\multicolumn{1}{c}{$S_{5}$}&\multicolumn{1}{c}{\AFB}&\multicolumn{1}{c}{$S_{7}$}&\multicolumn{1}{c}{$S_{8}$}&\multicolumn{1}{c}{$S_{9}$}\\
\hline
       \FL  &       $\phantom{-}1\phantom{.00}$&       $0.02$&       $-0.01$&       $0.06$&       $-0.08$&       $-0.02$&       $-0.07$&       $0.04$\\
       $S_{3}$       &  &       $\phantom{-}1\phantom{.00}$&       $0.02$&       $-0.06$&       $-0.01$&       $-0.03$&       $0.07$&       $0.02$\\
       $S_{4}$       &       &  &       $\phantom{-}1\phantom{.00}$&       $0.00$&       $-0.06$&       $0.10$&       $-0.05$&       $-0.00$\\
       $S_{5}$       &       &       &  &       $\phantom{-}1\phantom{.00}$&       $0.01$&       $-0.07$&       $0.00$&       $-0.11$\\
       \AFB       &       &       &       &  &       $\phantom{-}1\phantom{.00}$&       $0.05$&       $0.06$&       $-0.16$\\
       $S_{7}$       &       &       &       &       &  &       $\phantom{-}1\phantom{.00}$&       $0.26$&       $-0.14$\\
       $S_{8}$       &       &       &       &       &       &  &       $\phantom{-}1\phantom{.00}$&       $-0.09$\\
       $S_{9}$       &       &       &       &       &       &       &  &       $\phantom{-}1\phantom{.00}$\\
\end{tabular}
\end{table} 

\begin{table}[!t]
\caption{
Correlation matrix for the \CP-averaged observables \FL, \AFB and $S_{3} \text{--} S_{9}$ from the maximum-likelihood fit in the interval $4.0<\qsq<6.0\gevgevcccc$. 
\label{tab:correlationsSBin3}
}
\centering 
\begin{tabular}{l|rrrrrrrr} 
&\multicolumn{1}{c}{\FL}&\multicolumn{1}{c}{$S_{3}$}&\multicolumn{1}{c}{$S_{4}$}&\multicolumn{1}{c}{$S_{5}$}&\multicolumn{1}{c}{\AFB}&\multicolumn{1}{c}{$S_{7}$}&\multicolumn{1}{c}{$S_{8}$}&\multicolumn{1}{c}{$S_{9}$}\\
\hline
       \FL  &       $\phantom{-}1\phantom{.00}$&       $0.20$&       $-0.09$&       $-0.09$&       $0.07$&       $0.01$&       $0.16$&       $-0.03$\\
       $S_{3}$       &  &       $\phantom{-}1\phantom{.00}$&       $-0.08$&       $-0.10$&       $0.03$&       $0.11$&       $0.17$&       $0.03$\\
       $S_{4}$       &       &  &       $\phantom{-}1\phantom{.00}$&       $-0.08$&       $-0.15$&       $0.07$&       $-0.04$&       $0.05$\\
       $S_{5}$       &       &       &  &       $\phantom{-}1\phantom{.00}$&       $-0.17$&       $-0.02$&       $0.09$&       $-0.02$\\
       \AFB       &       &       &       &  &       $\phantom{-}1\phantom{.00}$&       $-0.04$&       $-0.03$&       $-0.01$\\
       $S_{7}$       &       &       &       &       &  &       $\phantom{-}1\phantom{.00}$&       $0.09$&       $0.09$\\
       $S_{8}$       &       &       &       &       &       &  &       $\phantom{-}1\phantom{.00}$&       $-0.08$\\
       $S_{9}$       &       &       &       &       &       &       &  &       $\phantom{-}1\phantom{.00}$\\
\end{tabular}
\end{table} 

\begin{table}[!t]
\caption{
Correlation matrix for the \CP-averaged observables \FL, \AFB and $S_{3} \text{--} S_{9}$ from the maximum-likelihood fit in the interval $6.0<\qsq<8.0\gevgevcccc$. 
\label{tab:correlationsSBin4}
}
\centering 
\begin{tabular}{l|rrrrrrrr} 
&\multicolumn{1}{c}{\FL}&\multicolumn{1}{c}{$S_{3}$}&\multicolumn{1}{c}{$S_{4}$}&\multicolumn{1}{c}{$S_{5}$}&\multicolumn{1}{c}{\AFB}&\multicolumn{1}{c}{$S_{7}$}&\multicolumn{1}{c}{$S_{8}$}&\multicolumn{1}{c}{$S_{9}$}\\
\hline
       \FL  &       $\phantom{-}1\phantom{.00}$&       $0.26$&       $-0.01$&       $0.07$&       $0.01$&       $-0.04$&       $0.06$&       $0.05$\\
       $S_{3}$       &  &       $\phantom{-}1\phantom{.00}$&       $0.01$&       $-0.03$&       $-0.05$&       $0.08$&       $-0.04$&       $-0.00$\\
       $S_{4}$       &       &  &       $\phantom{-}1\phantom{.00}$&       $0.35$&       $0.02$&       $-0.05$&       $-0.03$&       $-0.10$\\
       $S_{5}$       &       &       &  &       $\phantom{-}1\phantom{.00}$&       $0.02$&       $-0.11$&       $-0.07$&       $-0.17$\\
       \AFB       &       &       &       &  &       $\phantom{-}1\phantom{.00}$&       $-0.05$&       $-0.19$&       $-0.13$\\
       $S_{7}$       &       &       &       &       &  &       $\phantom{-}1\phantom{.00}$&       $-0.10$&       $-0.06$\\
       $S_{8}$       &       &       &       &       &       &  &       $\phantom{-}1\phantom{.00}$&       $0.04$\\
       $S_{9}$       &       &       &       &       &       &       &  &       $\phantom{-}1\phantom{.00}$\\
\end{tabular}
\end{table} 

\begin{table}[!t]
\caption{
Correlation matrix for the \CP-averaged observables \FL, \AFB and $S_{3} \text{--} S_{9}$ from the maximum-likelihood fit in the interval $11.0<\qsq<12.5\gevgevcccc$. 
\label{tab:correlationsSBin5}
}
\centering 
\begin{tabular}{l|rrrrrrrr} 
&\multicolumn{1}{c}{\FL}&\multicolumn{1}{c}{$S_{3}$}&\multicolumn{1}{c}{$S_{4}$}&\multicolumn{1}{c}{$S_{5}$}&\multicolumn{1}{c}{\AFB}&\multicolumn{1}{c}{$S_{7}$}&\multicolumn{1}{c}{$S_{8}$}&\multicolumn{1}{c}{$S_{9}$}\\
\hline
       \FL  &       $\phantom{-}1\phantom{.00}$&       $0.09$&       $0.03$&       $0.09$&       $-0.44$&       $-0.09$&       $-0.13$&       $-0.08$\\
       $S_{3}$       &  &       $\phantom{-}1\phantom{.00}$&       $-0.08$&       $-0.13$&       $-0.08$&       $-0.04$&       $-0.04$&       $-0.19$\\
       $S_{4}$       &       &  &       $\phantom{-}1\phantom{.00}$&       $0.08$&       $0.06$&       $-0.05$&       $-0.09$&       $0.12$\\
       $S_{5}$       &       &       &  &       $\phantom{-}1\phantom{.00}$&       $-0.30$&       $0.05$&       $-0.04$&       $-0.10$\\
       \AFB       &       &       &       &  &       $\phantom{-}1\phantom{.00}$&       $0.10$&       $0.11$&       $0.15$\\
       $S_{7}$       &       &       &       &       &  &       $\phantom{-}1\phantom{.00}$&       $0.05$&       $-0.07$\\
       $S_{8}$       &       &       &       &       &       &  &       $\phantom{-}1\phantom{.00}$&       $-0.07$\\
       $S_{9}$       &       &       &       &       &       &       &  &       $\phantom{-}1\phantom{.00}$\\
\end{tabular}
\end{table} 

\begin{table}[!t]
\caption{
Correlation matrix for the \CP-averaged observables \FL, \AFB and $S_{3} \text{--} S_{9}$ from the maximum-likelihood fit in the interval $15.0<\qsq<17.0\gevgevcccc$. 
\label{tab:correlationsSBin6}
}
\centering 
\begin{tabular}{l|rrrrrrrr} 
&\multicolumn{1}{c}{\FL}&\multicolumn{1}{c}{$S_{3}$}&\multicolumn{1}{c}{$S_{4}$}&\multicolumn{1}{c}{$S_{5}$}&\multicolumn{1}{c}{\AFB}&\multicolumn{1}{c}{$S_{7}$}&\multicolumn{1}{c}{$S_{8}$}&\multicolumn{1}{c}{$S_{9}$}\\
\hline
       \FL  &       $\phantom{-}1\phantom{.00}$&       $0.19$&       $0.04$&       $0.07$&       $-0.28$&       $-0.06$&       $-0.13$&       $-0.07$\\
       $S_{3}$       &  &       $\phantom{-}1\phantom{.00}$&       $-0.09$&       $-0.06$&       $0.04$&       $0.01$&       $-0.06$&       $0.01$\\
       $S_{4}$       &       &  &       $\phantom{-}1\phantom{.00}$&       $0.27$&       $0.07$&       $0.10$&       $0.06$&       $0.14$\\
       $S_{5}$       &       &       &  &       $\phantom{-}1\phantom{.00}$&       $-0.15$&       $0.09$&       $-0.06$&       $-0.13$\\
       \AFB       &       &       &       &  &       $\phantom{-}1\phantom{.00}$&       $0.07$&       $-0.02$&       $0.16$\\
       $S_{7}$       &       &       &       &       &  &       $\phantom{-}1\phantom{.00}$&       $0.23$&       $0.02$\\
       $S_{8}$       &       &       &       &       &       &  &       $\phantom{-}1\phantom{.00}$&       $0.00$\\
       $S_{9}$       &       &       &       &       &       &       &  &       $\phantom{-}1\phantom{.00}$\\
\end{tabular}
\end{table}

\begin{table}[!t]
\caption{
Correlation matrix for the \CP-averaged observables \FL, \AFB and $S_{3} \text{--} S_{9}$ from the maximum-likelihood fit in the interval $17.0<\qsq<19.0\gevgevcccc$. 
\label{tab:correlationsSBin7}
}
\centering 
\begin{tabular}{l|rrrrrrrr} 
&\multicolumn{1}{c}{\FL}&\multicolumn{1}{c}{$S_{3}$}&\multicolumn{1}{c}{$S_{4}$}&\multicolumn{1}{c}{$S_{5}$}&\multicolumn{1}{c}{\AFB}&\multicolumn{1}{c}{$S_{7}$}&\multicolumn{1}{c}{$S_{8}$}&\multicolumn{1}{c}{$S_{9}$}\\
\hline
       \FL  &       $\phantom{-}1\phantom{.00}$&       $-0.10$&       $-0.02$&       $-0.10$&       $-0.10$&       $0.07$&       $-0.10$&       $-0.01$\\
       $S_{3}$       &  &       $\phantom{-}1\phantom{.00}$&       $-0.08$&       $0.06$&       $0.09$&       $0.01$&       $0.08$&       $-0.02$\\
       $S_{4}$       &       &  &       $\phantom{-}1\phantom{.00}$&       $-0.06$&       $-0.07$&       $0.00$&       $0.06$&       $0.04$\\
       $S_{5}$       &       &       &  &       $\phantom{-}1\phantom{.00}$&       $-0.19$&       $-0.03$&       $0.09$&       $0.02$\\
       \AFB       &       &       &       &  &       $\phantom{-}1\phantom{.00}$&       $0.17$&       $0.01$&       $-0.07$\\
       $S_{7}$       &       &       &       &       &  &       $\phantom{-}1\phantom{.00}$&       $-0.17$&       $0.10$\\
       $S_{8}$       &       &       &       &       &       &  &       $\phantom{-}1\phantom{.00}$&       $-0.19$\\
       $S_{9}$       &       &       &       &       &       &       &  &       $\phantom{-}1\phantom{.00}$\\
\end{tabular}
\end{table} 

\begin{table}[!t]
\caption{
Correlation matrix for the \CP-averaged observables \FL, \AFB and $S_{3} \text{--} S_{9}$ from the maximum-likelihood fit in the interval $1.1<\qsq<6.0\gevgevcccc$. 
\label{tab:correlationsSBinA7}
}
\centering 
\begin{tabular}{l|rrrrrrrr} 
&\multicolumn{1}{c}{\FL}&\multicolumn{1}{c}{$S_{3}$}&\multicolumn{1}{c}{$S_{4}$}&\multicolumn{1}{c}{$S_{5}$}&\multicolumn{1}{c}{\AFB}&\multicolumn{1}{c}{$S_{7}$}&\multicolumn{1}{c}{$S_{8}$}&\multicolumn{1}{c}{$S_{9}$}\\
\hline
       \FL  &       $\phantom{-}1\phantom{.00}$&       $0.17$&       $-0.00$&       $-0.02$&       $0.01$&       $0.04$&       $0.08$&       $0.06$\\
       $S_{3}$       &  &       $\phantom{-}1\phantom{.00}$&       $-0.01$&       $-0.02$&       $-0.02$&       $0.04$&       $-0.03$&       $-0.05$\\
       $S_{4}$       &       &  &       $\phantom{-}1\phantom{.00}$&       $-0.03$&       $0.06$&       $-0.02$&       $0.19$&       $-0.01$\\
       $S_{5}$       &       &       &  &       $\phantom{-}1\phantom{.00}$&       $0.01$&       $0.14$&       $0.04$&       $0.04$\\
       \AFB       &       &       &       &  &       $\phantom{-}1\phantom{.00}$&       $-0.05$&       $0.04$&       $0.05$\\
       $S_{7}$       &       &       &       &       &  &       $\phantom{-}1\phantom{.00}$&       $0.17$&       $-0.02$\\
       $S_{8}$       &       &       &       &       &       &  &       $\phantom{-}1\phantom{.00}$&       $-0.01$\\
       $S_{9}$       &       &       &       &       &       &       &  &       $\phantom{-}1\phantom{.00}$\\
\end{tabular}
\end{table} 

\begin{table}[!t]
\caption{
Correlation matrix for the \CP-averaged observables \FL, \AFB and $S_{3} \text{--} S_{9}$ from the maximum-likelihood fit in the interval $15.0<\qsq<19.0\gevgevcccc$. 
\label{tab:correlationsSBinB}
}
\centering 
\begin{tabular}{l|rrrrrrrr} 
&\multicolumn{1}{c}{\FL}&\multicolumn{1}{c}{$S_{3}$}&\multicolumn{1}{c}{$S_{4}$}&\multicolumn{1}{c}{$S_{5}$}&\multicolumn{1}{c}{\AFB}&\multicolumn{1}{c}{$S_{7}$}&\multicolumn{1}{c}{$S_{8}$}&\multicolumn{1}{c}{$S_{9}$}\\
\hline
       \FL  &       $\phantom{-}1\phantom{.00}$&       $0.13$&       $-0.05$&       $-0.02$&       $-0.17$&       $-0.02$&       $0.03$&       $-0.05$\\
       $S_{3}$       &  &       $\phantom{-}1\phantom{.00}$&       $-0.07$&       $-0.00$&       $-0.02$&       $0.12$&       $0.10$&       $-0.05$\\
       $S_{4}$       &       &  &       $\phantom{-}1\phantom{.00}$&       $0.05$&       $-0.14$&       $0.06$&       $0.05$&       $-0.02$\\
       $S_{5}$       &       &       &  &       $\phantom{-}1\phantom{.00}$&       $0.05$&       $-0.07$&       $0.07$&       $-0.05$\\
       \AFB       &       &       &       &  &       $\phantom{-}1\phantom{.00}$&       $-0.10$&       $-0.03$&       $0.10$\\
       $S_{7}$       &       &       &       &       &  &       $\phantom{-}1\phantom{.00}$&       $0.15$&       $-0.01$\\
       $S_{8}$       &       &       &       &       &       &  &       $\phantom{-}1\phantom{.00}$&       $-0.09$\\
       $S_{9}$       &       &       &       &       &       &       &  &       $\phantom{-}1\phantom{.00}$\\
\end{tabular}
\end{table}

\clearpage
\section{Correlation matrices for the \boldmath{$P_{i}^{(\prime)}$} observables}\label{sec:correlationsP}

Correlation matrices between the \CP-averaged observables $P_i$ in the different \qsq intervals are provided in Tables~\ref{tab:correlationsPBin0}--\ref{tab:correlationsPBinB}. Correlations between observables measured with different folds are obtained using using the bootstrapping technique~\cite{efron:1979}. The different $\qsq$ intervals are statistically independent.

\begin{table}[!ht]
\caption{
Correlation matrix for the \CP-averaged observables \FL and $P_{i}^{(\prime)}$ from the maximum-likelihood fit in the interval $0.10<\qsq<0.98\gevgevcccc$.
\label{tab:correlationsPBin0}
}
\centering 
\begin{tabular}{l|rrrrrrrr}
&\multicolumn{1}{c}{\FL}&\multicolumn{1}{c}{$P_{1}$}&\multicolumn{1}{c}{$P_{2}$}&\multicolumn{1}{c}{$P_{3}$}&\multicolumn{1}{c}{$P_{4}^{\prime}$}&\multicolumn{1}{c}{$P_{5}^{\prime}$}&\multicolumn{1}{c}{$P_{6}^{\prime}$}&\multicolumn{1}{c}{$P_{8}^{\prime}$}\\
\hline
       \FL  &       $\phantom{-}1\phantom{.00}$&       $-0.14$&       $0.02$&       $-0.18$&       $-0.03$&       $0.00$&       $0.12$&       $-0.01$\\
       $P_{1}$       &  &       $\phantom{-}1\phantom{.00}$&       $-0.00$&       $-0.01$&       $-0.03$&       $0.17$&       $0.01$&       $0.03$\\
       $P_{2}$       &       &  &       $\phantom{-}1\phantom{.00}$&       $0.02$&       $-0.08$&       $0.09$&       $0.19$&       $-0.24$\\
       $P_{3}$       &       &       &  &       $\phantom{-}1\phantom{.00}$&       $0.06$&       $-0.03$&       $-0.21$&       $0.04$\\
       $P_{4}^{\prime}$       &       &       &       &  &       $\phantom{-}1\phantom{.00}$&       $-0.22$&       $-0.23$&       $0.15$\\
       $P_{5}^{\prime}$       &       &       &       &       &  &       $\phantom{-}1\phantom{.00}$&       $0.18$&       $-0.18$\\
       $P_{6}^{\prime}$       &       &       &       &       &       &  &       $\phantom{-}1\phantom{.00}$&       $-0.25$\\
       $P_{8}^{\prime}$       &       &       &       &       &       &       &  &       $\phantom{-}1\phantom{.00}$\\
\end{tabular}
\end{table} 
\begin{table}[!ht]
\caption{
Correlation matrix for the \CP-averaged observables \FL and $P_{i}^{(\prime)}$ from the maximum-likelihood fit in the interval $1.1<\qsq<2.5\gevgevcccc$.
\label{tab:correlationsPBin1}
}
\centering 
\begin{tabular}{l|rrrrrrrr}
&\multicolumn{1}{c}{\FL}&\multicolumn{1}{c}{$P_{1}$}&\multicolumn{1}{c}{$P_{2}$}&\multicolumn{1}{c}{$P_{3}$}&\multicolumn{1}{c}{$P_{4}^{\prime}$}&\multicolumn{1}{c}{$P_{5}^{\prime}$}&\multicolumn{1}{c}{$P_{6}^{\prime}$}&\multicolumn{1}{c}{$P_{8}^{\prime}$}\\
\hline
       \FL  &       $\phantom{-}1\phantom{.00}$&       $0.03$&       $0.02$&       $-0.01$&       $-0.06$&       $-0.01$&       $0.05$&       $-0.08$\\
       $P_{1}$       &  &       $\phantom{-}1\phantom{.00}$&       $-0.05$&       $-0.01$&       $0.06$&       $-0.09$&       $-0.03$&       $0.04$\\
       $P_{2}$       &       &  &       $\phantom{-}1\phantom{.00}$&       $-0.05$&       $0.15$&       $0.12$&       $0.10$&       $0.13$\\
       $P_{3}$       &       &       &  &       $\phantom{-}1\phantom{.00}$&       $-0.08$&       $-0.07$&       $-0.02$&       $-0.13$\\
       $P_{4}^{\prime}$       &       &       &       &  &       $\phantom{-}1\phantom{.00}$&       $0.03$&       $-0.01$&       $0.22$\\
       $P_{5}^{\prime}$       &       &       &       &       &  &       $\phantom{-}1\phantom{.00}$&       $0.09$&       $-0.08$\\
       $P_{6}^{\prime}$       &       &       &       &       &       &  &       $\phantom{-}1\phantom{.00}$&       $-0.01$\\
       $P_{8}^{\prime}$       &       &       &       &       &       &       &  &       $\phantom{-}1\phantom{.00}$\\
\end{tabular}
\end{table} 
\begin{table}[!ht]
\caption{
Correlation matrix for the \CP-averaged observables \FL and $P_{i}^{(\prime)}$ from the maximum-likelihood fit in the interval $2.5<\qsq<4.0\gevgevcccc$.
\label{tab:correlationsPBin2}
}
\centering 
\begin{tabular}{l|rrrrrrrr}
&\multicolumn{1}{c}{\FL}&\multicolumn{1}{c}{$P_{1}$}&\multicolumn{1}{c}{$P_{2}$}&\multicolumn{1}{c}{$P_{3}$}&\multicolumn{1}{c}{$P_{4}^{\prime}$}&\multicolumn{1}{c}{$P_{5}^{\prime}$}&\multicolumn{1}{c}{$P_{6}^{\prime}$}&\multicolumn{1}{c}{$P_{8}^{\prime}$}\\
\hline
       \FL  &       $\phantom{-}1\phantom{.00}$&       $0.00$&       $0.00$&       $0.02$&       $-0.02$&       $-0.03$&       $-0.06$&       $-0.03$\\
       $P_{1}$       &  &       $\phantom{-}1\phantom{.00}$&       $0.00$&       $0.04$&       $0.04$&       $0.00$&       $-0.04$&       $-0.06$\\
       $P_{2}$       &       &  &       $\phantom{-}1\phantom{.00}$&       $0.07$&       $-0.01$&       $0.04$&       $0.04$&       $-0.03$\\
       $P_{3}$       &       &       &  &       $\phantom{-}1\phantom{.00}$&       $-0.03$&       $0.02$&       $0.06$&       $-0.01$\\
       $P_{4}^{\prime}$       &       &       &       &  &       $\phantom{-}1\phantom{.00}$&       $0.07$&       $0.06$&       $0.08$\\
       $P_{5}^{\prime}$       &       &       &       &       &  &       $\phantom{-}1\phantom{.00}$&       $-0.02$&       $-0.09$\\
       $P_{6}^{\prime}$       &       &       &       &       &       &  &       $\phantom{-}1\phantom{.00}$&       $0.21$\\
       $P_{8}^{\prime}$       &       &       &       &       &       &       &  &       $\phantom{-}1\phantom{.00}$\\
\end{tabular}
\end{table} 
\begin{table}[!t]
\caption{
Correlation matrix for the \CP-averaged observables \FL and $P_{i}^{(\prime)}$ from the maximum-likelihood fit in the interval $4.0<\qsq<6.0\gevgevcccc$.
\label{tab:correlationsPBin3}
}
\centering 
\begin{tabular}{l|rrrrrrrr}
&\multicolumn{1}{c}{\FL}&\multicolumn{1}{c}{$P_{1}$}&\multicolumn{1}{c}{$P_{2}$}&\multicolumn{1}{c}{$P_{3}$}&\multicolumn{1}{c}{$P_{4}^{\prime}$}&\multicolumn{1}{c}{$P_{5}^{\prime}$}&\multicolumn{1}{c}{$P_{6}^{\prime}$}&\multicolumn{1}{c}{$P_{8}^{\prime}$}\\
\hline
       \FL  &       $\phantom{-}1\phantom{.00}$&       $0.16$&       $-0.10$&       $0.02$&       $-0.02$&       $-0.08$&       $0.02$&       $0.08$\\
       $P_{1}$       &  &       $\phantom{-}1\phantom{.00}$&       $-0.03$&       $0.02$&       $-0.08$&       $0.03$&       $0.03$&       $0.08$\\
       $P_{2}$       &       &  &       $\phantom{-}1\phantom{.00}$&       $0.04$&       $-0.12$&       $-0.14$&       $-0.03$&       $-0.05$\\
       $P_{3}$       &       &       &  &       $\phantom{-}1\phantom{.00}$&       $-0.02$&       $-0.02$&       $-0.05$&       $0.09$\\
       $P_{4}^{\prime}$       &       &       &       &  &       $\phantom{-}1\phantom{.00}$&       $-0.11$&       $-0.01$&       $-0.10$\\
       $P_{5}^{\prime}$       &       &       &       &       &  &       $\phantom{-}1\phantom{.00}$&       $-0.04$&       $0.07$\\
       $P_{6}^{\prime}$       &       &       &       &       &       &  &       $\phantom{-}1\phantom{.00}$&       $0.05$\\
       $P_{8}^{\prime}$       &       &       &       &       &       &       &  &       $\phantom{-}1\phantom{.00}$\\
\end{tabular}
\end{table} 
\begin{table}[!t]
\caption{
Correlation matrix for the \CP-averaged observables \FL and $P_{i}^{(\prime)}$ from the maximum-likelihood fit in the interval $6.0<\qsq<8.0\gevgevcccc$.
\label{tab:correlationsPBin4}
}
\centering 
\begin{tabular}{l|rrrrrrrr}
&\multicolumn{1}{c}{\FL}&\multicolumn{1}{c}{$P_{1}$}&\multicolumn{1}{c}{$P_{2}$}&\multicolumn{1}{c}{$P_{3}$}&\multicolumn{1}{c}{$P_{4}^{\prime}$}&\multicolumn{1}{c}{$P_{5}^{\prime}$}&\multicolumn{1}{c}{$P_{6}^{\prime}$}&\multicolumn{1}{c}{$P_{8}^{\prime}$}\\
\hline
       \FL  &       $\phantom{-}1\phantom{.00}$&       $0.11$&       $-0.10$&       $0.01$&       $-0.03$&       $0.05$&       $-0.05$&       $-0.00$\\
       $P_{1}$       &  &       $\phantom{-}1\phantom{.00}$&       $-0.04$&       $0.01$&       $0.01$&       $-0.02$&       $0.02$&       $-0.06$\\
       $P_{2}$       &       &  &       $\phantom{-}1\phantom{.00}$&       $0.12$&       $-0.01$&       $0.02$&       $-0.05$&       $-0.17$\\
       $P_{3}$       &       &       &  &       $\phantom{-}1\phantom{.00}$&       $0.04$&       $0.12$&       $0.00$&       $-0.04$\\
       $P_{4}^{\prime}$       &       &       &       &  &       $\phantom{-}1\phantom{.00}$&       $0.25$&       $-0.03$&       $-0.01$\\
       $P_{5}^{\prime}$       &       &       &       &       &  &       $\phantom{-}1\phantom{.00}$&       $-0.08$&       $-0.06$\\
       $P_{6}^{\prime}$       &       &       &       &       &       &  &       $\phantom{-}1\phantom{.00}$&       $-0.05$\\
       $P_{8}^{\prime}$       &       &       &       &       &       &       &  &       $\phantom{-}1\phantom{.00}$\\
\end{tabular}
\end{table} 
\begin{table}[!t]
\caption{
Correlation matrix for the \CP-averaged observables \FL and $P_{i}^{(\prime)}$ from the maximum-likelihood fit in the interval $11.0<\qsq<12.5\gevgevcccc$.
\label{tab:correlationsPBin5}
}
\centering 
\begin{tabular}{l|rrrrrrrr}
&\multicolumn{1}{c}{\FL}&\multicolumn{1}{c}{$P_{1}$}&\multicolumn{1}{c}{$P_{2}$}&\multicolumn{1}{c}{$P_{3}$}&\multicolumn{1}{c}{$P_{4}^{\prime}$}&\multicolumn{1}{c}{$P_{5}^{\prime}$}&\multicolumn{1}{c}{$P_{6}^{\prime}$}&\multicolumn{1}{c}{$P_{8}^{\prime}$}\\
\hline
       \FL  &       $\phantom{-}1\phantom{.00}$&       $-0.05$&       $0.35$&       $-0.09$&       $0.00$&       $0.04$&       $-0.06$&       $-0.14$\\
       $P_{1}$       &  &       $\phantom{-}1\phantom{.00}$&       $-0.05$&       $0.17$&       $-0.09$&       $-0.14$&       $-0.03$&       $-0.02$\\
       $P_{2}$       &       &  &       $\phantom{-}1\phantom{.00}$&       $-0.15$&       $0.12$&       $-0.14$&       $-0.00$&       $0.07$\\
       $P_{3}$       &       &       &  &       $\phantom{-}1\phantom{.00}$&       $-0.09$&       $0.06$&       $0.07$&       $0.09$\\
       $P_{4}^{\prime}$       &       &       &       &  &       $\phantom{-}1\phantom{.00}$&       $0.04$&       $-0.03$&       $-0.10$\\
       $P_{5}^{\prime}$       &       &       &       &       &  &       $\phantom{-}1\phantom{.00}$&       $0.05$&       $-0.01$\\
       $P_{6}^{\prime}$       &       &       &       &       &       &  &       $\phantom{-}1\phantom{.00}$&       $0.06$\\
       $P_{8}^{\prime}$       &       &       &       &       &       &       &  &       $\phantom{-}1\phantom{.00}$\\
\end{tabular}
\end{table} 
\begin{table}[!t]
\caption{
Correlation matrix for the \CP-averaged observables \FL and $P_{i}^{(\prime)}$ from the maximum-likelihood fit in the interval $15.0<\qsq<17.0\gevgevcccc$.
\label{tab:correlationsPBin6}
}
\centering 
\begin{tabular}{l|rrrrrrrr}
&\multicolumn{1}{c}{\FL}&\multicolumn{1}{c}{$P_{1}$}&\multicolumn{1}{c}{$P_{2}$}&\multicolumn{1}{c}{$P_{3}$}&\multicolumn{1}{c}{$P_{4}^{\prime}$}&\multicolumn{1}{c}{$P_{5}^{\prime}$}&\multicolumn{1}{c}{$P_{6}^{\prime}$}&\multicolumn{1}{c}{$P_{8}^{\prime}$}\\
\hline
       \FL  &       $\phantom{-}1\phantom{.00}$&       $0.07$&       $0.15$&       $-0.09$&       $0.08$&       $0.09$&       $0.00$&       $-0.09$\\
       $P_{1}$       &  &       $\phantom{-}1\phantom{.00}$&       $0.01$&       $-0.05$&       $0.00$&       $-0.01$&       $-0.00$&       $-0.06$\\
       $P_{2}$       &       &  &       $\phantom{-}1\phantom{.00}$&       $-0.23$&       $0.10$&       $-0.06$&       $0.07$&       $-0.03$\\
       $P_{3}$       &       &       &  &       $\phantom{-}1\phantom{.00}$&       $-0.15$&       $0.10$&       $-0.03$&       $0.02$\\
       $P_{4}^{\prime}$       &       &       &       &  &       $\phantom{-}1\phantom{.00}$&       $0.27$&       $0.09$&       $0.05$\\
       $P_{5}^{\prime}$       &       &       &       &       &  &       $\phantom{-}1\phantom{.00}$&       $0.09$&       $-0.07$\\
       $P_{6}^{\prime}$       &       &       &       &       &       &  &       $\phantom{-}1\phantom{.00}$&       $0.21$\\
       $P_{8}^{\prime}$       &       &       &       &       &       &       &  &       $\phantom{-}1\phantom{.00}$\\
\end{tabular}
\end{table} 
\begin{table}[!t]
\caption{
Correlation matrix for the \CP-averaged observables \FL and $P_{i}^{(\prime)}$ from the maximum-likelihood fit in the interval $17.0<\qsq<19.0\gevgevcccc$.
\label{tab:correlationsPBin7}
}
\centering 
\begin{tabular}{l|rrrrrrrr}
&\multicolumn{1}{c}{\FL}&\multicolumn{1}{c}{$P_{1}$}&\multicolumn{1}{c}{$P_{2}$}&\multicolumn{1}{c}{$P_{3}$}&\multicolumn{1}{c}{$P_{4}^{\prime}$}&\multicolumn{1}{c}{$P_{5}^{\prime}$}&\multicolumn{1}{c}{$P_{6}^{\prime}$}&\multicolumn{1}{c}{$P_{8}^{\prime}$}\\
\hline
       \FL  &       $\phantom{-}1\phantom{.00}$&       $-0.10$&       $0.09$&       $0.07$&       $0.02$&       $-0.10$&       $0.06$&       $-0.08$\\
       $P_{1}$       &  &       $\phantom{-}1\phantom{.00}$&       $0.06$&       $0.04$&       $-0.10$&       $0.02$&       $-0.01$&       $0.06$\\
       $P_{2}$       &       &  &       $\phantom{-}1\phantom{.00}$&       $0.07$&       $-0.07$&       $-0.16$&       $0.13$&       $-0.00$\\
       $P_{3}$       &       &       &  &       $\phantom{-}1\phantom{.00}$&       $-0.08$&       $0.03$&       $-0.08$&       $0.17$\\
       $P_{4}^{\prime}$       &       &       &       &  &       $\phantom{-}1\phantom{.00}$&       $-0.08$&       $-0.03$&       $0.05$\\
       $P_{5}^{\prime}$       &       &       &       &       &  &       $\phantom{-}1\phantom{.00}$&       $0.00$&       $0.08$\\
       $P_{6}^{\prime}$       &       &       &       &       &       &  &       $\phantom{-}1\phantom{.00}$&       $-0.12$\\
       $P_{8}^{\prime}$       &       &       &       &       &       &       &  &       $\phantom{-}1\phantom{.00}$\\
\end{tabular}
\end{table} 
\begin{table}[!t]
\caption{
Correlation matrix for the \CP-averaged observables \FL and $P_{i}^{(\prime)}$ from the maximum-likelihood fit in the interval $1.1<\qsq<6.0\gevgevcccc$.
\label{tab:correlationsPBinA}
}
\centering 
\begin{tabular}{l|rrrrrrrr}
&\multicolumn{1}{c}{\FL}&\multicolumn{1}{c}{$P_{1}$}&\multicolumn{1}{c}{$P_{2}$}&\multicolumn{1}{c}{$P_{3}$}&\multicolumn{1}{c}{$P_{4}^{\prime}$}&\multicolumn{1}{c}{$P_{5}^{\prime}$}&\multicolumn{1}{c}{$P_{6}^{\prime}$}&\multicolumn{1}{c}{$P_{8}^{\prime}$}\\
\hline
       \FL  &       $\phantom{-}1\phantom{.00}$&       $0.11$&       $-0.19$&       $0.01$&       $-0.01$&       $-0.02$&       $0.02$&       $0.08$\\
       $P_{1}$       &  &       $\phantom{-}1\phantom{.00}$&       $-0.05$&       $0.07$&       $0.01$&       $-0.01$&       $0.00$&       $-0.04$\\
       $P_{2}$       &       &  &       $\phantom{-}1\phantom{.00}$&       $-0.06$&       $0.04$&       $0.00$&       $-0.05$&       $0.01$\\
       $P_{3}$       &       &       &  &       $\phantom{-}1\phantom{.00}$&       $0.01$&       $-0.04$&       $0.01$&       $0.01$\\
       $P_{4}^{\prime}$       &       &       &       &  &       $\phantom{-}1\phantom{.00}$&       $-0.03$&       $-0.02$&       $0.18$\\
       $P_{5}^{\prime}$       &       &       &       &       &  &       $\phantom{-}1\phantom{.00}$&       $0.14$&       $0.04$\\
       $P_{6}^{\prime}$       &       &       &       &       &       &  &       $\phantom{-}1\phantom{.00}$&       $0.17$\\
       $P_{8}^{\prime}$       &       &       &       &       &       &       &  &       $\phantom{-}1\phantom{.00}$\\

\end{tabular}
\end{table} 
\begin{table}[!t]
\caption{
Correlation matrix for the \CP-averaged observables \FL and $P_{i}^{(\prime)}$ from the maximum-likelihood fit in the interval $15.0<\qsq<19.0\gevgevcccc$.
\label{tab:correlationsPBinB}
}
\centering 
\begin{tabular}{l|rrrrrrrr}
&\multicolumn{1}{c}{\FL}&\multicolumn{1}{c}{$P_{1}$}&\multicolumn{1}{c}{$P_{2}$}&\multicolumn{1}{c}{$P_{3}$}&\multicolumn{1}{c}{$P_{4}^{\prime}$}&\multicolumn{1}{c}{$P_{5}^{\prime}$}&\multicolumn{1}{c}{$P_{6}^{\prime}$}&\multicolumn{1}{c}{$P_{8}^{\prime}$}\\
\hline
       \FL  &       $\phantom{-}1\phantom{.00}$&       $-0.00$&       $0.03$&       $-0.01$&       $0.00$&       $0.01$&       $0.03$&       $0.05$\\
       $P_{1}$       &  &       $\phantom{-}1\phantom{.00}$&       $0.01$&       $0.04$&       $-0.06$&       $0.04$&       $0.03$&       $0.08$\\
       $P_{2}$       &       &  &       $\phantom{-}1\phantom{.00}$&       $-0.07$&       $-0.13$&       $-0.00$&       $-0.12$&       $-0.03$\\
       $P_{3}$       &       &       &  &       $\phantom{-}1\phantom{.00}$&       $0.03$&       $0.04$&       $0.02$&       $0.08$\\
       $P_{4}^{\prime}$       &       &       &       &  &       $\phantom{-}1\phantom{.00}$&       $-0.00$&       $0.12$&       $0.04$\\
       $P_{5}^{\prime}$       &       &       &       &       &  &       $\phantom{-}1\phantom{.00}$&       $-0.09$&       $0.07$\\
       $P_{6}^{\prime}$       &       &       &       &       &       &  &       $\phantom{-}1\phantom{.00}$&       $0.17$\\
       $P_{8}^{\prime}$       &       &       &       &       &       &       &  &       $\phantom{-}1\phantom{.00}$\\
\end{tabular}
\end{table}

\clearpage
\section{Systematic uncertainties}\label{sec:systematics}

The systematic uncertainties are determined for each observable in each \qsq interval. Table~\ref{tab:systematics} summarises the sizes of the systematic effects by giving the maximum value for each systematic uncertainty studied.

The larger systematic uncertainties of the $P_{i}^{(\prime)}$ observables compared to the $S_{i}$ observables arise due to an additional scale factor in the definition of the $P_{i}^{(\prime)}$ observables, which depends on the value of \FL for a given \qsq interval.

\begin{table}[h]
\caption{Maximum values for each source of systematic uncertainty.}
\begin{center}
\begin{tabular}{r|ccccc}
Source &\makebox[1.4cm]{\FL} &\makebox[1.4cm]{\AFB} &\makebox[1.4cm]{$S_{3}\text{--}S_9$} & \makebox[1.4cm]{$P_1$} &\makebox[1.4cm]{$P_2\text{--}P_8^\prime$} \\ 
\hline 
Size of the simulation sample &$<0.03$  &$<0.03$ &$<0.04$ &$<0.06$ &$<0.08$\\
Data-simulation differences   &$<0.04$  &$<0.01$ &$<0.04$ &$<0.13$ &$<0.17$\\
Acceptance polynomial order   &$<0.05$  &$<0.04$ &$<0.06$ &$<0.09$ &$<0.10$\\
S-wave fraction constraint    &$<0.05$  &$<0.02$ &$<0.03$ &$<0.13$ &$<0.14$\\
$m(\KS\pip\mumu)$ model       &$<0.01$  &$<0.01$ &$<0.01$ &$<0.06$ &$<0.02$\\
Peaking background veto       &$<0.01$  &$<0.01$ &$<0.01$ &$<0.07$ &$<0.04$\\
Angular resolution            &$<0.01$  &$<0.01$ &$<0.01$ &$<0.01$ &$<0.01$\\
Background model              &$<0.01$  &$<0.01$ &$<0.02$ &$<0.06$ &$<0.12$\\
Trigger simulation            &$<0.01$  &$<0.01$ &$<0.01$ &$<0.03$ &$<0.03$\\
Fit bias at best-fit values   &$<0.04$  &$<0.05$ &$<0.05$ &$<0.28$ &$<0.13$\\
\end{tabular}
\end{center}
\label{tab:systematics}
\end{table}

\clearpage
\section{Yields of signal candidates per \lowercase{\boldmath{\qsq}} interval}

\begin{table}[h]
  \caption{
	Yields of signal candidates in the ten \qsq intervals. They are obtained from extended maximum-likelihood fits to the $m(\KS\pip\mumu)$ distribution. The total number corresponds to the sum of the eight nominal \qsq intervals.
	}
	\begin{center}
	\begin{tabular}{c|r}
    \qsq [\gevgevcccc] & Signal yield\\
    \hline
    $\phz[0.1, 0.98]$&$102 \pm 12$\\
    $[1.1, 2.5]$		& $ 49 \pm 10$\\
    $[2.5, 4.0]$		& $ 42 \pm 10$\\
    $[4.0, 6.0]$		& $109 \pm 13$\\
    $[6.0, 8.0]$		& $105 \pm 14$\\
    $[11.0, 12.5]$	& $111 \pm 13$\\
    $[15.0, 17.0]$	& $144 \pm 13$\\
    $[17.0, 19.0]$	& $ 76 \pm 10$\\
    \hline
    $[1.1, 6.0]$	& $200 \pm 19$\\
    $[15.0, 19.0]$	& $220 \pm 17$\\
    \hline
    Total & $737\pm 34$\\
    \end{tabular}
    \label{tab:binyields}
    \end{center}
\end{table}

\clearpage
\section{Projections of data and fit model}

\clearpage %put clearpage above last set of Figs to include bibliography on the same page as Fig8
\clearpage

\addcontentsline{toc}{section}{References}
%\setboolean{inbibliography}{true}
\bibliographystyle{LHCb}
\bibliography{main,standard,LHCb-PAPER,LHCb-CONF,LHCb-DP,LHCb-TDR}
 
\newpage
% LHCb collaboration author list
% Data extracted on October 15th, 2020 at 11:20pm for reference date 13-Oct-2020
\centerline
{\large\bf LHCb collaboration}
\begin
{flushleft}
\small
R.~Aaij$^{32}$,
C.~Abell{\'a}n~Beteta$^{50}$,
T.~Ackernley$^{60}$,
B.~Adeva$^{46}$,
M.~Adinolfi$^{54}$,
H.~Afsharnia$^{9}$,
C.A.~Aidala$^{85}$,
S.~Aiola$^{26}$,
Z.~Ajaltouni$^{9}$,
S.~Akar$^{65}$
% ...

{\footnotesize \it

% ...
$ ^{9}$Universit{\'e} Clermont Auvergne, CNRS/IN2P3, LPC, Clermont-Ferrand, France\\
% ...
$ ^{26}$INFN Sezione di Milano, Milano, Italy\\
% ...
$ ^{32}$Nikhef National Institute for Subatomic Physics, Amsterdam, Netherlands\\
% ...
$ ^{46}$Instituto Galego de F{\'\i}sica de Altas Enerx{\'\i}as (IGFAE), Universidade de Santiago de Compostela, Santiago de Compostela, Spain\\
% ...
$ ^{50}$Physik-Institut, Universit{\"a}t Z{\"u}rich, Z{\"u}rich, Switzerland\\
% ...
$ ^{54}$H.H. Wills Physics Laboratory, University of Bristol, Bristol, United Kingdom\\
% ...
$ ^{60}$Oliver Lodge Laboratory, University of Liverpool, Liverpool, United Kingdom\\
% ...
$ ^{65}$University of Cincinnati, Cincinnati, OH, United States\\
% ...
$ ^{68}$Syracuse University, Syracuse, NY, United States\\
% ...
$ ^{85}$University of Michigan, Ann Arbor, United States, associated to $^{68}$\\
\bigskip
% ...
\medskip
}
\end{flushleft}

\end{document}